%% file: Thesis_LorenzoBongiovanni.tex
\documentclass[12pt,a4paper,english,titlepage,fleqn]{book}

\usepackage{amsmath,amsfonts,amssymb,babel,makeidx,textcomp,slashed,subfigure,comment}
\usepackage{pstricks,pst-grad}
\usepackage{phonetic}
\usepackage{rotating}
\usepackage{slashed}
\usepackage[breaklinks=true]{hyperref}
\usepackage{xr}
\usepackage{ifpdf}
\ifpdf
  \usepackage{epstopdf}
\fi

\definecolor{darkblue}{RGB}{0,0,102}

\def\Id{\hbox{$1\hskip -1.2pt\vrule depth 0pt height 1.6ex width 0.7pt
\vrule depth 0pt height 0.3pt width 0.12em$}}
\newcommand{\Tr}{\mbox{Tr}}

\def\SLN{SL($N,\mathbb{C}$)}
\def\SUN{SU($N$)}
\def\SL3{SL($3,\mathbb{C}$)}
\def\SU3{SU($3$)}
\def\SL2{SL($2,\mathbb{C}$)}
\def\SU2{SU($2$)}
\newcommand{\dd}{\mbox{\hausad}}

\def\SLN{SL($N,\mathbb{C}$)}
\def\SUN{SU($N$)}
\def\SLtree{SL($3,\mathbb{C}$)}
\def\SUtree{SU(3)}
\def\SL2{SL($2,\mathbb{C}$)}
\def\SU2{SU($2$)}
\def\t{$\theta$}
\def\ti{$\theta_I$}
\def\tr{$\theta_R$}
\def\tl{$\theta_L$}
\def\ttwo{$\theta^2$}
\def\rE{$\rho(E)$}
\def\aE{$a(E_i)$}

\def\be{\begin{equation}}
\def\ee{\end{equation}}
\def\ba{\begin{array}{lll}}
\def\ea{\end{array}}
\def\beq{\begin{eqnarray}}
\def\eeq{\end{eqnarray}}

\begin{document}

\begin{titlepage}

\begin{center}

{{\Large{\textsc{Swansea University}}}} \rule[0.1cm]{15.8cm}{0.1mm}
\rule[0.5cm]{15.8cm}{0.6mm}
{\small{\bf Physics Department}}
\end{center}
\vspace{15mm}
\begin{center}
{\LARGE{\bf Numerical methods for the sign problem}}\\
\vspace{1.5mm}
{\LARGE{\bf in }}\\
\vspace{2.5mm}
{\LARGE{\bf Lattice Field Theory}}\\
\vspace{19mm} {\large{ Submitted to Swansea University in fulfilment of the requirements for the Degree of Doctor of Philosophy}}
\end{center}

\begin{figure}[h]
\centering
\includegraphics[scale=0.7]{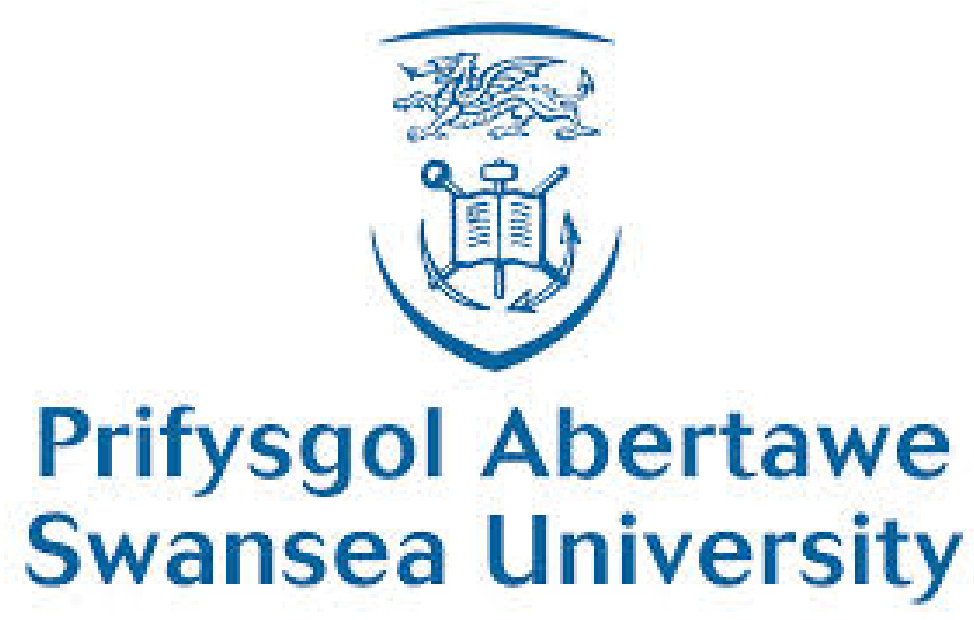}
\end{figure}

\vspace{20mm}
\par
\noindent
\begin{center}
\Large{\bf Lorenzo Bongiovanni}
\end{center}
\vspace{1mm} 
\begin{center}
{\large{ Swansea University 2015}}
\end{center}

\end{titlepage}

\input{NonChaptersStuff/TitleBack.tex}
\clearpage\input{NonChaptersStuff/Abstract.tex}
\tableofcontents

\input{cap_INTRO/Introduction.tex}

\input{cap_Lattice/QFTLattice.tex}

\input{cap_Langevin/CLangevin.tex}

\input{cap_THETA/ThetaTerm.tex}

\input{cap_LEFSCHETZ/Lefschetz_thimbles.tex}

\input{cap_DENSITY/Density_of_States.tex}

\input{cap_Conclusions/Conclusions.tex}

\listoffigures



\input{Bibliography/Bibliography.tex}

\end{document}

%% file: NonChaptersStuff/TitleBack.tex
\thispagestyle{empty}

\hfill

\vfill

\noindent Lorenzo Bongiovanni: \textit{Numerical methods for the sign problem\\ 
\phantom{x} \hspace{3.7cm}  in Lattice Field Theory}

\bigskip

\noindent\textbf{Supervisors} : \\
Gert Aarts\\
Biagio Lucini

\medskip

\noindent\textbf{Date of Submission} : \\
25/9/2015 \hfill 

%% file: NonChaptersStuff/Abstract.tex
\end{comment}

\chapter*{Abstract}
The great majority of algorithms employed in the study of lattice field theory are based on Monte Carlo's importance sampling method, i.e. on probability interpretation of the Boltzmann weight.  Unfortunately in many theories of interest one cannot associated a real and positive weight to every configuration, that is because their action is explicitly complex or because the weight is multiplied by some non positive term. In this cases one says that the theory on the lattice is affected by the \textit{sign problem}. An outstanding example of sign problem preventing a quantum field theory to be studied, is QCD at finite chemical potential.   

Whenever the sign problem is present, standard Monte Carlo methods are problematic to apply and, in general, new approaches are needed to explore the phase diagram of the complex theory. Here we will review three of the main candidate methods to deal with the sign problem, namely complex Langevin dynamics, Lefschetz thimbles and density of states method.

We will first study complex Langevin dynamics, combined with the gauge cooling method, on the one-dimensional Polyakov line model, and then we will apply it to pure gauge Yang-Mills theory with a topological $\theta$ term. It follows a comparison between complex Langevin dynamics and the Lefschetz thimbles method on three toy models, which are the quartic model, the U(1) one-link model with a $\mu$ dependent determinant, and the SU(2) non abelian one-link model with complex $\beta$ parameter.

Lastly, we introduce the density of state method, based on the LLR algorithm, and we will employ it in the study of the relativistic Bose gas at finite chemical potential.


%% file: cap_INTRO/Introduction.tex
\end{comment}

\addcontentsline{toc}{chapter}{\protect\numberline{}Introduction}
\chapter*{Introduction}

\vspace{0.5 cm}
\begin{figure}[!h]
\begin{center}
\includegraphics[scale=0.8, angle =0]{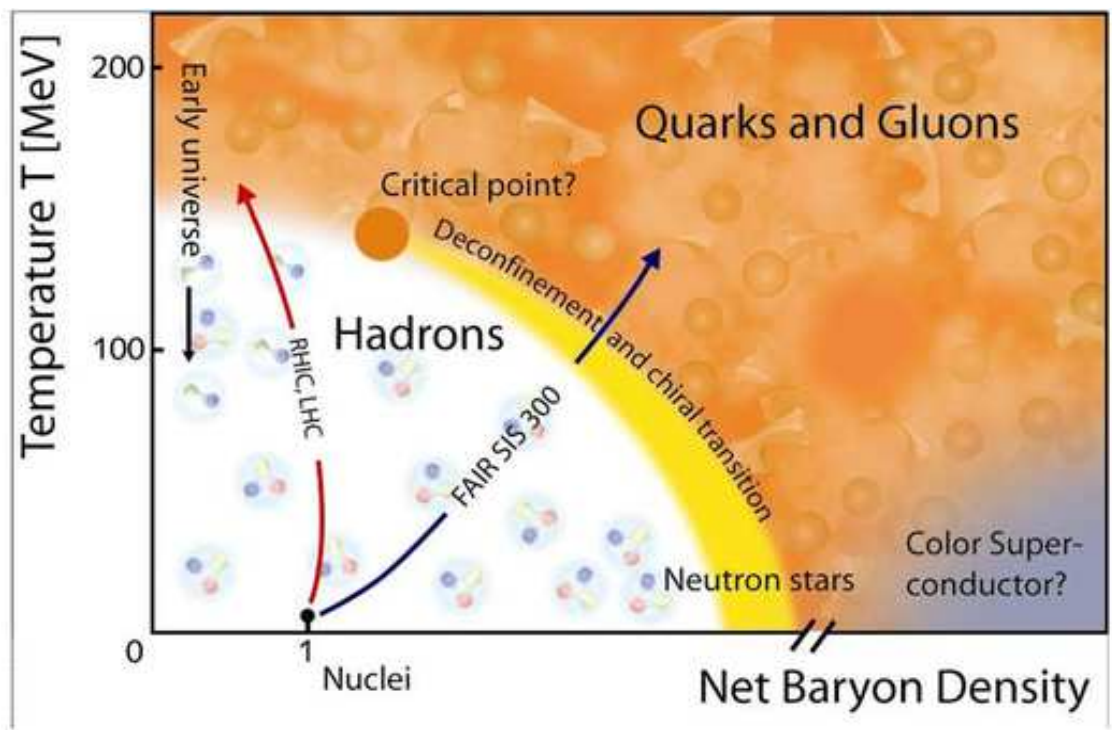}
\end{center}
 \caption{Qualitative representation of quantum chromodynamics (QCD) phase diagram.}
 \label{fig:PhaseDyagram}
\end{figure}

Starting from the 1980s, lattice field theory has been developed and it has been proven to be a formidable tool to study quantum field theory (QFT). In general, whenever a theory manifests a non-perturbative behaviour at some energy scale, analytic quantitative solutions are very hard or impossible to obtain.

Discretization on the lattice is a well-established non-perturbative approach for Euclidian time gauge theories \cite{Creutz:1979zg,Creutz:1980zw,Weingarten:1981jy,Marchesini:1991ch}. In particular, much progress has been made in the study of quantum chromodynamics (QCD), i.e. the theory of quarks and gluons, which describes a big part of the high energy physics as we know it. The success of lattice field theory is due to the fact that it can be mapped into a Statistical Mechanics ensemble and, therefore, all the techniques developed for the latter are available.  The most acknowledged class of methods, successfully employed to study a vast number of models, is based on Monte Carlo's importance sampling. This is a probabilistic way  of exploring the space of configurations of a system, based on the action $S$ of each configuration. More specifically, the Boltzmann weight $e^{-S}$ of a configuration is interpreted as the unrenormalized probability of the configuration itself and, therefore, it will contribute with this weight to the partition function. In this way one is able to build a finite representative sample of the full configuration space, and use it to compute average values of observables. 
 
Unfortunately this way of proceeding is not applicable any more whenever the exponential of the action of the system cannot be interpreted as a probabilistic weight, that is when it is not real and positive. This phenomenon is commonly referred to as \emph{sign problem}. 

In this case, a way around is to employ methods that, even though they are still  based on Monte Carlo, allow to extrapolate some information even when the action is complex. Some of these are :

\begin{itemize}
\item  re-weighting \cite{Munger:1989sp}, where the complex part $e^{i \phi }$ of the weight is incorporated in the observable, while the expectation value is computed only using the real part $\vert \rho \vert$ of the weight $e^{-S}=\vert \rho \vert\; e^{i \phi}$, namely the \textit{phase quenched} theory
\begin{equation}
\langle O \rangle =\dfrac{\langle O e^{i \phi }\rangle_{pq}}{ \langle e^{i \phi }\rangle_{pq}}
\end{equation}

\item Taylor expansion in the parameters that trigger the sign problem, so that the quantities one has to compute appear as coefficients, in a theory without sign problem.

\item Analytic continuation from a part of the phase diagram where the action is real, typically where the parameters that trigger the sign problem are imaginary \cite{Scalettar:1989ig,Lombardo:1999cz,Philipsen:2001ws,deForcrand:2002yi}.

\item Strong coupling expansion, in which one exploits the fact that the truncated series of character expansion of the functional integral is a spin-like system with a much milder sign problem than the original one \cite{Munster:1981es,Langelage:2007pi,Langelage:2010yr,Fromm:2011qi}.

\item Mean field theory \cite{Green:1983sd,Aarts:2009hn,Akerlund:2013fsa}
\end{itemize}

However, those methods are typically  restricted to some areas of the theory where the parameters that trigger the sign problem are not too large. There are other approaches which aim to solve the sign problem all together. They are mostly of recent development so they sometimes lack the control which Monte Carlo methods achieved over many years. Some of those approaches are :
\begin{itemize}
\item complex Langevin dynamics : a stochastic quantization of the fields is adopted. The degrees of freedom are complexified and a real probability distribution for the observable is generated in the complex space (solution of the Fokker-Planck equation) \cite{Damgaard:1987rr,Aarts:2011ax,Seiler:2012wz}.

\item Lefschetz thimble : new manifolds, equivalent to the original domain of integration, are found in the complexified space, along which the imaginary part of the action is constant and, therefore, the integral is (mostly) real \cite{Witten:2010zr,Cristoforetti:2012su,Fujii:2013sra}.

\item Density of states : the density of states of a system is computed and the original path integral is reduced to a one-dimensional oscillating integral \cite{Anagnostopoulos:2001yb,Fodor:2007vv,Langfeld:2012ah}.

\item Canonical approach : the canonical partition function at fixed number of particles (or in general conserved charge) is computed, and it is connected with the general partition function via the fugacity expansion \cite{Kratochvila:2004wz,Alexandru:2005ix,Kratochvila:2005mk}. 

\item Reformulation of the theory in term of dual variables : in some theories it is possible an exact map into another theory with other variables where the sign problem is milder or absent\cite{Chandrasekharan:2008gp,Mercado:2013yta}.
\end{itemize}

In this thesis we will discuss in detail some of the last group, mainly focusing on complex Langevin dynamics.

There are many theories affected by the sign problem, not only in quantum field theory but also in real time quantum mechanics and condensed matter. One of the most important and challenging is QCD at \textit{finite density} \cite{Aarts:2013naa,deForcrand:2010ys}. Here, in Euclidian time, the sign problem is generated by the fermionic determinant
\begin{equation}\label{Zintro}
Z=\int [DU] e^{- S_{YM}} \det M(\mu),
\end{equation}
 where the links $U$ represent the gluonic degrees of freedom on the lattice, and $\det M(\mu)$ is the result of the Grassman integral over the fermionic part $\det M(\mu)=\int d\psi d \overline{\psi} \exp\left[ \overline{\psi} (\slashed D+m+\gamma_0 \mu) \psi\right]$. It can be shown that
 \begin{equation}
\left[ \det M(\mu)\right]^* = \det M(-\mu^*),
 \end{equation}
 which allows it to be real only if $\mu$ is zero or completely imaginary. At finite $\mu$, then, the weight is complex and the sign problem arises together with cancellations in \eqref{Zintro}. Those are responsible, for example, for an interesting phenomenon called \textit{Silver Blaze}, for which, at $T=0$, all the thermodynamic observables have to be independent from the chemical potential, up to the nucleon mass, even though the determinant explicitly depends on $\mu$. In particular, the critical baryon chemical potential is the nucleon mass minus the nuclear binding energy. The qualitative explanation for this behaviour is that, before the critical chemical potential, there is not enough energy in the system to create a nucleon at $T=0$, at $T\neq 0$, the nucleon has still a chance to be created but it will be suppressed by a Boltzmann factor $\sim\exp (- (m_N-\mu_B)/T)$, where $m_N$ is the nucleon mass and $\mu_B$ is the critical baryon chemical potential. This behaviour is totally spoiled \cite{Stephanov:1996ki} if the oscillations are neglected by taking the \textit{phase quenched} theory, i.e. substituting $\det M(\mu)$ with its absolute value $\vert\det M(\mu) \vert$ in the partition function. 
 
 In practise, the QCD phase diagram in Fig.\ref{fig:PhaseDyagram} can only be explored, with standard Monte Carlo methods, close to the $y$ axis, i.e. where the ratio $(\mu/T)$ is small. In that region, one can still extrapolate information by Taylor expansion in  $(\mu/T)$, reweighting methods or by analytic continuation from imaginary $\mu$ where the theory is real \cite{Lombardo:1999cz,D'Elia:2002gd}.
 
Recently some progress has been made in exploring the QCD phase diagram, deep into the oscillating region, thanks to \textit{complex Langevin dynamics}  \cite{Sexty:2014zya,Fodor:2015doa}.\\\\
 
One of the other major open problem in quantum chromodynamics is the \textit{strong CP} problem. The theory allows the pure gauge action to have an extra gluonic term   
\begin{equation}\label{Stheta_intro}
S=S_{YM}+i \theta \dfrac{g^2}{32 \pi^2}\int dx\; \Tr \left( G_{\mu \nu}\tilde{G}^{\mu \nu} \right),
\end{equation}
which is proportional to the topological charge $Q_{top}=\frac{1}{32 \pi^2}\int dx\; G_{\mu \nu}^a\tilde{G}_{\mu \nu}^a$. The fact that in nature experimental evidence constrain 
\begin{equation}
\theta < 10^{-10} ,
\end{equation}
makes fine tuning problems arise.  Some axion models for a dynamical solution to this problem have been proposed \cite{Peccei:1977hh,Peccei:1977ur,Berkowitz:2015aua}, but still a non-perturbative investigation of the theory at finite $\theta$ is required. Unfortunately, the topological $\theta$-term is imaginary, that makes the action \eqref{Stheta_intro} complex. Again, the sign problem prevents standard Monte Carlo methods to explore the whole $\theta-T$ plane.

Results have been achieved by analytic continuation from imaginary $\theta$  \cite{Panagopoulos:2011rb,D'Elia:2012vv,D'Elia:2013eua}, and recently the problem at real $\theta$ has been studied with complex Langevin  dynamics \cite{Bongiovanni:2014rna,Bongiovanni:2013nxa}.
 
 Many other systems affected by the sign problem have been studied, we will see some of those in the following, and some successfully solved. The main challenge, however, remains to solve the QCD related sign problem.

\vspace{1cm} 
This thesis is divided in five chapters, and all the results shown come from my work in first person. 

 The first chapter reviews the discretization of a field theory on the lattice. In the second, we introduce the stochastic quantization and analyse the requirements for good control of complex Langevin dynamics. We also discuss the gauge cooling method and show results of convergence in some models.

The third chapter is about the $\theta$-term. The first half, is a review of the instantons' theory and of the discretization of topology on the lattice. In the second part, we discuss the application of complex Langevin dynamics to this theory and show some results. I would like to thank D{\'e}nes Sexty for providing the base code which I developed to get the results shown in this chapter, and also Ben J{\"a}ger and Felipe Attanasio for their help and discussions along the development of the same code.

In the fourth chapter, we introduce the Lefschetz thimbles method and compare it with complex Langevin dynamics in the study of some toy models.

The fifth, and last, chapter is about the density of states method. We introduce the method and discuss its application to the relativistic Bose gas at finite density. I would like to thank Roberto Pellegrini for his collaboration in developing the code we have been using to get the results shown in this chapter.

\begin{comment}

%% file: cap_Lattice/QFTLattice.tex
\tableofcontents
\end{comment}

\chapter{Quantum field theory on the lattice}
Making predictions for a non trivial Quantum Field Theory (QFT) is never an easy task. Furthermore, if it allows non perturbative interaction between the fields, analytical methods are usually not readily available. The most acknowledged way to overcome these problems is, up until now, to adopt a non-perturbative regularization of the theory on a discrete lattice of points in Euclidian space-time. Proposed by Wilson in 1974 \cite{Wilson:1974sk}, lattice QFT has become the most reliable tool for strongly interacting systems.\\
In this chapter we'll briefly review the main concepts behind it.

\vspace{2cm}
\section{Path Integral approach to quantum theory}
The path integral approach to QFT, introduced by Feynman in 1948 \cite{Feynman:1948ur}, is one of the most powerful tools when it comes to non-classical calculations. It is essential not only in order to perform most of the perturbative calculations but also to study non perturbative physics.  Thanks to this formulation, in fact, a mapping of a regularized QFT into a statistical mechanic system has been made possible. More precisely, the quantum degrees of freedom on a \textit{discretized} space-time can be identified as the ones in the canonical ensemble at temperature $T$  and the system is statistically allowed to visit every configuration. 

We will now recall the basic ideas behind the path integral and its connection with statistical mechanics when Euclidian time is introduced.  Let us consider a non relativistic quantum system ($(0+1)$ QFT) described by a Hamiltonian $\mathcal{H}(p,x)$. The matrix element relative to the evolution from the point $x_a(t_a)$ to  $x_b(t_b)$ is given by
\begin{equation}\label{matrice di trasf 1}
T_{x_a,x_b} \ = \ \langle x_a | e^{-i\frac{\mathcal{H}}{\hbar}(t_b-t_a)} | x_b \rangle  \ ; 
\end{equation}
one can now insert somewhere in between $t_a$ and $t_b$ the operator identity $I\ = \ \int \ dx_c \ |x_c(t_c) \rangle \langle x_c(t_c) |$ that leaves, of course, unchanged the probability of  transition from    
$|x_a\rangle$ to $|x_b\rangle$
\be
T_{x_a,x_b} \ = \int_{x_c} dx_c \ \langle x_a | e^{-i\frac{\mathcal{H}}{\hbar}(t_c-t_a)} | x_c \rangle \langle x_c | e^{-i\frac{\mathcal{H}}{\hbar}(t_b-t_c)} | x_b \rangle \ .
\ee
One can think of repeating this process an infinite number of times by inserting the identity operator $I$ at every time between $t_a$ and $t_b$, integrating over all the possible values of the field $x(t)$ at that time $t$. Furthermore, if the Hamiltonian is quadratic in the momenta $\mathcal{H}=\frac{1}{2}p^2+V(x)$, it is possible to carry out the Gaussian integral over the momenta $p$ and express the eq.\eqref{matrice di trasf 1} in the path integral form form :
\be \label{matrice di trasf 2}
T_{x_a,x_b} \ \rightarrow \ C \int_{x_a}^{x_b} Dx(t) \ e^{\frac{i}{\hbar}\int_{t_a}^{t_b} dt \mathcal{L}} 
\ee
where $\mathcal{L}$ is the \textit{Lagrangian} of the system
\begin{equation}
\mathcal{L}(\dot{x},x)=\dfrac{1}{2} \dot{x}^2 + V(x) \ .
\end{equation}
Physically, this procedure has the meaning of interpreting the probability of transition of a system from a state to another as \textit{the sum over all possible intermediate paths} weighted with the oscillating phase given by their action along each one of these paths . \ As a result, one is not surprised to see the biggest contribution coming from the paths with stationary phase, called \textit{semi-classical approximation}, which, for values of the parameters of the theory comparable with the ones in our everyday life, will be reduced into the well known classical equation of motion.\\

Let us go back now to the analogy with statistical mechanics. If we consider a canonical system described by a Hamiltonian $\mathcal{H}$ and in thermal equilibrium with a heat bath at temperature $T$, we can write the partition function of the system as
\be \label{Z_mec_1}
Z(\beta) \ = \ \Tr(e^{-\beta H})\ = \ \sum_n \langle n| e^{-\beta H} |n \rangle = \ \sum_n e^{-\beta E_n} \ , 
\ee
where $\beta = 1 / k_B T$ and $|n\rangle$ are the eigenstates of the Hamiltonian with eigenvalues the values of the energy $E_n$ .  Moreover, since the trace of an operator does not depend on the basis on which it is computed, one can rewrite the trace in \eqref{Z_mec_1} choosing as the base the position $|x\rangle$ 
\be \label{Z_mec_2}
Z(\beta) \ = \ \Tr(e^{-\beta H}) = \int_{-\infty}^\infty dx \ \langle x | e^{-\beta H} | x \rangle \ 
\ee 
where the similarity with \eqref{matrice di trasf 2} is now fairly evident . The last step we have to do in order to achieve the exact analogy of the two expressions is to rotate the eq.\eqref{matrice di trasf 2} into \textit{Euclidian time} (\textit{Wick rotation})
\be
\left\lbrace\!
\begin{split}
&x_i \rightarrow x_i  \\
&t \rightarrow  -i \  \tau \ ,
\end{split}
\!\right.
\ee 
to identify the inverse of the temperature $\beta$ with $\frac{t_b-t_a}{\hbar} $ and to restrict ourselves only to periodic paths $ x(t_a) = x(t_b) = x$. With these devices the eq.\eqref{matrice di trasf 2} becomes
\begin{equation}
Z(\beta) \ = \ \mathcal{N} \int_{x(0)}^{x(\beta\hbar)=x(0)} Dx(\tau) e^{-\int_0^{\beta\hbar}d\tau \ \mathcal{L}_E} \ .
\end{equation}
The integral over the time of the euclidian lagrangian $\mathcal{L}_E$ is referred to as the euclidian \textit{action} of the theory
\begin{equation}
\mathcal{S}_E(\dot{x},x)  = \int_0^{\beta\hbar}d\tau \ \mathcal{L}_E(\dot{x},x)  \ ,
\end{equation}  
so that the most common way to express the partition function of a quantum mechanical system in euclidian time is
\begin{equation}\label{Z}
Z \ = \ \mathcal{N} \int Dx(\tau) \ e^{-S_E} \ .
\end{equation}
Thanks to this formulation, it is now possible to assign to every quantum trajectory $x(\tau)$ a probability based on the action of the system along that path
\begin{equation}\label{P quant}
P(x(\tau)) \ = \ \dfrac{e^{-S_E(x(\tau))}}{Z} \ . 
\end{equation}
When dealing with number of dimensions higher that 0, the above formulation of a QFT is often still affected by ultraviolet divergences and, therefore, it needs some regularization before physically relevant quantities can be computed. One possible regulator, largely used for non perturbative calculations, is the \textit{lattice} discretization of the space time. This particular regularization has the property of mapping a $d$ spatial and 1 time dimension quantum field theory in an equilibrium $(d+1)$ dimensions statistical mechanics model.  Therefore, using the equivalent of \eqref{P quant}, it is possible to evaluate the probability the system is likely to be in a certain configuration given the euclidian action of the configuration itself.
More generally this map allows us to use all the well developed tools of statistical mechanics to study the QFT \ .

\vspace{2cm}
\section{Monte Carlo simulations}\label{Monte Carlo simulations}
Most of the physical relevant problems in Quantum Field theory are such that the fields in the theory are coupled to each other in a non trivial and non-perturbative way. In a more formal way this is equivalent to saying that it is impossible to compute the path integral
\be \label{int_O}
 Z=\int D\phi_i \ \mathcal{O}(\phi_i)  \ e^{-S(g_i;\partial \phi_j,\phi_k)}
\ee  
exactly or perturbatively because of the interactions in the action $S(g_i;\partial \phi_j,\phi_k)$; here we called the fields $\phi_i$, their derivatives $\partial \phi_j$, and the couplings of the theory $g_i$; also from now on we will always work in Euclidean space so we will omit the label $E$ at the bottom of the operators, in this case for example $S=S_E$.

The only way to compute \eqref{int_O} is then the numerical approach. There is more than one class of numerical methods that are, in principle, able to compute the correct estimation of \eqref{int_O} but, if we assume for now that $S$ is \textit{real}, the most popular and well developed are for sure the Monte Carlo methods. To be precise, what a Monte Carlo algorithm is able to compute is not exactly \eqref{int_O} but actually the \textit{average value} of the observable $\mathcal{O}$
\be
\langle \mathcal{O}\rangle \ =\ \dfrac{ \int D\phi_i \ \mathcal{O}(\phi_i)  \ e^{-S(g_i;\partial \phi_j,\phi_k)}}{\int D\phi_i  \ e^{-S(g_i;\partial \phi_j,\phi_k)}} \ .
\ee 
The idea at the base of Monte Carlo methods is to \textit{sample} the configuration's space of the fields $\phi_i$ guided by the probability $e^{-S(\phi_i)}$, in such a way that the configurations with smaller action are visited more often and vice versa. This procedure assigns a weight to the configurations, based on the frequency $f_{\lbrace \phi_i \rbrace}$ they have been visited, and allows one to build a good sample of the configuration space that can be used to compute any observable
\be
 \langle \mathcal{O}\rangle \ \sim \ \ \sum_{\lbrace \phi_i \rbrace} f_{\lbrace \phi_i \rbrace} \mathcal{O} (\lbrace \phi_i \rbrace) \ .
\ee 
Of course the bigger the number of configurations sampled, the more accurate is the estimation of $\langle \mathcal{O}\rangle$; more precisely, the error scales like the inverse of the square root of the number N of measurements (as long as these are independent from each other) 
\be 
\delta  \langle \mathcal{O}\rangle \sim \dfrac{1}{\sqrt{N}} \ .
\ee
It is crucial to observe how important is the hypothesis that the action $S(g_i;\partial \phi_j,\phi_k)$ is \textit{real}, because if this is no longer true one cannot interpret $e^{-S}$ as a weight to sample the configuration space and the very premises of the Monte Carlo methods fail.\\\\

Lastly, we observe that in general the fields of a theory live in a space-time continuum that, in practice, is impossible to be represented compatibly with the application of any numerical methods. The solution to this problem is to adopt a \textit{discretized space-time} commonly called \textit{Lattice}, where each continuous direction is replaced with a discrete multiple of a unit of length \textit{a} called \textit{lattice spacing}  
\be 
\begin{split}
&x \ =\ n_x \ a_x, \ \ \ \ n_x =0,1,..,N_x \\
&y \ =\ n_y \ a_y, \ \ \ \ n_y =0,1,..,N_y \\
&\ \ \ \ \ . \\
&\ \ \ \ \ .
\end{split}
\ee
and the boundary of this lattice are periodic in each direction (periodic boundary conditions) so that 
\begin{equation}
\phi(n_i=N_i)\ =\ \phi(n_i=0) \ .
\end{equation}
The original volume is now replaced by a lattice of points evenly spaced in each direction but, of course, to recover the real value of $\langle \mathcal{O} \rangle$ one has to correctly estimate the \textit{continuum limit} which we're going to discuss later. 

Discretization of the space-time has an important impact on the momentum space, in fact if the shortest wavelength can't be less then the lattice spacing $x_{min}=a$, it means the large momenta can't be bigger than $p_{max}=\pi/a$. It follows that the lattice introduces a \textit{cut-off in the momentum space} 
\begin{equation}
\int_{-\infty}^{\infty} \dfrac{dk}{2 \pi} \tilde{f}(k) e^{ikx} \ \rightarrow \ \int_{-\frac{\pi}{a}}^{\frac{\pi}{a}} \dfrac{dk}{2 \pi} \tilde{f}(k) e^{ikna}
\end{equation}
restricting all the integrals on the momenta to the first Brillouin zone $[-\pi/a,\pi/a]$. Hence, all the loop integrals are finite and the functional integrals are high dimensional standard integrals.
Eventually the cut-off has to be removed while approaching the continuum limit and this process determines the \textit{renormalization group flow} of the theory of the observables. 
From the point of view of the statistical mechanics, the lattice structure does not play a crucial role if all the physical length scales are much larger than the the lattice spacing, which means
\begin{equation}
\xi\ \gg\ a
\end{equation}
where $\xi$ is the correlation length of the system. In this regime the physical quantities are not sensitive to changes in lattice spacing and therefore one can safely claim the lattice system to be in the continuum limit. Such a regime is reached in correspondence of a \textit{second order phase transition} of the theory.

\vspace{2cm}
\section{An example: the Scalar Field}
In this section we're going to briefly review the case of the scalar field both because it is a quite popular and instructive example of discretization of a field theory on the Lattice and because  we'll come across it again describing the \textit{density of state} method. Also from now on we will work in the Planck units
\begin{equation}
\hbar=k_B=c=1\ ,
\end{equation}
which are the natural units of measurement to use in quantum field theory.\\
A possible scalar field theory, expressed in the new units of measurement, could be represented by the action  
\begin{equation}
S \ = \ \int_0^{\beta} d\tau \int_V d^3x \  \left[ \frac{1}{2} \left( \partial_\mu \varphi \partial^\mu \varphi + m^2 \varphi^2 \right) + g_3 \varphi^3 + \lambda \varphi^4 + ... \right] ,
\end{equation} 
where $\varphi$ is the scalar field and $g_3,\lambda$,etc are the couplings.   In general the action can be quite complicated, including the coupling with non perturbative global charges, however the purpose of this section is just to illustrate the process of discretization of a field theory; therefore we will limit ourself to the case of the free theory
\begin{equation}\label{free scalar theory}
S \ = \ \int_0^{\beta} d\tau \int_V d^3x \ \frac{1}{2} \left( \partial_\mu \varphi \partial^\mu \varphi + m^2 \varphi^2 \right) \ .
\end{equation} 
As we mentioned in the last section the analogy with statistical mechanics requires the adoption of \textit{periodic boundary condition} in the time direction; as long as bosonic degrees of freedom are involved, that is formally equivalent to imposing
\begin{equation}
\varphi(\overrightarrow{x},\tau) \ = \ \varphi(\overrightarrow{x},\tau+\beta) \ 
\end{equation}
on our scalar field .\\
The next step is to discretise the space-time volume 
\begin{equation}
x_\mu \rightarrow a n_\mu \ ,
\end{equation} 
where $a$ has the dimensions of a length or, alternatively, of the inverse of an energy and $n_{\mu}$ is the integer that indicates the distance, in multiples of $a$, in the $\mu$ direction.  Since our volume has to be finite, for practical reasons, there is going to be an $N_{\mu}$ such that
\begin{equation}
aN_\mu = L_\mu \ , 
\end{equation}
where $L_\mu$ is the edge of our volume in the  $\mu$ direction, so that we have the condition
\begin{equation}
0<n_\mu \leq N_\mu \ .
\end{equation}
The most common way (but not the only one) to cope with the finite volume is to impose periodic boundary conditions also in the space directions, so that
\be 
\varphi(n_k + N_k) = \varphi(n_k)\ , \ \ \ k =1,2,3...,$space directions$ 
\ee
and to take in account the finite size effects by studying the scaling of the observables for increasing lattice sizes, a process known as \textit{thermodynamic limit}.

It is good practice, then, to rescale every quantity in units of $a$ so that everything on the lattice is adimensional except the lattice spacing. The mass $m$ has, of course, the dimensions of an energy, while for the fields it usually depends on the space-time dimensions the theory is defined on; in our case, like most of the times, we're working with the 4 dimensions so $\varphi$ has the dimensions of an energy. The lattice variable will then be defined as
\begin{equation} \label{discr1}
\begin{split}
&m \rightarrow \dfrac{\widehat{m}}{a} \ ,\\
&\varphi(x) \rightarrow \frac{\phi(n)}{a} \ , 
\end{split}
\end{equation} 
while the integral over the continuous 4 dimensions will be replaced by a sum over the lattice sites times the fundamental cube 
\begin{equation}\label{discr2}
\int d^4x \rightarrow \sum_n a^4 \ .
\end{equation}   
Also, for the derivative the discratization is quite straightforward
\begin{equation} \label{discr3}
\partial_\mu \phi(x) \rightarrow \frac{\phi(n + \overrightarrow{\mu}) - \phi(n)}{a}  \ ,
\end{equation}
recalling the original value in the limit $a \rightarrow 0$; the notation $\varphi(n + \overrightarrow{\mu})$ means one is considering the nearest neighbour of $\varphi(n)$ in the $\overrightarrow{\mu}$ direction . It is easy to see that the way we choose to discretize the derivative is not unique and there are at least 3 equivalent ways of defining a first order discretization of the derivative
\begin{itemize}
 \item \textbf{forward} derivative
 \begin{equation}
  \partial_\mu \phi(x) \rightarrow \frac{\phi(n + \overrightarrow{\mu}) - \phi(n)}{a}
 \end{equation}
 \item  \textbf{backward} derivative
 \begin{equation}
 \partial_\mu \phi(x) \rightarrow \frac{\phi(n) - \phi(n - \overrightarrow{\mu})}{a} 
 \end{equation}
 \item \textbf{symmetric}  derivative
 \begin{equation}\label{deriv simm}
 \partial_\mu \phi(x) \rightarrow \frac{\phi(n + \overrightarrow{\mu}) - \phi(n - \overrightarrow{\mu} )}{2a} . 
\end{equation}
\end{itemize} 
Let us note that, out of the three, only the last one maintains, on the lattice, the anti-Hermiticity proper of its continuous version; the other two transform one into the other under the Hermitain conjugate operation. For the scalar field this is however not a problem since the square of the derivative in the action makes everything Hermitian again anyway. 

The last thing to be mentioned is that also the integral measure will, of course, be affected by the discretization of space-time. The integral over all possible paths will be replaced by the product of the differential of the fields on each point of the lattice
\be \label{discr4}
\int_V [\mathcal{D}\varphi(x)] \ \rightarrow \int_{n\ \in\ Latt } \left (\prod_n \ d\phi_n \right) \ .
\ee

Following step by step all the points discussed before (\ref{discr1}-\ref{discr4}) , we can now write down the lattice version of the free theory \eqref{free scalar theory} 
\begin{equation}\label{S_L scalarfield}
\begin{split}
S_L \ = \ \frac{1}{2} &\sum_n \ \left[ (\widehat{m}^2+2d)\ \phi^2(n) - \sum_{\mu = 1}^4 \phi(n)\phi(n + \overrightarrow{\mu})   \right] \ '\\
&=\ \frac{1}{2}\ \phi_n M_{nl} \phi_l \ ,
\end{split}
\end{equation}
where $d$ are the dimensions ($d=4$ in our case) and
\begin{equation}
M_{nm}\ =\ (\widehat{m}^2+2d)\delta_{nm} \ - \ \frac{1}{2} \sum_{\mu=1}^4 (\delta_{n+\overrightarrow{\mu},m} + \delta_{m,n-\overrightarrow{\mu}})\ ,
\end{equation}
having expressed the $\sum_{\mu = 1}^4 \phi(n)\phi(n + \overrightarrow{\mu})$ in \eqref{S_L scalarfield}  in a symmetric way.
The 2-point function
\begin{equation}
\langle \phi_n \phi_m \rangle\ =\ M^{-1}_{nm}\ ,
\end{equation}
is easily computed using 
\begin{equation}\label{M_nm Id}
\sum_l M_{nl}M^{-1}_{lm}\ =\ \delta_{nm}
\end{equation}
and the Fourier transform of $M_{nm}$
\begin{equation}\label{M(k)}
\tilde{M}(k)\ =\ \int_{-\pi}^\pi \dfrac{d^4k}{(2 \pi)^4}M_{nm}\ e^{i(n-m)k} \ =\  \widehat{m}^2+4\sum_{\mu=1}^4 sin^2(\frac{k_\mu}{2})\ ,
\end{equation}
where, as always on the lattice, we are using the adimensional momentum $k_\mu=p_\mu a$ . If we now take the Fourier transform of \eqref{M_nm Id}, using \eqref{M(k)}, we're able to obtain an expression for 
\begin{equation}\label{propag Four}
M^{-1}_{nm} \ = \ \int_{-\pi}^\pi \dfrac{d^4k}{(2\pi)^4} \dfrac{e^{ik(n-m)}}{\widehat{m}^2+4\sum_\mu sin^2(\frac{k_\mu}{2})} \ .
\end{equation}
We're now interested in taking the limit of the lattice 2-point function $M^{-1}_{nm}$ for $a\rightarrow 0$, holding $m$, $\varphi$, $x=na$ and $y=ma$ fixed. In general, the mass $m$ could not be held fixed because it would take cut-off dependent contributions from the self energy and, to perform the continuum limit, one would have to take in account its renormalization. For our free theory, however, that is not the case and one can naively send  $a\rightarrow 0$ in
\begin{equation}\label{propag continuo}
M^{-1}(x-y) \ = \ \int_{-\frac{\pi}{a}}^{\frac{\pi}{a}} \dfrac{d^4p}{(2\pi)^4} \dfrac{e^{ip(x-y)}}{m^2+4\sum_\mu \frac{sin^2(p_\mu a/2)}{a^2}} \ .
\end{equation}
We just need to see, sending $a\rightarrow 0$, that the quantity  $\frac{sin^2(p_\mu a/2)}{a^2}$ remains finite only for $sin^2(p_\mu a/2)\rightarrow (p_\mu a/2)^2$, and we recover for the \eqref{propag continuo} the well-known expression for the theory in the continuum
 \begin{equation}\label{propag scal}
\langle \phi(x) \phi(y) \rangle \ = \ \int_{- \infty}^\infty \dfrac{d^4p}{(2\pi)^4}\dfrac{e^{ip(x-y)}}{m^2 + p^2} \ .
\end{equation}

\vspace{2 cm}
\section{Gauge Theories on the Lattice}
In this section we are going to briefly review the regularization on the lattice of a widely studied category of physical theories, namely the \textit{Gauge Theories}. 

Let's consider a $N$ dimensional matter field $\phi$ defined on the sites of our lattice, for example a set of $N$ equivalent scalar fields of the section before, that transforms under local unitary group $\mathcal{G} \in $ SU($N$) as
\begin{equation}
\phi(x)\ \rightarrow \phi'(x)\ =\ \mathcal{G}(x) \phi(x) \ ;
\end{equation}
we now require its action 
\begin{equation}
S_L \ = \ \frac{1}{2}\ \sum_n \left[ (\phi(n+\mu)-\phi(n)) (\phi(n+\mu)-\phi(n))^\dag  + m^2 \phi \phi^\dag )    \right] \
\end{equation}
to be left invariant by such transformation. We clearly see that the part 
\begin{equation}
S_{deriv}\ =\  \sum_{\mu = -4}^4 \phi^\dag(n)\phi(n + \overrightarrow{\mu}) \ ,
\end{equation}
coming from the derivative, cannot be invariant under the action of $\mathcal{G}$ at least in a trivial way. The way to make up for it, in analogy with the continuum case, is to introduce a field $U_\mu(x)$, namely the \textit{gauge} field. Its function is to \textit{parallel transport} the action of $\mathcal{G}$ from $\phi(n + \overrightarrow{\mu})$ to $\phi(n)$ , along the geodesic connecting the two fields.  The latter, on the lattice, is the straight line connecting the site $n$ with the one $n + \overrightarrow{\mu}$. The gauge field transforms
\begin{equation}\label{gaugeU_transfom}
U_\mu(x) \ \rightarrow \ U_\mu'(x) = \mathcal{G}(x) U_\mu(x) \mathcal{G}^{-1}(x+\mu)\ ,
\end{equation}
so that the gauged part of action
\begin{equation}
S^{g}_{deriv}\ =\  \sum_{\mu = -4}^4 \phi^\dag(n) U_\mu(x) \phi(n + \overrightarrow{\mu}) \ ,
\end{equation}
has now been made invariant under the action of $\mathcal{G}$. It is clear from the \eqref{gaugeU_transfom} that $U(x)$ is itself an element of SU($N$) and can therefore be written in the form
\begin{equation}
 U_\mu(n)\ =\ e^{igaA_\mu(n)}
\end{equation} 
where $g$ is the coupling constant of the gauge field, $a$ is the lattice spacing and  $A_\mu = A^a_\mu \lambda^a$ is an element of the Lie algebra of SU($N$). The generators of the algebra $\lambda_a$ obey the commutation relations
\begin{equation}
\begin{split}
& \Tr(\lambda_a \lambda_b)=\dfrac{1}{2} \delta_{ab}\ , \\
&[\lambda_a,\lambda_b]=i \sum_{c=1}^{N^2-1}f_{abc} \lambda_c \ , \\
&\lbrace\lambda_a,\lambda_b\rbrace=\dfrac{1}{N}\delta_{ab}+\dfrac{1}{2}\sum_{c=1}^{N^2-1}d_{abc} \lambda_c \ ,
\end{split}
\end{equation}
where $f_{abc}$ are completely antisymmetric tensors, called the \textit{structure constants} of the group, while $d_{abc}$ are completely symmetric tensors.\\
The way we introduced the SU($N$) gauge fields follows what has been originally the idea from Yang and Mills, i.e. they started from a fermionic matter field to point out that the gauge field has to be provided with a dynamics and, therefore a lagrangian of its own. That means the gauge field has the right be considered by itself regardless of the interaction with any other field.  Whether or not we consider the continuum theory or its discretizion on the lattice, the new gauge action has, of course, to satisfy our initial requirement of being gauge invariant .
 The simplest gauge invariant object that can be constructed on the lattice with just gauge fields is the \textit{plaquette}
\begin{equation}\label{Plaquette}
\Pi_{\mu\nu} = U_\mu(n) U_\nu(n+\overrightarrow{\mu}) U_\mu^\dag(n+\overrightarrow{\nu}) U_\nu^\dag(n) \ ,
\end{equation}
which, because of \eqref{gaugeU_transfom}, corresponds to the smallest possible closed loop. It is indeed the plaquette that is the basic element of the Wilson formulation of the lattice pure gauge action
\begin{equation}\label{Wilson action}
S_W \ = \ \beta_G\sum_n \sum_{\mu\nu} \left( 1 - \dfrac{1}{2 N} \left[ \Tr(\Pi_{\mu\nu})+\Tr(\Pi_{\mu\nu})^\dag \right] \right) \ ,
\end{equation}
where 
\begin{equation}
\beta_G\ =\ \dfrac{2N}{g^2} \ .
\end{equation}
Using the Baker-Hausdorff lemma and expanding the fields at the first order in $a$ around the point $n$ 
\begin{equation}
A_\mu(n+\nu)=A_\mu(n)+a\partial_\nu A_\mu(n) + O(a^2)\ ,
\end{equation}
it can be shown that the plaquette depends on the strength tensor $F_{\mu \nu}$ at the first order in $a$
\begin{equation}
\Pi_{\mu\nu}\ =\ e^{i a^2 g F_{\mu \nu} + O(a^4)}\ ,
\end{equation}
that correctly allows the Wilson action \eqref{Wilson action} to reduce to the well-known pure gauge action in the limit $a \rightarrow 0$
\begin{equation}\label{pureGaugeAction}
S_W\ \rightarrow \dfrac{1}{4} \int d^4x \ \Tr( F_{\mu \nu} F^{\mu \nu}) \ +O(a^2).
\end{equation}
The Wilson action \eqref{Wilson action} is just one of the available choices and any lattice action based on closed loops that succeeds in recovering the correct naive continuum limit \eqref{pureGaugeAction} can, in principle, be used.  Pure gauge actions based on more complicated closed loops are often used to achieve a weaker dependence on  the lattice spacing, in order to obtain a better scaling of the observables in the continuum limit.

\vspace{2cm}
\section{The continuum limit}
Performing the continuum limit of a theory is a more subtle process than just consider small lattice spacings. As we mentioned before there is a very large number of lattice actions that correspond to the same continuum formulation for $a=0$, however that is not sufficient to claim that the regularized theory processes the correct continuum limit. In particular, all the dimensional quantities which will be proportional to a non-zero power of $a$ will go to zero or infinity.

As we mentioned in section \ref{Monte Carlo simulations}, the essential requirement for the continuum limit to exist is that every correlation length  $\widehat{\xi}$, defined in lattice units, has to diverge compared to the lattice spacing. Hence, the continuum field theory can only be realized at a critical point $\lbrace g^*_i \rbrace$ in the parameter space of the discrete theory 
\begin{equation}
\widehat{\xi}(g_i)_{\ \ \overrightarrow{\ \ \lbrace g_i \rbrace \rightarrow \lbrace g^*_i \rbrace \ \ }}\ \ \infty \ .
\end{equation}
For the Wilson regularization \eqref{Wilson action} of an SU($N$) pure gauge theory, the only parameter is the bare coupling $g$ so that, in this case, the renormalization group is determined by just one equation. One can find it by requiring that any physical quantity $\mathcal{O}(a,g)$ should not depend on the regulator of the lattice spacing
\begin{equation}
\left[ a \dfrac{\partial}{\partial a} - \beta(g) \dfrac{\partial}{\partial g} \right] \mathcal{O}(a,g)\ =\ 0\ ,
\end{equation}
where
\begin{equation}\label{beta function}
\beta(g)\ =\ -a \dfrac{\partial g}{\partial a}
\end{equation}
is called $\beta-$\textit{function}. If one was able to know $\beta(g)$ then it would be possible to integrate \eqref{beta function} to obtain $g(a)$ and basically know the renormalization group of the theory for every value of $a$. Of course $\beta(g)$ cannot be known exactly, but it can still be calculated in perturbation theory around the critical point
\begin{equation}
\beta(g^*)\ =\ 0 \ .
\end{equation}
In the proximity of the fixed point we can expand the $\beta-$\textit{function} in powers of $(g-g^*)$
\begin{equation}
a \dfrac{\partial g}{\partial a}\  =\beta_1 (g-g^*)^3+\beta_2 (g-g^*)^5+ ... \ .
\end{equation}
For SU($N$), the first non zero term that can be compute in perturbation theory is $\beta_1=\frac{11N}{48\pi}$. Furthermore, we know from asymptotic freedom that  $g^*=0$. We are now able to predict, at least at the first order, the dependence of the lattice spacing from the coupling
\begin{equation}
a\ =\ \dfrac{1}{\Lambda_L} e^{- \dfrac{1}{2 \beta_0 g^2}} \ ,
\end{equation}
where $\Lambda_L$ in an integration constant with the dimensions of a mass and, in general, depends on the renormalization scheme that has been adopted.\\
The standard way to approach the continuum limit is to consider the ratio of dimensionful quantities of the theory. Let us suppose we know a particular mass $m_0$ from the experiments
\begin{equation}
m_0\ =\ \dfrac{1}{a(g) \xi(g)}\ ,
\end{equation}
where $\xi_{phy}=a(g) \xi(g)$ is the physical correlation length corresponding to the inverse of the mass $m_0$; we can use this information to set the scale of the theory on the lattice so that every other mass $m_i$ can be computed in relation to it
\begin{equation}\label{RG_massRatio}
\dfrac{m_i}{m_0}\ =\ \lim_{g \rightarrow g^*} \dfrac{\xi_0(g)}{\xi_i(g)} \ .
\end{equation}
In practice one can consider to have reached the continuum limit when the above ratio \eqref{RG_massRatio} holds constant within the statistical error of the measurement.

\begin{comment}

%% file: cap_Langevin/CLangevin.tex
\tableofcontents

\end{comment}

\externaldocument{cap_Langevin/CLangevin.tex}
\chapter{Complex Langevin dynamics}\label{cap:Langevin}
In this chapter we are going to review the main concepts behind complex Langevin dynamics along with its recent developments and successes. The main advantage  of this method, compared to standard Monte Carlo ones, is that it does not rely on the action to assign a weight to the field configurations. This approach, as we shall see, is therefore not affected by the \textit{sign problem} at all when it comes to numerical simulations.

\vspace{2cm}
\section{Stochastic dynamics}
The path integral approach described in the previous chapter, however successful, is just one way to quantize a QFT. In general, other choices are possible and, following Damgaard and Huffel \cite{Damgaard:1987rr}, we are going to review one of the most robust alternatives, namely \textit{stochastic quantization}. The idea was first introduced by Parisi and Wu \cite{Parisi:1980ys} in 1980 and consists of considering the Euclidian QFT as an equilibrium limit of a system governed by a stochastic process. The system evolves in an additional time $t_L$ under the effect of some drift force, determined by the system, together with random noise. When the equilibrium is reached, for $t_L \rightarrow \infty$, stochastic averages become identical to ordinary Euclidean vacuum expectation values.

The oldest and best known stochastic equation, and the one we are interested in, is the \textit{Langevin equation} \cite{P.Langevin:1908} 
\begin{equation}\label{Lang eq classic}
m \dfrac{d}{dt} \textbf{v}(t)\ =\ - \alpha \textbf{v}(t) + \boldsymbol{\eta }(t) \ ,
\end{equation}
introduced in 1908 to describe the Brownian motion of a particle of mass $m$ in a fluid with viscosity $\alpha$ that randomly collides with other particles of the fluid with intensity and direction $\boldsymbol{\eta }$. The latter is represented by a gaussian distributed random noise 
\begin{equation}\label{<eta>}
\begin{split}
&\langle \eta_i(t) \rangle\ =\ 0 \ , \\
&\langle \eta_i(t) \eta_j(t')\rangle\ =\ 2 \lambda \delta_{i j} \delta(t-t') \ ,
\end{split}
\end{equation}
with mean $0$ and variance $2 \lambda$ . This old model, however simple and classical,  is worth brief analysis to gain some insight into the Langevin stochastic process. The Langevin equation \eqref{Lang eq classic} is a non-homogeneous first order linear dfferential equation and therefore can be analytically solved as the Green's function
\begin{equation}\label{vi(t) Lang}
v_i(t)\ =\ \exp\left( -\dfrac{\alpha}{m}\ t\right) v_i(0) + \dfrac{1}{m} \int_0^t \exp\left(  -\dfrac{\alpha}{m}\ (t-\tau)\right) \eta_i(\tau) \ d\tau \ .
\end{equation}
We notice that the dependence on the initial conditions $v(0)$ is lost exponentially fast with time so that we might as well assume that $v(0)=0$ without losing any generality. Having an equation for $v(t)$ we want to calculate some physical quantity from it, for example the average kinetic energy of our Brownian particle
\begin{equation}
\begin{split}
\dfrac{1}{2} m\ \langle v^2(t) \rangle \ &=\ \dfrac{1}{2 m} \int_0^t d\tau \int_0^t d\tau' \exp\left( -\dfrac{\alpha}{m}\ (2t-\tau-\tau') \right) \langle \eta_i(\tau) \eta_j(\tau')\rangle \\
&=\dfrac{3 \lambda}{2 \alpha} \left[ 1- \exp \left(  -\ \dfrac{2 \alpha}{m}\ t\right) \right] \ .
\end{split}
\end{equation}
We note that by taking $\lambda=kT\alpha$ the correct value for the average kinetic energy $E=\frac{3}{2}kT$ is recovered for $t\rightarrow \infty. $\\
Since eventually we are going to be interested in numerical simulations, it is essential for us to know the behaviour of the probability distribution governing the Langevin stochastic process. Let us set $\lambda=1$, in this case, the observables will be functions of the velocity $v(t)$
\begin{equation}\label{avgO Lang}
\begin{split}
 \langle \mathcal{O}(v(t))\rangle_\eta \ & \equiv \ \int D\eta\ \exp\left(-\frac{1}{4} \int_0^t \eta^2(\tau) d\tau \right) \mathcal{O}(v(t)) \\
&=  \int dv\ \mathcal{O}(v) P(v,t) \ \equiv \ \langle \mathcal{O}(v(t))\rangle_P \ ,
 \end{split}
 \end{equation} 
where we used the fact that averages can be computed either over the noise $\eta$ or over the probability distribution $P(v)$ ,
\begin{equation}
\langle f \rangle_\eta\ =\ \langle f \rangle_P \ .
\end{equation}
Taking the time derivative of \eqref{avgO Lang} and using the Langevin equation \eqref{Lang eq classic} (with $m=\alpha=1$), one gets
\begin{equation}\label{FP-eq classic1}
  \langle\ \dfrac{\partial \mathcal{O}(v)}{\partial v}\  \dfrac{d v}{d t}\  \rangle =  \langle\ \dfrac{\partial \mathcal{O}(v) }{\partial v}\ (-v+\eta)\ \rangle\ =   \int dv\ \mathcal{O}(v) \dfrac{\partial P(v,t)}{\partial t} \ .
 \end{equation} 
Furthermore, using eq.\ \eqref{avgO Lang} and integrating by parts, we can write\\
\begin{equation}\label{FP-eq classic2}
\begin{split}
\langle\ \dfrac{\partial \mathcal{O}(v) }{\partial v}\eta\ \rangle\ =\ & \int D\eta\ \left[ \dfrac{\partial}{\partial \eta(t)}\exp\left(-\frac{1}{4} \int_0^t \eta^2(\tau) d\tau \right) \right] \dfrac{\partial \mathcal{O}(v) }{\partial v}\ =\\
& \ \ \  2 \langle\ \dfrac{\partial^2 \mathcal{O}(v) }{\partial v^2} \dfrac{\partial v}{\partial \eta} \rangle \ =\ \langle\ \dfrac{\partial^2 \mathcal{O}(v) }{\partial v^2}  \rangle,
\end{split}
\end{equation}
where in the last step we used the expression \eqref{vi(t) Lang}  for $v(t)$
\begin{equation}
\dfrac{\partial v(t)}{\partial \eta(t)}\ =\ \dfrac{\partial}{\partial \eta(t)} \int_0^\infty \theta(t-\tau) e^{  -\ (t-\tau)} \eta_i(\tau) \ d\tau  = \theta(0) = \dfrac{1}{2}
\end{equation}
adopting the middle point prescription for the Heaviside step function $\theta(t-\tau)$. At this point, we can use \eqref{FP-eq classic2} and, after integrating by part, we can rewrite \eqref{FP-eq classic1}  as
\begin{equation}
\int dv \ \mathcal{O}(v)\ \left[ \dfrac{\partial}{\partial v}  \left( v + \dfrac{\partial}{\partial v} \right) \right] P(v,t) \ =\  \int dv\ \mathcal{O}(v) \dfrac{\partial P(v,t)}{\partial t} \ .
\end{equation}
This gives an equation for the evolution of the probability distribution $P(v,t)$ that goes under the name of \textit{Fokker-Planck} equation
\begin{equation}\label{Fokker-Planck eq Brownian}
\dfrac{\partial P(v,t)}{\partial t}\ =\ \dfrac{\partial}{\partial v}  \left( v + \dfrac{\partial}{\partial v} \right) P(v,t) \ .
\end{equation}
We notice that the stationary solution $\frac{\partial}{\partial t} P =0$ of  eq.\eqref{Fokker-Planck eq Brownian}, after requiring the condition $P(v)=P(-v)$, leads to the Boltzmann distribution for the Brownian particle in equilibrium with the system 
\begin{equation}
P^{eq}\ \sim \ \exp\left( -\frac{v^2}{2} \right)\ .
\end{equation}

The Fokker-Planck equation, as we shall see, will play a crucial role in the numerical application of the Langevin dynamics in the case of a complex field theory.

\section{Stochastic quantization of a field theory}
The idea behind stochastic quantization is to formulate the equivalent of the Langevin equation \eqref{Lang eq classic} for a field theory in such a way that the associated Fokker-Planck distribution may have the Euclidian Boltzmann distribution $\exp(-S_E)$ as the unique stationary solution.

The first step is the introduction of an additional fictitious time $t_L$ in which the stochastic systems evolves 
\begin{equation}
\phi(x_0,..,x_n) \rightarrow \phi(x_0,..,x_n,t_L) \ .
\end{equation}
From now on we are going to call the Langevin time $t_L$ just $t$, having in mind it is different from the Euclidian time $x_0$ \ .\\
The second requirement is that the evolution of fields be described by the Langevin stochastic equation
\begin{equation}\label{Lang eq field}
\dfrac{\partial}{\partial t} \phi(x,t) \ =\ - \dfrac{\delta S}{\delta \phi(x,t)} + \eta(x,t) \ ,
\end{equation}
where $S$ is the Euclidian action of the field theory, which also depends on the Langevin time ,
\begin{equation}
S\ =\ \int dt d^n x \ \mathcal{L}\left(\phi(x,t),\dfrac{\partial}{\partial t}\phi(x,t) \right) \ , 
\end{equation}
and $\eta(x,t)$ is again Gaussian white noise 
\begin{equation}
\begin{split}
&\langle \eta(x,t) \rangle = 0 \ , \\
&\langle \eta(x,t) \eta(x',t')\rangle = 2  \ \delta^n(x-x') \delta(t-t') \ .
\end{split}
\end{equation}
In the same way as in the classical case, equation \eqref{Lang eq field} is associated with a probability distribution function $P(\phi,t)$ for the fields at the Langevin time $t$
\begin{equation}
\langle \phi(x_1,t)...\phi(x_n,t)\rangle_\eta \ =\ \int D\phi\ P(\phi,t) \ \phi(x_1)...\phi(x_n) \ ,
\end{equation}
which satisfies the Fokker-Planck equation, corresponding to \eqref{Fokker-Planck eq Brownian}, generalized for the field theory
\begin{equation}\label{Fokker-Planck}
 \dfrac{\partial P(\phi,t)}{\partial t}\ =\ \int d^nx \ \dfrac{\delta}{\delta \phi(x,t)}  \left(  \dfrac{\delta S}{\delta \phi(x,t)} + \dfrac{\delta}{\delta \phi(x,t)} \right) P(\phi,t) \ .
 \end{equation} 
As a final remark we would like to show that the former equation \eqref{Fokker-Planck} leads $ P(\phi,t)$ to converge to the Euclidian Boltzmann weight of the standard path integral quantization exponentially fast with Langevin time. Let us consider, for simplicity, one degree of freedom $x$. The partition function of this system reads
\begin{equation}
Z\ =\ \int dx\; e^{-S(x)} 
\end{equation}
and the corresponding Langevin equation is
\begin{equation}\label{Lang eq x}
\dfrac{d x}{dt} \ =\ - \partial_x S(x) + \eta \ .
\end{equation}
The time evolution of the associated probability distribution function $P(x)$ 
\begin{equation}\label{PDF_FP}
\langle \mathcal{O}(x) \rangle \ =\ \int dx\; \mathcal{O}(x) P(x,t)
\end{equation}
is determined by the Fokker-Planck equation
\begin{equation}\label{FP(x)}
\partial_t P(x,t) =\ \partial_x \left[ \partial_x + \partial_x S(x) \right] P(x,t) \ ,
\end{equation}
whose stationary point is easily found to be $P(x) \sim e^{-S(x)}$. Moreover, we can rewrite eq.\eqref{FP(x)} upon the transformation
\begin{equation}
P(x,t)\ =\ \psi(x,t) \ e^{- \frac{1}{2} S(x)} \ ,
\end{equation}
into the Schr{\"o}dinger-like equation
\begin{equation}
\dot{\psi}(x,t)\ =- \ H_{FP} \ \psi(x,t)
\end{equation}
where 
\begin{equation}\label{H_FP}
 H_{FP} \ =\ \left( -\partial_x + \frac{1}{2} S'(x) \right) \left( \partial_x + \frac{1}{2} S'(x) \right) \ .
\end{equation}
The operator \eqref{H_FP} is self-adjoint and, if $\vert \lim_{x \rightarrow \infty} S'(x) \rightarrow \infty \vert$, the spectrum of its eigenvalues is non negative and discrete
\begin{equation}
H \Psi_n \ =\ E_n \Psi_n \ ,
\end{equation}
with the ground state $\Psi_0 \sim e^{- \frac{1}{2} S(x)}$ annihilating  eq.\ \eqref{H_FP}. Therefore we can rewrite $\psi(x,t)$ on the base of the eigenvectors of $H_{FP}$
\begin{equation}
\psi(x,t)\ =\ c_0\ e^{- \frac{1}{2} S(x)} \ +\ \sum_n c_n \Psi_n(x)\ e^{- E_n t} \ \rightarrow \ c_0\ e^{- \frac{1}{2} S(x)} 
\end{equation}
and the correct distribution $P(x)\sim e^{-S(x)}$ is reached exponentially fast.

\vspace{2cm}
\section{Complex Langevin dynamics}
We already discussed how a complex weight prevents the application of standard Monte Carlo methods. On the other hand, stochastic processes do not rely on importance sampling, which makes them good candidates to deal with the sign problem. In this section we shall see how Langevin dynamics can be generalized to the case of complex actions $S(x)$, examining in detail the careful steps that make this method successful.\\
A straightforward adaptation of \eqref{Lang eq x} for a complex action is still possible and would lead to the Langevin equation
\begin{equation}
  \dot{x}\ =\ - \partial_x S(x) + \eta
  \end{equation}  
and, consequently, to the FP equation
\begin{equation}\label{Complex FP x}
\partial_t \rho(x,t) \ =\  L_0^T \rho(x,t) \ ,
\end{equation}
where $L_0^T$ is the usual Fokker-Planck operator $L_0^T\ =\ \partial_x \left( \partial_x + \partial_x S_c(x) \right) $ that now is complex. Eq.\eqref{Complex FP x} is expected to have the desired complex weight  
\begin{equation}
\rho(x) \sim e^{- S(x)}
\end{equation}
as a stationary solution. However, being complex-valued, $\rho(x)$ is not suitable to be regarded as a probability distribution function (PDF) as in \eqref{PDF_FP}. Furthermore, the associated FP Hamiltonian $H_{FP}(z)$, the complex equivalent of \eqref{H_FP}, is not self-adjoint any more, so that a proof of exponentially fast convergence to the unique solution cannot be provided. The way to proceed then \cite{Aarts:2011ax,Aarts:2013uxa}, is to consider the real and imaginary parts of the complexified variables $z\rightarrow x +i y$ as new and independent  degrees of freedom  
\begin{equation}\label{Lang x e y}
\begin{split}
&\dot{x}\ =\ K_x+\sqrt{N_R}\ \eta_R \\
&\dot{y}\ =\ K_y+\sqrt{N_I}\ \eta_I \ ,
\end{split}
\end{equation}
with the two drifts
\begin{equation}\label{drifts KxKy}
K_x\ =\ - \Re e \ \partial_z S(z), \ \ \ \ \ \ \ \ K_x\ =\ -\Im m\ \partial_z S(z) \ .
\end{equation}
The correlators between the noises $\eta_R$ and $\eta_I$ derive from the original prescription \eqref{<eta>} on the complex noise $\eta=\eta_R+i \eta_I$ and read
\begin{equation}\label{eta noise}
\begin{split}
&\langle \eta_R(t) \eta_R(t') \rangle\ =\ 2 N_R \delta(t-t') \\
&\langle \eta_I(t) \eta_I(t') \rangle\ =\  2 N_I \delta(t-t') \\
&\langle \eta_R(t) \eta_I(t') \rangle\ =\ 0 
\end{split}
\end{equation}
where $N_R-N_I=1$ and $N_I \geq 0$. 

 The complexification of the Fokker-Planck equation
\begin{equation}
 \dot{P}(z,t)\ =\ \partial_z (N_R \partial_z -K_z ) P(z,t)
 \end{equation} 
 can be written, for \textit{holomorphic} observables, in terms of the two independent variables $x(t)$ and $y(t)$ in \eqref{Lang x e y}
\begin{equation}\label{Real FP xy}
\dot{P}(x,y,t)\ =\ \left[ \partial_x (N_R \partial_x -K_x )  +  \partial_y (N_I \partial_y -K_y ) \right] P(x,y,t)
\end{equation}
and has the form of a continuity equation with the probability density $P(x,y;t)$ being the charge
\begin{equation}
\dot{P}(x,y,t)\ =\ \partial_x J_x + \partial_y J_y \ ,
\end{equation}
and 
\begin{equation}
J_x \ =\  (N_R \partial_x -K_x ) P \ , \ \ \ \ \ \ \  J_y\ =\ (N_I \partial_y -K_y ) P \ ,
\end{equation}
being the currents. Eq.\eqref{Real FP xy}  generates a \textit{real}  PDF for the holomorphic observables
\begin{equation}
\langle \mathcal{O} \rangle_{P(t)} \ =\ \int dx dy\ P(x,y,t) O(x+iy) \ ,
\end{equation}
which is, in fact, the main idea of complex Langevin  (CL) dynamics, i.e. to reformulate a $d-$dimensional complex system into a $2d-$dimensional real one. One requirement is to consider the holomorphic continuations of the observables $\langle \mathcal{O}(x)\rangle \rightarrow\langle\mathcal{O}(z)\rangle=\langle\mathcal{O}(x+i y)\rangle$. In this sense neither the quantity $\langle \mathcal{O}(x)\rangle$ nor $\langle \mathcal{O}(y)\rangle$ have, by themselves, any meaning in the complexified space.   
\\

The reason why complex Langevin dynamics was not largely employed immediately after its introduction in the 80's is that the equation \eqref{Real FP xy}, even keeping the same form of the real case,  is much harder to solve or to be proved convergent to the appropriate stationary distribution $\rho(x) \sim e^{-S(x)}$ \cite{Klauder:1985ks,Ambjorn:1985iw,Ambjorn:1986fz}. On top of that, unstable solutions of \eqref{drifts KxKy} can be found in the complex plane and that was believed to inevitably spoil the dynamics when solved numerically. These two problems have been more recently addressed, allowing CL to become one of the most acknowledged methods when it comes to system affected by the sign problem.\\
The issue of instabilities on the lattice was the first to be successfully and consistently solved \cite{Aarts:2009dg}. The discretized CL equations for the field $\phi$ are
\begin{equation}
\begin{split}
&\phi_x^R(n+1)\ =\ \phi_x^R(n)+\epsilon K_x^R + \sqrt{\epsilon N_R}\ \eta^R_x(n) \\
&\phi_x^I(n+1)\ =\ \phi_x^I(n)+\epsilon K_x^I+ \sqrt{\epsilon N_I}\ \eta^I_x(n)  \ ,
\end{split}
\end{equation}
where $\epsilon$ is the discrete Langevin time step and $x$ labels the sites of the lattice.  When the system is brought near an unstable trajectory $\tau(\phi^R,\phi^I)$, the drifts $K^R(\tau)$ and $K^I(\tau)$ can potentially lead the fields to infinity, in a finite Langevin time $t_L=\epsilon n$. It turns out that careful integration in the form of adaptive stepsize $\epsilon_n$ along those trajectories is enough to completely remove the problem. The idea is to keep the product $\epsilon_n \mathcal{K}$ constant, where $\mathcal{K} = f(\sqrt{K_R^2+K_I^2})$ is a function of the drift to be chosen optimally depending on the system, in order to greatly reduce the stepsize $\epsilon_n$ along the unstable trajectories and allow the real component of the random noise $\eta^R$ to kick the system away from such trajectories. For this purpose, the imaginary component of the random noise $\eta^I$ is, in general, counter-productive so that is usually preferable to get rid of it. This is in perfect accord with the prescriptions \eqref{eta noise} and corresponds to the choice of parameters $N_I=0$ and $N_R=1$.  \\\\
The problem of convergence of CL is much more complicated to address.  Although no definitive solution has been found yet, fundamental progress has been made to fully understand this issue. In particular \textit{proofs of convergence} have been found to infer, from the distribution of the observables, whether the CL is expected to converge to the right result or not \cite{Aarts:2009uq,Aarts:2011ax,Aarts:2011sf,Aarts:2012ft,Aarts:2013uza}. In the following, we are going to review the main arguments behind these criteria. \\
We can rewrite eq. \eqref{Real FP xy} in a more compact way
\begin{equation}\label{Real FP xy L^T}
\dfrac{\partial}{\partial t} P(x,y;t)\ =\ L^T P(x,y;t)
\end{equation}
where
\begin{equation}
 L^T \ =\ \partial_x (N_R \partial_x -K_x )  +  \partial_y (N_I \partial_y -K_y ) \ .
 \end{equation} 
It is easy to see that the adjoint of the operator $L^T$ determines the time evolution of a function $f$ along a solution $z(t)=x(t)+iy(t)$ of the CL equation \eqref{Lang x e y}. Let us consider the scalar product :
\begin{equation}
\langle P, f \rangle \ =\ \int f(x,y) P(x,y) \ dxdy
\end{equation}
where the time evolution operator can be moved from the probability density $P(x,y;t)$ to the function $f$ in the usual way
\begin{equation}
 \langle L^T P, f \rangle \ =\ \langle P, L f \rangle  \ ,
 \end{equation}
so that the time evolution of  $f$ is described by
\begin{equation}
\dfrac{d}{dt} \langle f(x(t),y(t))\rangle\ =\ \langle L \ f(x(t),y(t))  \rangle \ ,
\end{equation}
where the brackets mean averages over the stochastic noise \eqref{eta noise} and the  Langevin operator $L$ has the form
\begin{equation}
L\ =\  (N_R \partial_x + K_x ) \partial_x +  (N_I \partial_y + K_y ) \partial_y  \ .
\end{equation}
Our purpose is to obtain some conditions under which the analytically continued holomorphic observables $\mathcal{O}(x+iy)$, distributed according to the real-valued PDF \eqref{Real FP xy} $P(x,y;t)$, retain the same average values as if defined on the real manifold $\mathcal{O}(x)$, weighted with the complex function $\rho(x)$ in \eqref{Complex FP x}, i.e.   
\begin{equation}
\begin{split}
&\langle \mathcal{O}\rangle_{P(t)} \ =\ \dfrac{\int \mathcal{O}(x+iy) P(x,y;t) dxdy}{\int P(x,y;t) dxdy} \ ,\\
&\langle \mathcal{O}\rangle_{\rho(t)} \ =\ \dfrac{\int \mathcal{O}(x) \rho(x;t) dx}{\int \rho(x;t) dx} \ .
\end{split}
\end{equation}
In other words, one would like
\begin{equation}\label{O_P=O_rho}
\langle \mathcal{O}\rangle_{P(t)} \ =\ \langle \mathcal{O} \rangle_{\rho(t)} 
\end{equation}
provided the two are identical at the start $ \langle \mathcal{O} \rangle_{P(0)} \ =\ \langle \mathcal{O} \rangle_{\rho(0)}$, which is easily assured if
\begin{equation}\label{P(0) e rho(0)}
P(x,y;0)\ =\ \rho(x;0)\delta(y) \ .
\end{equation}
In order to link the two expressions for the average values of $\mathcal{O}$ in \eqref{O_P=O_rho}, we define for $0 \leq \tau \leq t$ the function
\begin{equation}
F(t,\tau)\ =\ \int P(x,y;t-\tau) \mathcal{O}(x+iy;\tau) dx dy
\end{equation}
which interpolates between the two definitions
\begin{equation}
F(t,0)= \langle \mathcal{O} \rangle_{P(t)},\ \ \ \ \ \ \ F(t,t)=\langle \mathcal{O} \rangle_{\rho(t)}\ .
\end{equation}
While the first equality is straightforward, the second can be seen using the prescription for the initial conditions \eqref{P(0) e rho(0)} and integrating by part
\begin{equation}
\begin{split}
F(t,t)\ & =\  \int P(x,y;0) \left( e^{t L} \mathcal{O} \right) (x+iy,0) dxdy\\
& =\ \int \rho(x;0) \left( e^{t L_0} \mathcal{O} \right) (x,0)  dx    \\
&  =\ \int \mathcal{O}(x,0) \left( e^{t L^T_0} \rho \right) (x,0)  dx \\
&  =\ \int \mathcal{O}(x,0) \rho(x;t) \ =\ \langle \mathcal{O} \rangle_{\rho(t)} \ .
\end{split}
\end{equation}
We only had to assume that no boundary contribution is introduced by the integration by parts, which is a very standard requirement on $\rho(x)$ without which it cannot possibly be integrated anyway. 
One could now obtain the \eqref{O_P=O_rho} on the condition that the interpolating function $F(t,\tau)$ is independent on $\tau$ .
\begin{equation}
\begin{split}
\dfrac{\partial}{\partial \tau} F(t,\tau) \ =\ - & \int \left( L^T P(x,y;t-\tau)\right) \mathcal{O}  (x+iy;\tau) dxdy \\
& +  \int P(x,y;t-\tau) \left( L \mathcal{O}\right) (x+iy;\tau) dxdy
\end{split}
\end{equation}
It is evident, after integration by part, that the two pieces of  the former equation are equal and opposite except for a possible \textit{boundary contribution coming from infinity in the complex plane}. 

The one above is the core argument for the \textit{criteria of correctness}, that can be summarized like this :\\ 
the formal argument for which complex Langevin is expected to converge to the right result, might fail when the distribution of the observables
\begin{equation}\label{PO}
P(x,y;t) \mathcal{O}(x+iy) 
\end{equation}  
does not \textit{decay fast enough in the complex plane}, giving rise to boundary terms. 
On the other hand, if the distribution \eqref{PO} is localized enough, one expects CL to converge to the right result (detailed studies of those conditions can be found here
\cite{Aarts:2013uza,Aarts:2012ft}). In general, more care is needed when the drift $\partial_z S(z)$ has poles, since the hypothesis of holomorphicity drops. In those cases one has to study in detail how the dynamics is influenced by the poles and whether the results are correct or not will depend on the specific case \cite{Mollgaard:2013qra,Greensite:2014cxa} .

Let us conclude by saying that, when the criteria of correctness are satisfied, the stationary Fokker-Plank probability  $P(x,y)$ should be positive-definite, since it is the distribution built up during the Langevin process. E.g. by binning the process one sees that in a bin with size $dx dy, P(x,y)$ is either $0$ or $>0$. This assumes the relation between the Langevin process \eqref{Lang x e y} and the FP distribution $P(x,y)$ holds, which it should when the stochastic process converges. Whether it is normalisable, depends on how the distribution goes to zero at large $x$ and $y$. In numerical simulations, we always check the behaviour of $P(x,y)$ at large $x$ and $y$, since it is also necessary for the criteria of correctness, and it always seems to be fast enough. However one might still argue that long tails might not be adequately sampled. In practice, we always find that if the complex Langevin converges, the stationary Fokker-Plank probability is normalisable.

\vspace{2cm}
\section{Gauge Theories and Gauge Cooling}
\begin{figure}[t] 
\centering 
\includegraphics[width=0.4\textwidth]{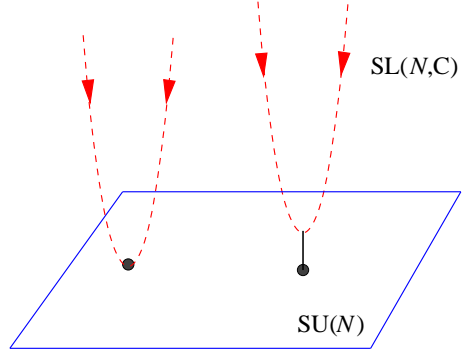} 
\caption{Gauge cooling in SL$(N,\mathbb{C})$ brings the link as close as possible to SU$(N)$. The orbit on the left is equivalent to a SU$(N)$ configuration, while the one on the right is not. } 
\label{fig:SUN orbit} 
\end{figure}

We will now start to describe the application of Langevin dynamics to gauge theories with sign problem. The most recent developments allowed the use of CL to investigate some of the most challenging and fundamental problems in the physics of strong interactions, such as finite density heavy quark QCD \cite{Seiler:2012wz,Aarts:2013uxa,Bongiovanni:2013nxa,Aarts:2013nja}, full QCD \cite{Sexty:2013ica,Aarts:2014bwa,Aarts:2014kja,Aarts:2014cua,Sexty:2014zya} and QCD in the presence of a topological $\theta-$term related to the strong CP problem \cite{Bongiovanni:2013nxa,Bongiovanni:2014rna}. We will review CL dynamics in the context of gauge theories, introduce the \textit{gauge cooling} method and analyze in some detail how this helps to control the criteria of convergence.\\
For nonabelian SU($N$) gauge theories on the lattice, the Langevin equation for the link $U$ reads
\begin{equation}\label{Gauge CL}
U_{\mu x}(t+\epsilon) = R_{\mu x}(t)U_{\mu x}(t), \ \ \ \ \ \ R_{\mu x}(t)= \exp\left[ -i \sum_a \lambda_a  \left( \epsilon D_a S[U_{\mu x}] + \sqrt{\epsilon} \eta_{a \mu x} \right)  \right] \ ,
\end{equation}
where $\epsilon$ is the discrete Langevin time step used during the simulations, $\lambda_a$ are the Gell-Mann matrices ($a=1,...,N-1$) and $D_a$ is the \SUN Lie derivative 
\begin{equation}
D_a f(U)\ =\ \dfrac{\partial}{\partial \omega} f\left( e^{i \omega \lambda_a} U\right)\vert_{\omega=0} \ .
\end{equation}
As usual $\eta_a$ is \textit{real} Gaussian noise satisfying the relations
\begin{equation}
\begin{split}
&\langle \eta_{a \mu x} \rangle \ =\ 0 \ ,\\
&\langle \eta_{a \mu x}(t)\; \eta_{b \mu x}(t') \rangle  = 2  \delta(t-t') \delta_{ab} \ . 
\end{split}
\end{equation}
Looking at eq.\eqref{Gauge CL}, one immediately realizes that for complex actions $S[U]$ the operator $R(t)$ takes values into the group \SLN. Consequently, even if at the beginning the links $U$ were in \SUN , the whole dynamic will be naturally enlarged in the bigger \SLN\ . The element of this group retain the property $\det U=1$ but lose the unitarity : $U^{-1} \neq U^\dag$. To preserve analyticity then, every observable will have to be expressed in terms of $U$ and $U^{-1}$, rather than $U^\dag$. For example the correct analytical continued equivalent of the scalar quantity $\Tr(U U^\dag )$, equal to $N$ in \SUN , would be $\Tr(U U^{-1})$  which, in fact, is still equal to $N$ for every matrix $U\in$ \SLN. The operator $\Tr(U U^\dag )$ itself, however, can now take any real value bigger than $N$, as a consequence of the non compactness of \SLN. In particular this can be seen since every element $U \in $ \SLN \ can be written (polar decomposition) as
\begin{equation}
U=PV
\end{equation}
with $V \in$ \SUN\ and $P=P^\dag$ a positive definite matrix with $  \det P =1$. This property of $\Tr(U U^\dag )$ allows us to define the \textit{Unitarity Norm} (UN) $\dd$, a quantity that measures how deep in the \SLN\ manifold is the configuration
\begin{equation}\label{UN}
\dd \ =\ \dfrac{1}{N}\Tr \left( UU^\dag-\Id \right) \geq 0 \ ,
\end{equation}
where evidently \dd = 0 for \SUN . It is possible to define the UN in many other ways that still allows to quantify the distance to real manifold \SUN, but the one in eq.\eqref{UN} is the simplest and the one we are going to use in the following. \\
Gauge invariance is still present in \SLN 
\begin{equation}
U_\mu(x) \rightarrow U'_\mu(x)=\Omega(x)  U_\mu(x) \Omega^{-1}(x+\mu), \ \ \ \ \ \Omega \in $ \SLN$ \ ,
\end{equation}
however the gauge transformation $\Omega(x)$ does not leave the UN \eqref{UN} invariant, while its analytical continuation Tr$(UU^{-1})$ is preserved. Let us note that the parameters of the gauge group are now twice as many as in the case of \SUN
\begin{equation}
 \Omega=e^{i \omega_a \lambda_a}, \ \ \ \ \omega_a \in \mathbb{C} \ 
 \end{equation} 
and, as we are about to learn, this abundance of gauge freedom is rather harmful for CL dynamics. 
It has been observed for many gauge models \cite{Berges:2007nr,Seiler:2012wz,Aarts:2013uxa} that the CL dynamics, if left alone, tend to explore the huge gauge freedom available, bringing the system deep in the complex manifold. That, as one might expect, eventually leads to a wide distribution that violates the criteria of correction mentioned in the previous section. \\
The way to deal with this problem, first introduced in \cite{Seiler:2012wz}, takes the name of \textit{gauge cooling} (gc.),  and consists in gauge transforming all links of a given configuration up to the point where the UN \eqref{UN} is minimal.  One can achieve this by choosing as a parameter of the gauge transformation the gradient of the UN itself along the
gauge orbit :
\begin{equation}\label{gc1}
\Omega(x) = e^{-\epsilon \alpha_{\rm gf}f_{ax} \lambda_a}, \quad\quad f_{ax} = 2 \Tr\left[ \lambda_a \sum_\mu \left(U_{x,\mu}U_{x,\mu}^\dag -U_{x-\hat\mu,\mu}^\dag U_{x-\hat\mu,\mu} \right)\right],
\end{equation}
where $\epsilon$ is of the order of magnitude of the stepsize used in the Langevin process and $\alpha_{gf}$ is a parameter which can still be chosen to optimize the gauge cooling. One can convince himself that the gauge cooling reduces the UN by looking at the effect an infinitesimal gauge transformation \eqref{gc1} has on \eqref{UN}. For convenience we chose the gauge cooling transformation to be active only on the even sites while being the $\Id$ on the odd ones
\begin{equation}
\begin{split}
&U_\mu(x)\rightarrow U'_\mu(x)=\Omega(x) U_\mu(x) \ , \\
&U_\mu(x-\mu)\rightarrow U'_\mu(x-\mu)= U_\mu(x-\mu) \Omega^{-1}(x)\ . 
\end{split}
\end{equation}
With this convention the change in UN along the gauge cooling trajectory can be easily calculated
\begin{equation}\label{d-d'}
\dd'-\dd=-\dfrac{\epsilon \alpha_{gf}}{N} (f_{ax})^2\ +\ \mathcal{O}((\epsilon \alpha_{gf})^2) \ .
\end{equation}
Depending only on $(f_{ax})^2$, the right hand side of the equation above is always negative, up to higher orders in $\epsilon \alpha_{gf}$, resulting in a monotonic decrease of the UN itself. If the original configuration is gauge-equivalent to a \SUN \ one, cooling will eventually transform the configuration into the unitary one. Once the unitary manifold is reached, $f_{ax}$ vanishes, as evident from \eqref{gc1}, so that gauge cooling no longer has any effect. On the other hand, if the starting configuration is not gauge-equivalent to \SUN, gauge cooling will bring it as close as possible, i.e. it will minimize $\dd$ (Fig.\ref{fig:SUN orbit}). 

Although it is rather difficult to show that the minimum of the unitarity norm is unique, in practice, we observed many times that if we operate a \SLtree\ random gauge transformation on a configuration and then we apply gauge cooling, the configuration is always brought back to the same minimum. That is a strong indication that the minimum is unique. However, even if the minimum wasn't unique, that wouldn't mine the usefulness of gauge cooling, which idea is to keep the dynamic close to the real manifold. 

It is also important to stress that, since the gauge cooling process is separate from the CL evolution, it is not equivalent to gauge fixing term, i.e. it is not derive from a gauge fixing of the action,  and, consequently, no Fadeev-Popov determinant is needed.

Let us remark that the complex Langevin dynamics is gauge invariant, in the sense that having an infinite precision machine the results with or without gauge cooling would agree. The effect of gauge cooling is to minimizing the rounding errors by keeping the links as 'small' as possible with a gauge transformation.

\vspace{1cm}
\subsubsection{One link model}\label{One link model}
Let us have a look at the gauge cooling process in a 1-link \SLN \ model \cite{Aarts:2013uxa}, where analytical insight is possible. Since gauge cooling is completely disconnected from the CL dynamics itself, the form of the action is irrelevant to the following. After taking the continuous Langevin time limit ($\epsilon \rightarrow 0$), the eq.\eqref{d-d'} becomes, to leading order in $\alpha_{gc}$,
\begin{equation}\label{dd dot}
\dot{\dd}=-\dfrac{\alpha_{gf}}{N} (f_{a})^2\
\end{equation}
with
\begin{equation}
f_a=2 \Tr \left[ \lambda_a \left( UU^\dag - U^\dag U \right)\right]\ ,
\end{equation}
where the $\mu$ and $x$ indices can be dropped since there is only one link.
Using the identity
\begin{equation}
\lambda^{ij}_a \lambda^{kl}_a=2 \left( \delta_{il}\delta_{jk}-\dfrac{1}{N}  \delta_{ij}\delta_{kl}\right) \ ,
\end{equation}
eq.\ \eqref{dd dot} can be written as
\begin{equation}\label{dd dot 2}
\dot{\dd}=- \dfrac{16 \alpha_{\rm gf}}{N} \Tr \left( UU^\dag \left[U,U^\dag \right] \right).
\end{equation}
In the case of SU(2) this expression can be further simplified and it is written as
\begin{equation}
\dot{\dd} = -8 \alpha_{\rm gf} \left(\dd^2+2(1-|c|^2)\dd+c^2+c^{*2}-2|c|^2\right)\ ,
\end{equation}
where  $c=\frac{1}{2} \Tr\, U$ and $c^*=\frac{1}{2} \Tr\, U^\dag$ (the case $c=1$ refers to $U=\Id$ which is trivial). When $c=c^* $, $U$ is gauge-equivalent to an element of SU(2) and eq.\ \eqref{dd dot 2} simplifies to
\be
\dot{\dd} = -8 \alpha_{\rm gf} \left(\dd+2(1-c^2)\right)\dd.
\ee
This equation indeed has a  {\em unique} fixed point at  $\dd=0$, which is reached exponentially fast,
\be\label{1link-fixpoint}
 \dd(t)\sim 2 (1-c^2) e^{-16\alpha_{gf}(1-c^2)t} \rightarrow 0.
 \ee
On the other hand, if $c \neq c^*$, $U$ cannot be gauge-equivalent to a SU(2) matrix and the stationary point is 
\be
\dd(t \rightarrow \infty ) =  |c|^2 -1 + \sqrt{1-c^2-c^{*2}+|c|^4} > 0,
\ee
where the minimum distance is, as expected, larger than 0. This brief example and the analytical computations involved support what is shown in the sketch of Fig.\ref{fig:SUN orbit}. When it comes to more complicated models, the effect of gauge cooling can only be computed numerically but the behaviour remains the same.

\vspace{1cm}
\subsubsection{Adaptive gauge cooling for \SU2 Polyakov chain}
Most of the cases of interest are too complicated to be studied analytically so that numerical computations are required. Here we shall discuss the case of a one-dimensional chain of $N_l$ \SU2 gauge links (Polyakov chain), with the action
\begin{equation}\label{SPolChain}
S=-\dfrac{\beta}{2} \Tr\left( U_1U_2...U_{N_l}\right)\ , \ \ \ \ \beta \in \mathbb{C}
\end{equation}
and partition function 
\begin{equation}\label{ZPolChain}
Z=\int \prod_{k=1}^{N_l} DU_k \ e^{-S[U]} \ .
\end{equation}
One might notice that all the links except one could be transformed into the identity matrix $\Id$ using the gauge transformation
\begin{equation}
U_n\rightarrow U'_n=U_n^{-1} U_n  U_{n+1}\ .
\end{equation}
The problem is, then, equivalent to the one-link model with $S=-\dfrac{\beta}{2} \Tr\left( U \right)$, where the moments
\begin{equation}
\langle \Tr(U^n) \rangle=\int  DU \ e^{-S[U]}\  \Tr(U^n) \ =\ \dfrac{I_n(\beta)}{I_0(\beta)}
\end{equation}
are known in term of the modified Bessel functions of the first kind
\begin{equation}
I_n(\beta)=\dfrac{1}{\pi} \int_0^\pi d\theta\ e^{\beta \cos(\theta)} \cos(n \theta) \ .
\end{equation}
This property is very useful in that it allows us to know the expectation value of the observable $\Tr(U)$ and, at the same time, it does not lessen the interest in the original Polyakov chain model. In fact, from a numerical point of view, the full model in \eqref{SPolChain} still has as many degrees of freedom as the number of group parameters times the number of links, and can be used as a useful toy model to study the effect of gauge cooling in anticipation of proper four-dimensional gauge theories.\\
Before studying the actual dynamics of the system, we start by gauge transforming a \SU2 configuration into \SL2 and applying gauge cooling to it until it is back in the original real manifold. 
\begin{figure}[t] 
\begin{center}
\includegraphics[width=0.8\textwidth]{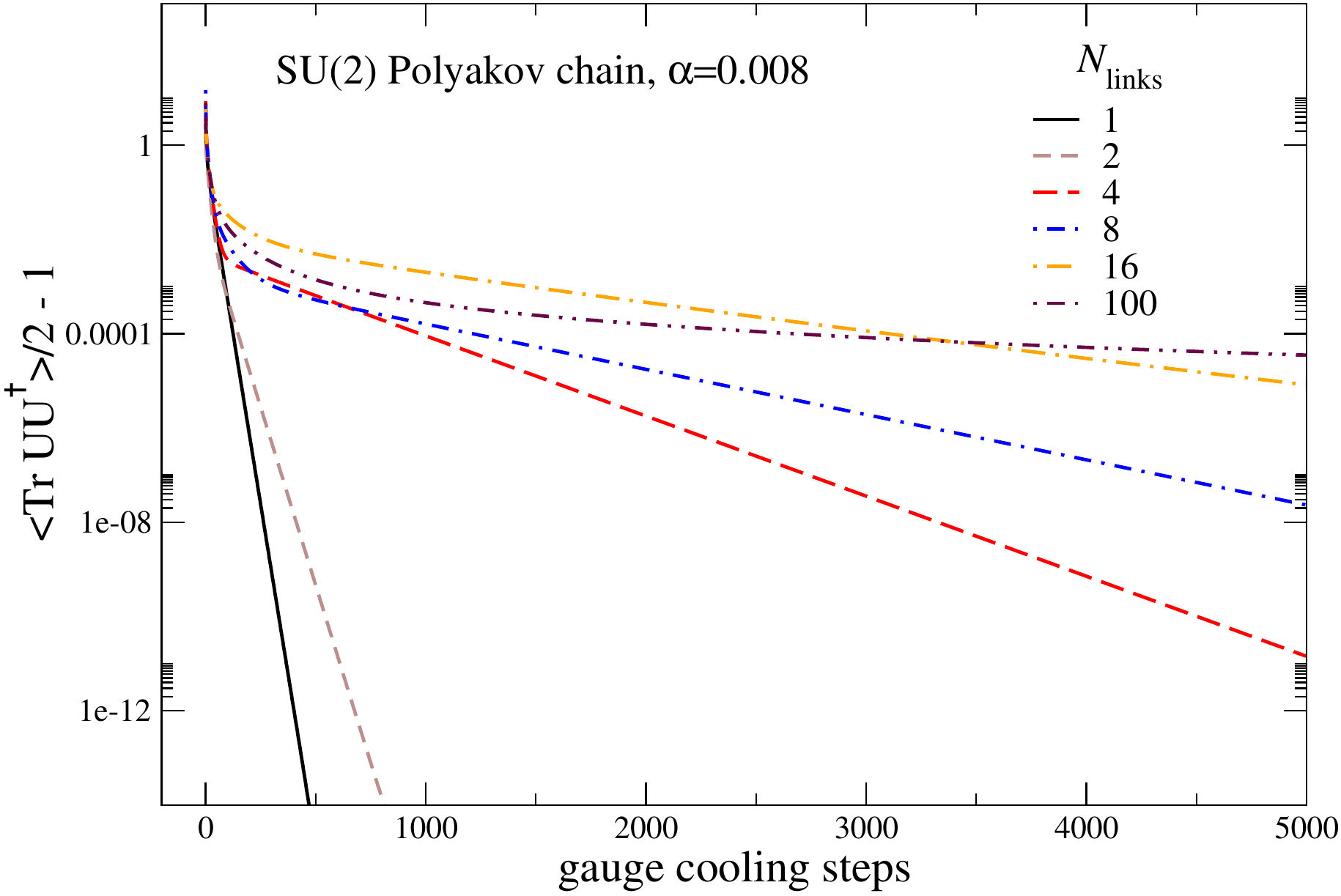} 
\end{center}
\vspace{-0.5cm}
\caption{SU(2) Polyakov chain. Effect of gauge cooling on Polyakov chains with a different number of links.} 
\label{fig:Nlink} 
\end{figure} 
Fig.\ \ref{fig:Nlink} shows, on a logarithmic scale, the effect of gauge cooling on the distance $\dd$ \eqref{UN}  increasing the number of links in the Polyakov chain. For $N_l=1$ one finds numerical agreement with our analytical calculation of the one link model, i.e. the exponential fall off of the UN to the \SU2 fixed point $\dd=0$. When $N_l > 1$ the exponential behaviour is slower and slower until it appears to become non-exponential for $N_l=100$.\\
In gauge theories the number of degrees of freedom usually grows with the volume, that is why it would be preferable to find an implementation of gauge cooling that can keep the fall-off of the UN as fast as possible given the number of degrees of freedom involved.  A possible solution is \textit{adaptive gauge cooling} \cite{Aarts:2013uxa}. The idea is, given a transformation $\Omega_{gc}(x) = e^{-\epsilon \alpha_{\rm gf}f_{ax} \lambda_a}$, to adapt the strength $a_{gf}$ of gauge cooling depending on the proximity to the stationary point of eq.\eqref{d-d'}. In this way gauge cooling is expected to proceed with bigger steps when far away from the minimum of the UN, while being more sensitive when closer to it. We define
\begin{equation}
\alpha_{ad}=\dfrac{\alpha_{gf}}{D\left( U,U^\dag \right)} ,
\end{equation}
where $D\left( U,U^\dag \right)$ is a scalar function of the gauge field that can be adapted to the model under investigation. Clearly for $D\left( U,U^\dag \right)=1$ it is the case of fixed (non adaptive) gauge cooling.\\
\begin{figure}[t] 
\begin{center}
\includegraphics[width=0.7\textwidth]{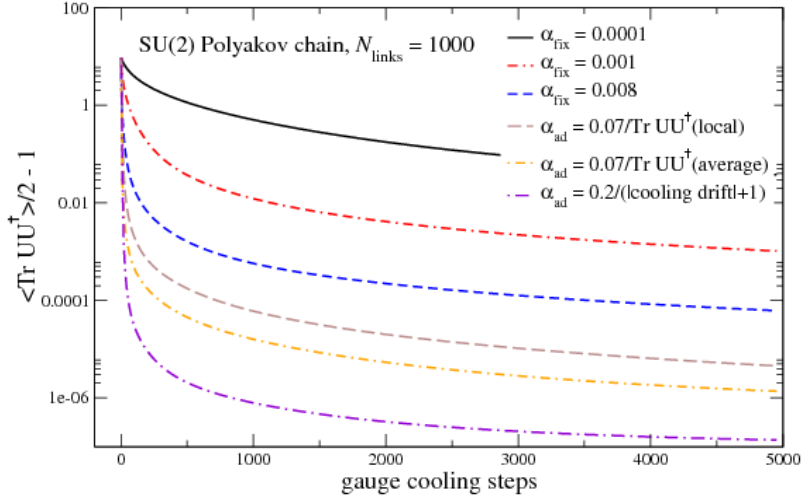} 
\end{center}
\vspace{-0.5cm}
\caption{SU(2) Polyakov chain. Effect of different implementations of adaptive gauge cooling on a Polyakov chain of length $N_\ell=1000$, plotted on a linear-log scale}
\label{fig:different alpha1} 
\end{figure} 
\begin{figure}[h!] 
\begin{center}
\includegraphics[width=0.7\textwidth]{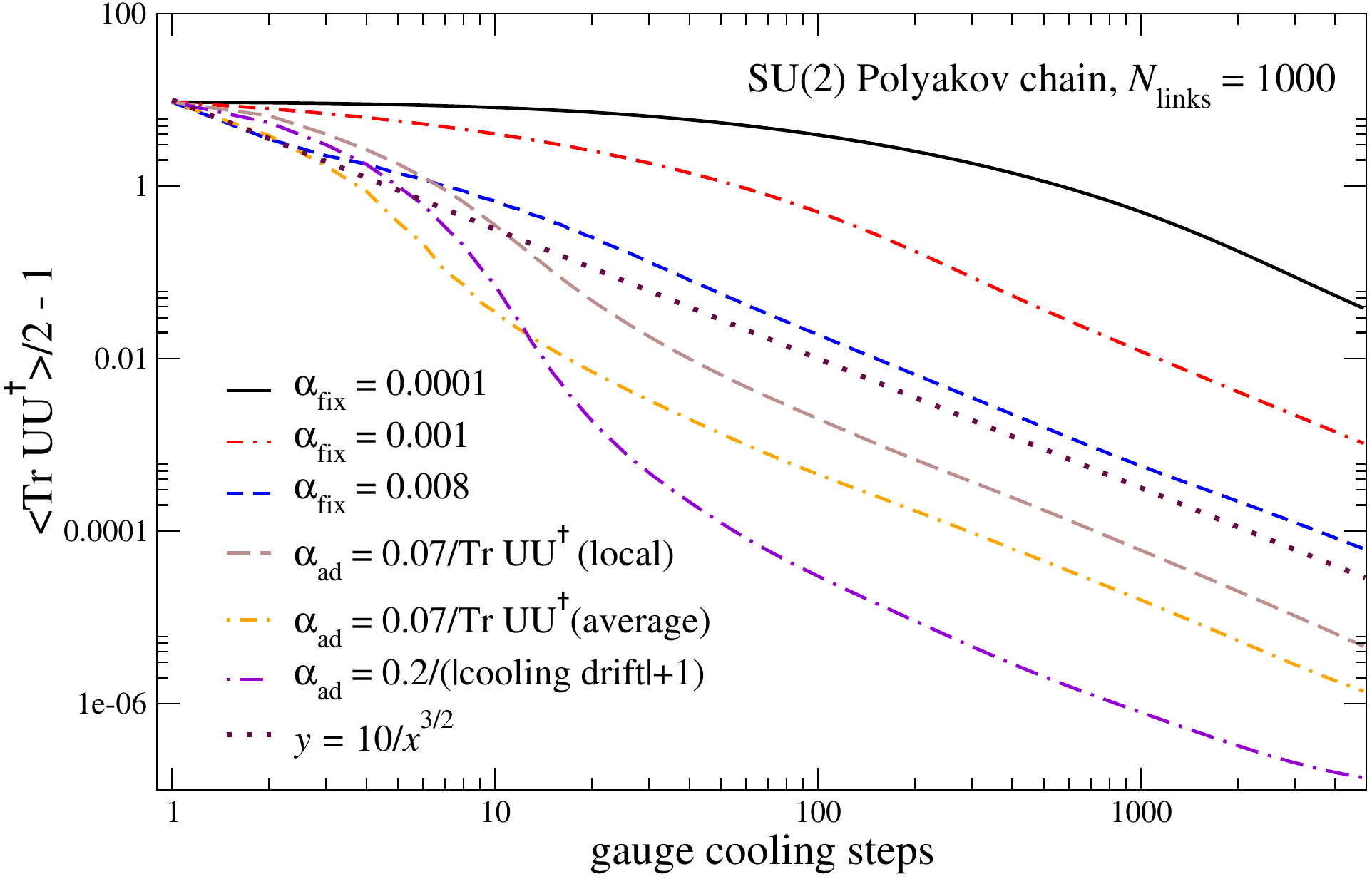} 
\end{center}
\vspace{-0.5cm}
\caption{Same as FIG. \ref{fig:different alpha1} but plotted on a log-log scale .} 
\label{fig:different alpha2} 
\end{figure} 
In Fig.\ \ref{fig:different alpha1} and Fig.\ \ref{fig:different alpha2} we show the effect of different kinds of adaptive gauge cooling applied to the same Polyakov chain.

 In particular we compare the cases
\begin{equation}
 \begin{split}
 &\alpha_{ad}(k)=\alpha_{gf}\ , \\
 &\alpha_{ad}(k)=\dfrac{\alpha_{gf}}{\frac{1}{N}\Tr\left( U_kU_k^\dag \right)}  \ , \\
 &\alpha_{ad}(k)=\dfrac{\alpha_{gf}}{\frac{1}{N}\langle \Tr\left( UU^\dag \right) \rangle} \ , \\
 &\alpha_{ad}(k)=\dfrac{\alpha_{gf}}{\frac{1}{N_l}\sum_{a,k}\vert f_{ak} \vert +1} \ .
 &\end{split}
 \end{equation}
Near the minimum of the UN, all the above definitions of $\alpha_{ad}(k)$ tend to an effective fixed $a^{eff}_{gf}$ ($\frac{1}{N}\Tr\left( U_kU_k^\dag \right) \rightarrow \Id$ \ and\ $f_{ak} \rightarrow 0$) so that, close to the fixed point of the configuration, we still expect a sub-exponential decay of the kind seen in Fig.\ \ref{fig:Nlink} for $N_l=100$. On the other hand, adaptive gauge cooling will be effective at the beginning of the cooling process when $\alpha_{ad}$ is truly adaptive and strongly dependent on the UN. All these consideration can be verified looking at Fig.\ \ref{fig:different alpha1}. Here we can see that, compared to the fixed gauge cooling the adaptive gauge cooling significantly helps the decrease of the UN in the first steps, while later on it stabilizes on an almost identical decay behaviour. This can be better seen looking at Fig.\ \ref{fig:different alpha2}, where the same data are plotted on a log-log scale and a common asymptotic powerlike decay of the order $t^\frac{3}{2}$ ($t$ being the number of gauge cooling steps) can be observed. In conclusion, we learned that with an adaptive implementation even just the first few cooling steps are enough to greatly reduce the distance from SU(2). In the remaining part of the chapter we are going employ this method in the actual dynamics of the system to help its convergence to the right result.\\

\begin{figure}[t] 
\begin{center}
\includegraphics[width=0.8\textwidth]{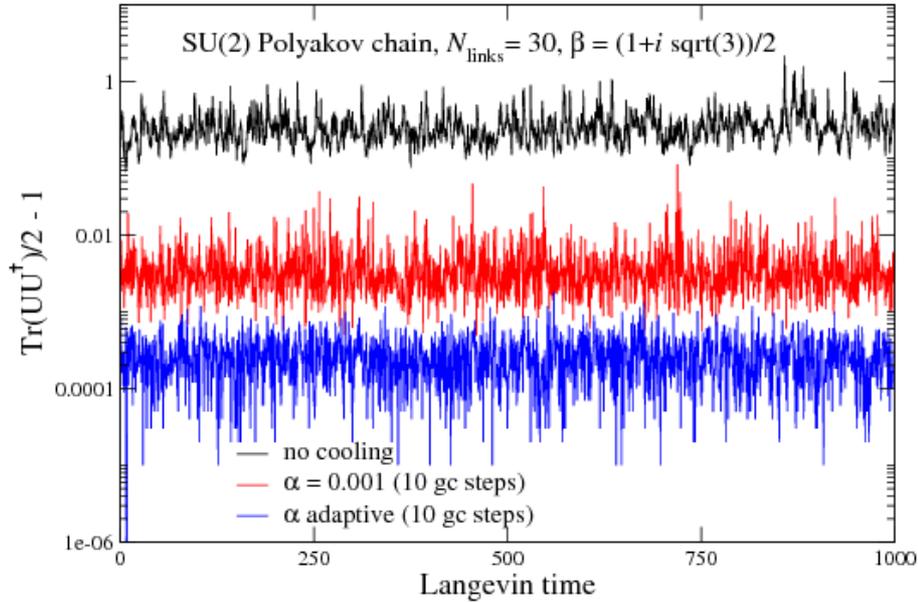} 
\end{center}
\vspace{-0.5cm}
\caption{SU(2) Polyakov chain : comparison between the history of unitarity norm without gauge cooling, with fixed gauge cooling and with adaptive gauge cooling} 
\label{fig:TrUU Pol.chain} 
\end{figure} 
The complex Langevin dynamics simulations are carried out in such a way that between two consecutive Langevin update steps, a certain number of gauge cooling steps is applied to the links. We would like to stress again that, since the action \eqref{SPolChain} is complex-valued for $\beta \in \mathbb{C}$, CL dynamics is required for the full model even if the exact results are available thanks to the equivalence with the one-link model.

We will now present results for a chain with $N_l=30$ links at $\beta=\frac{1}{2} (1+i \sqrt{3})$. First of all, looking at Fig.\ \ref{fig:TrUU Pol.chain}, one is completely convinced of the role of gauge cooling, i.e. keeping the distance \eqref{UN} from the configuration to the SU(2) manifold as small as possible during the whole CL evolution. Here the black line represents the UN as a function of langevin time during the CL evolution, without any gauge cooling being applied. One can then appreciate how the introduction of fixed gauge cooling (red line) has an important effect in reducing the UN of roughly two orders of magnitude. Even more, when adaptive gauge cooling is applied, the UN gets reduced by another order of magnitude. Of course the Polyakov chain, thanks to the complex action, will evolve in \SL2 and will not be gauge equivalent to an SU(2) configuration, so that one expects the UN to be, in general, different than 0.\\  
Although the UN is a strong indication of the compactness of the distribution in the complex \SL2 manifold, the exact condition that has to be satisfied in order to satisfy the criteria of correctness, is about the distribution of the observables itself. 
\begin{figure}[t] 
\begin{center}
\includegraphics[width=0.49\textwidth]{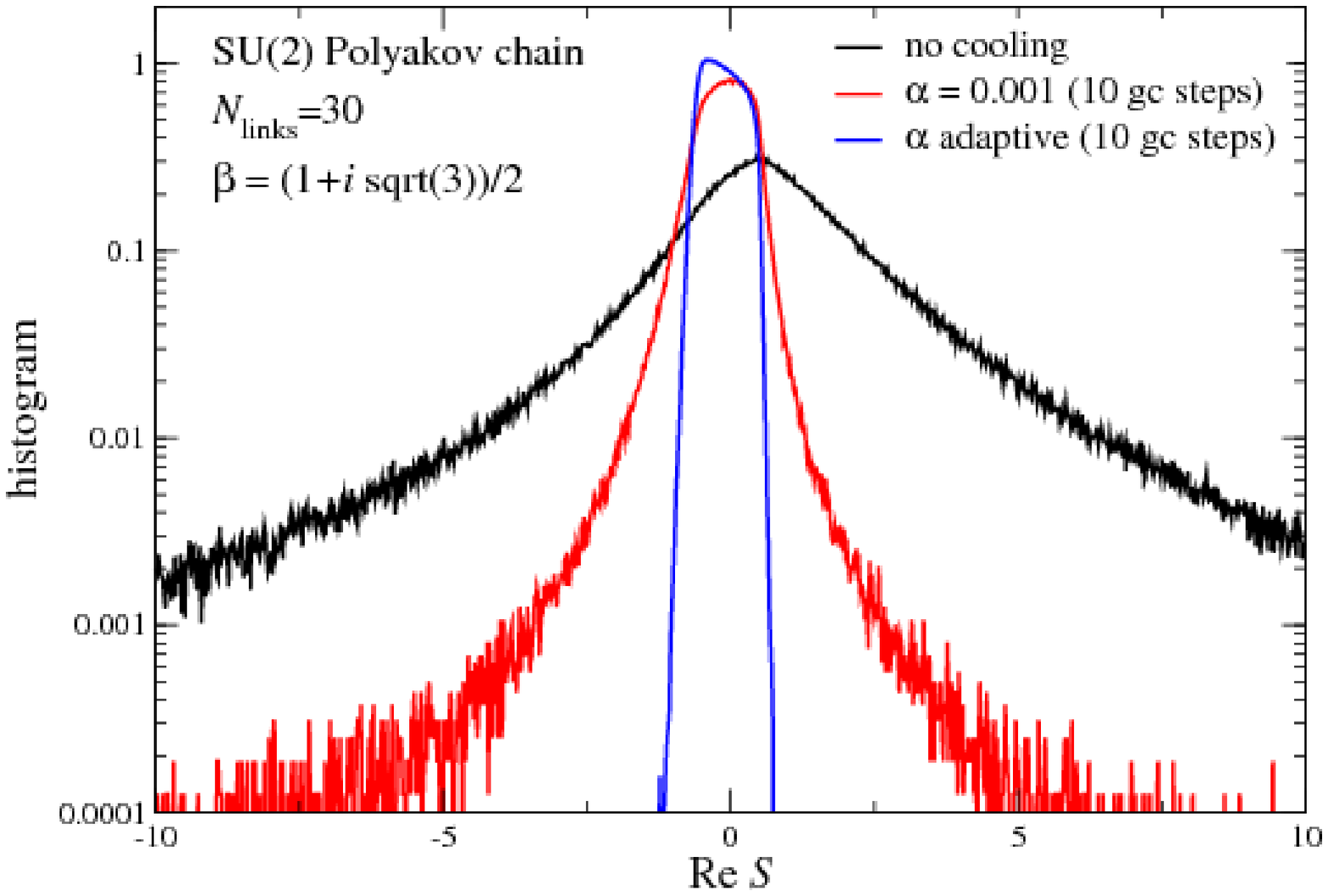} \includegraphics[width=0.49\textwidth]{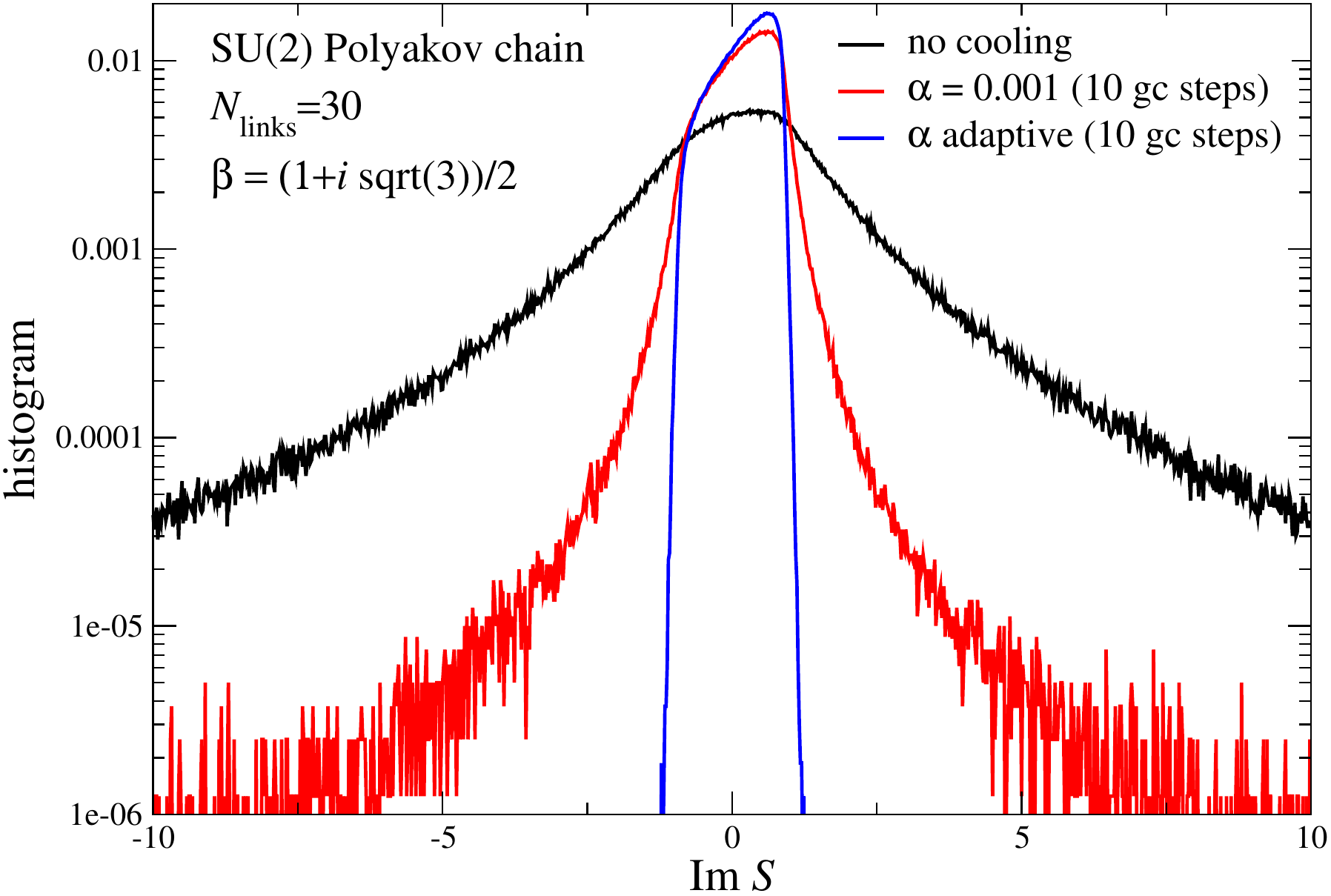} 
\end{center}
\vspace{-0.5cm}
\caption{Histogram of the distribution of the real and imaginary part of $\langle S \rangle$} 
\label{fig:ReS e ImS} 
\end{figure} 
\begin{figure}[t] 
\begin{center}
\includegraphics[width=0.8\textwidth]{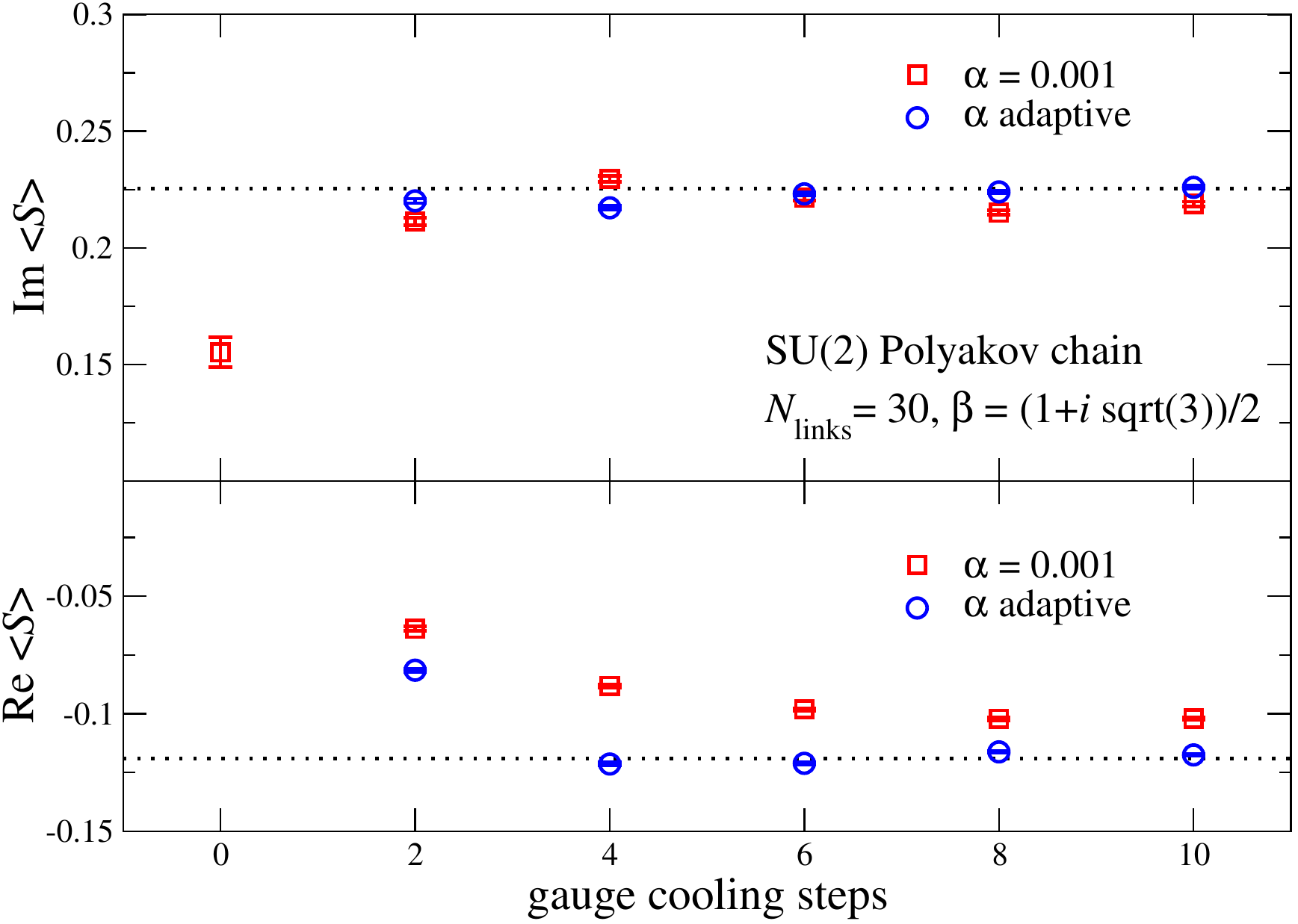} 
\end{center}
\vspace{-0.5cm}
\caption{Convergence of CL to the correct result after the application of fixed and adaptive gauge cooling in
the SU(2) Polyakov chain} 
\label{fig:S vs gcsteps} 
\end{figure} 
Fig.\ \ref{fig:ReS e ImS} shows an histogram of the distribution of the action $S=\frac{\beta }{N_l} \Tr\left( U_1U_2...U_{N_l} \right)$ in the complex plane ($\Re$e$(S)$ and $\Im$m$(S)$). We can see how the effect of gauge cooling in keeping the UN as small as possible also directly affects the distribution of $S$. The black line (no gauge cooling applied) clearly shows that the distribution of $S$ is quite widely spread and presents a sub-exponential decay in the complex direction. That corresponds to violation of the criteria of correctness and, as shown in Fig.\ \ref{fig:S vs gcsteps} for number of gauge cooling steps = 0, to the convergence of CL to the wrong results. The application of fixed gauge cooling, 10 fixed gauge cooling steps every CL step, improves considerably the situation and makes the distribution much more localised although skirts are still present. Finally, with the adaptive choice the distribution is localised very well and drops quickly to zero outside its main support.

In Fig.\ \ref{fig:S vs gcsteps} we show the value of $\langle S \rangle$ which CL dynamics converges to, as a function of the number of gauge cooling steps applied after every CL step. The black dotted lines are the correct analytical result $\frac{I_n(\beta)}{I_0(\beta)}$ for $\beta=\frac{1}{2} (1+i \sqrt{3})$. Looking at the red dots (fixed gauge cooling), we can see how increasing the number of gauge cooling steps pushes CL dynamics towards the right result. Furthermore, using adaptive gauge cooling (blue dots), both $\Re$e$\langle S \rangle$ and $\Im$m$\langle S \rangle$ stabilize on the correct value after just 4 steps.\\

\vspace{1cm}
In this chapter, we showed how complex Langevin dynamics is a powerful method to deal with the sign problem.  In particular, we discussed the existence of criteria of correctness, the meeting of which leads to the convergence of CL to the right results.  Then, we introduced the concept of gauge cooling and employed it in the study of the Polyakov chain model. We explicitly showed how the effect of gauge cooling results in the satisfaction of the criteria of correctness and, consequently, leads the CL dynamics to the right results.

In the next chapter we are going to apply those concepts in the study of a more complicated and physically interesting gauge theory, i.e. Yang-Mills pure gauge with a CP violating topological $\theta$ term.

\begin{comment}

%% file: cap_THETA/ThetaTerm.tex
\tableofcontents

\end{comment}

\externaldocument{cap_THETA/ThetaTerm.tex}
\chapter{Theta term}\label{cap:thetaTerm}
Topology in QCD started to get attention in the middle of the 70s with the famous paper by Polyakov al. \cite{Belavin:1975fg}. In particular, they discovered classical solutions with non-trivial topology to the Yang-Mills equation, called the \textit{instantons}. Those fields were found to represent, in Minkowsky space, tunneling events between degenerate classical vacua of the theory. Because proven to generate confinement in certain 3d models \cite{Polyakov:1976fu} instantons became immediately a very hot topic of research. However it was eventually realized that they could not be the cause in 4d Yang-Mills theories. Nevertheless, they were proven to be responsible for  very important phenomena such as the anomalous breaking of the U(1)$_A$ symmetry \cite{'tHooft:1976up} due to their interaction with the zero-modes of light quarks (index theorem). For the same reason they were found to be the cause of the large mass of the $\eta'$ meson by the mechanism proposed by Witten \cite{Witten:1979vv} and Veneziano \cite{Veneziano:1979ec}. In the same years, the $\theta$-vacua structure of QCD was introduced and systematically studied \cite{Jackiw:1976pf,Callan:1976je,Polyakov:1976fu}. It was realized that a CP violating topological $\theta$-term should be included in the QCD lagrangian; however, a calculation of the neutron magnetic dipole \cite{Crewther:1979pi} constrains this coupling to being negligible ($\theta < 10^{-10}$). A definitive mechanism to explain why the $\theta$-term has to be so small is still to be discovered and goes under the name of \textit{strong CP problem}.

One of the most elegant solutions would be the introduction of the QCD \textit{axion} particle. This particle would be a pseudo-Nambu-Goldstone boson which arises from the spontaneous breaking of a hidden chiral symmetry (Peccei-Quinn symmetry) of the Standard Model \cite{Peccei:1977hh,Peccei:1977ur}. Not only the broken symmetry would automatically suppress the CP violating part of the action, but the mass of the axion could be a candidate for the dark matter in the universe. 
\\\\
In this chapter we will review in some detail the instantons' picture and the structure of the QCD vacua (see \cite{Schafer:1996wv,Rajaraman:1982is} for extensive treatment). Later on, we will discuss the study of topology on the lattice and present new results using complex Langevin dynamics.

\section{Vacua of SU(N) and instantons}\label{sec:Top&inst_continuum}
To describe the classical vacua of the pure gauge Yang Mills theory one has to find the zero energy  field configurations. The gauge \SUN \ fields are defined in the usual way :
\begin{equation}
\begin{split}
&A_\mu=\sum_a A^a_\mu \dfrac{\lambda^a}{2}\ ,\\
&\left[ \lambda^a, \lambda^b\right]=2i f^{abc} \lambda^c , \ \ \ \ \ \Tr\left( \lambda^a \lambda^b\right) = 2 \delta^{ab}\ , 
\end{split}
\end{equation}
where $a=1...N \ \ \ \text{and}\ \ \  \lambda^a$  are the Gell-Mann matrices. The minimum of the action
\begin{equation}
S \ = \ \frac{1}{4g^2} \int d^4x \ G^a_{\mu\nu}G^a_{\mu\nu} ,
\end{equation}
is given by those fields that satisfy the condition
\begin{equation}
 G_{\mu\nu}=0 \ ,
\end{equation} 
i.e. the pure gauge fields of the form 
\begin{equation}
A_\mu= i U(x) \partial_\mu U^\dag(x) \ \ \ U\in \ $SU$(N) \ .
\end{equation}
A physical interpretation of their role can be given in the temporal gauge $A_0=0$, in which the fields of the vacuum have the time independent form
\begin{equation}\label{A_vacuum}
A_i(\textbf{x})= i U(\textbf{x}) \partial_i U^\dag(\textbf{x}) ,
\end{equation}
where $\textbf{x}$ indicates only the spatial directions. In this gauge, it is possible to identify \cite{Callan:1977gz} that the gauge transformations that can be connected with the trivial topological sector, where $A_i(\textbf{x})=0$, are only the ones that respect the condition 
\begin{equation}\label{U->1 all'infinito}
U(\textbf{x})\rightarrow 1 \ \ \text{for} \ \ |\textbf{x}|\rightarrow \infty \ .
\end{equation} 
This condition is equivalent, for what concerns $U(\textbf{x})$, to the compactification of $\mathbb{R}^3$ in the $S^3$ hyperspherical surface and, therefore, $U(\textbf{x})$ can be seen as an application that maps $S^3$ into SU(N).

Those maps can be divided into disjointed \textit{homotopy classes} and classified with an integer \textit{winding} (or Pontryagin) number  $n_w$
\begin{equation}
n_w =\dfrac{1}{24 \pi^2} \int d^3x\; \epsilon^{ijk} \Tr\left[ (U^\dag \partial_i U)(U^\dag \partial_j U)(U^\dag \partial_k U) \right] ,
\end{equation}
which, in terms of the corresponding gauge fields, is the Chern-Simons number
\begin{equation}\label{n_CS}
n_{CS}=\dfrac{1}{16\pi^2} \int d^3x\;  \epsilon_{ijk} \left( A_i^a \partial_j A_k^a + \frac{1}{3} f^{abc} A_i^a A_j^b A_k^k \right) .
\end{equation}
Let us remark that the winding number $n_w$ (or $n_{CS}$) is not gauge invariant, as it is defined in the temporal gauge, and in three dimensions, only. However, the difference between two winding numbers is gauge invariant because, as we shall see, it can be related to a gauge invariant quantity, i.e. the \textit{topological charge}. 

The picture is then that Yang-Mills theories have an infinite number of distinct classical configurations with zero energy, each one can be related to an integer number $n_w$. Those vacua, however, are not completely isolated from one another. In fact, there are fields configurations that tunnel between two vacua with a different winding number. Those fields are classical solutions of the euclidean equations of motion. In particular, using the identity 
\begin{equation}\label{S_YM indentity}
S \ = \ \frac{1}{4g^2} \int d^4x \ G^a_{\mu\nu}G^a_{\mu\nu} \ = \ \frac{1}{4g^2} \int d^4x \left[ \pm G^a_{\mu\nu}\tilde{G}^a_{\mu\nu} + \frac{1}{2} \left( G^a_{\mu\nu} \mp \tilde{G}^a_{\mu\nu} \right)^2\right] \ ,
\end{equation}
one can rewrite the action as a sum of the topological invariant $\int d^4x G_{\mu\nu} \tilde{G}_{\mu\nu}$, with a semi-positive definite part. The minimum, then, is the (anti)self-dual solution
\begin{equation}
G^a_{\mu\nu} \ = \ \pm \ \tilde{G}^a_{\mu\nu} \ .
\end{equation}
In Euclidian space, one can prove that the (anti)self-dual solutions automatically satisfy the Yang-Mills classical equations of motion
\begin{equation}
D_\mu G_{\mu \nu} =0.
\end{equation}
These configurations are called \textit{instantons} (\textit{anti-instantons}) and, 
to have a finite contribution from the action, they need to be pure gauge at infinity
\begin{equation}\label{inf_pg}
A_\mu(x)_{\ \overrightarrow{x \rightarrow \infty}}\ i U(x) \partial_\mu U(x)^\dag\ .
\end{equation}
We notice that this generates again a maps from the three sphere $S^3$ into the gauge group SU(N), which can labelled by an integer number $Q\in \mathbb{Z}$ called \textit{topological charge}. This can be better understood in the case of \SU2 where the group itself is homeomorphic to the $S^3$ sphere. In this case eq.\eqref{inf_pg}, or better its exponentialization, is an application that maps $S^3$ into $S^3$ and the corresponding homotopy group $\pi_3(S^3)$ is also isomorphic to the integers $\mathbb{Z}$, from which the winding number that represents how many times the first sphere wraps around the second one. For this reason two solutions with different topological charge $Q$, cannot be deformed one into the other with a continuous transformation without violating finiteness of the action. The same argument can be extended to every gauge group since \cite{Bott:1956} every continuous application $U:S^3 \rightarrow G$, with $G$ Lie group, can be deformed continuously in an application $U:S^3 \rightarrow$ \SU2$\subset G$.   

Differently from the case of the vacuum fields \eqref{A_vacuum}, this time the map arises naturally, not only in a specific gauge, which results in the topological charge $Q$ being a gauge invariant quantity. This can also be seen from its formal definition
\begin{equation}\label{Q_top}
\begin{split}
&Q=\int d^4x\ q(x)\ , \\
&q(x)=\dfrac{1}{32 \pi^2} \Tr \left[G_{\mu \nu}^a \tilde{G}^{a \mu \nu} \right]
\end{split}
\end{equation}
where $\tilde{G}^{a \mu \nu}$ is the Hodge dual of $G^{a \mu \nu}$ defined as $\tilde{G}^{a \mu \nu}=\frac{1}{2} \epsilon_{\mu \nu \rho \sigma} G^{a \rho \sigma}$. One can show that the topological charge density $q(x)$ is a total derivative
\begin{equation}\label{Chern-Simons}
\begin{split}
&q(x)=\partial_\mu K^\mu\ ,\\
&K^\mu=\dfrac{1}{32\pi^2}\epsilon_{\mu\alpha\beta\gamma} \left( A_\alpha^a \partial_\beta A_\gamma^a + \frac{1}{3} f^{abc} A_\alpha^a A_\beta^b A_k^\gamma \right) .
\end{split}
\end{equation}
$K^\mu$ describes, at $\vert x \vert \rightarrow \infty$, the infinitesimal winding of the $S^3$ sphere on the \SUN \ gauge group, that makes the quantity
\begin{equation}\label{Q_CS}
Q=\int_V d^4x\ \partial_\mu K^\mu \ = \int_\Sigma d\sigma_\mu  K^\mu ,
\end{equation}
precisely the integral of the total derivative that accounts for the topological charge. Going back to the temporal gauge, we notice that the only component of $K^{\mu}$ that survives in \eqref{Chern-Simons} is $K^0$, which is precisely the winding (or Chern-Simons) number defined before in \eqref{n_CS}. Then, the integral \eqref{Q_CS}
\begin{equation}
 Q = \int_{-\infty}^\infty dt \frac{d}{dt} \int d^3x K_0 = n_{CS}(t=\infty) - n_{CS}(t=-\infty) , 
\end{equation} 
shows that field configurations with $Q \neq 0$ connect different topological vacua.  From \eqref{S_YM indentity}, the action of those configurations is
\begin{equation}
S_{inst}=\dfrac{8 \pi^2 \vert Q\vert}{g^2}
\end{equation}
which generates a probability of tunneling
\begin{equation}
P_{tun.} \sim e^{-8 \pi^2 /g^2}\ ,
\end{equation}
where the coefficient of proportionality has to be evaluated using perturbation theory. 

A classic example of a \SU2 instanton with charge $Q=1$ is the \textit{BPST} (Belavin, Polyakov, Schwartz $ \& $ Tyupkin) instanton \cite{Belavin:1975fg}
\begin{equation}\label{Ist BPRS}
A^a_\mu(x) \ = \ 2 \eta_{a \mu \nu} \dfrac{(x-x_0)_\nu}{(x-x_0)^2 + \rho^2} \ ,
\end{equation}
where $x_0$ is the centre of the instanton, $\rho$ is its size and $\eta_{a \mu \nu}$ is the 't Hooft symbol :
\begin{equation}\label{eta 't Hooft}
\eta_{a \mu \nu} \ = \ \left\lbrace\!\begin{split}
&\epsilon_{a \mu \nu} \ \ \ \mu,\nu = 1,2,3 , \\
&\delta_{a\mu} \ \ \ \nu = 4 , \\
&-\delta_{a\nu} \ \ \ \mu = 4 .
\end{split}\!\right. 
\end{equation}
For the anti-instanton solution, with topological charge $Q=-1$,  one just needs to replace $\eta_{a \mu \nu}\rightarrow \overline{\eta}_{a \mu \nu} $, which is the same tensor as \eqref{eta 't Hooft} with just opposite signs in the temporal directions. 

Let us remark that, despite the BPST instanton being a long range field ($A_\mu \sim 1/x$), its field strength
\begin{equation}
G^a_{\mu \nu} G^{a \mu \nu}=\dfrac{192 \rho^4}{(x^2+\rho^2)^4}\ ,
\end{equation}
is well localized ($G^a_{\mu \nu}\sim 1/x^4$) in space and time, from which the name 'instanton' derives.

Let us briefly recall the properties of the two fields : 
\begin{itemize}
\item Fields of the vacuum :
\begin{equation}
A^{vacuum}_\mu(x)= i U(\overrightarrow{x}) \partial_\mu U^\dag(\overrightarrow{x}), \ \ \ \forall\ x \ ;
\end{equation}
those are the solutions which annihilate the strength tensor 
\begin{equation}
G_{\mu \nu}(A_\mu^{vacuum})=0\ ,
\end{equation}
and, therefore, have got a 0 action contribution.
\item Instantons :
\begin{equation}
A^{inst}_\mu(x)\xrightarrow[x \to \infty]{}\ i U(\overrightarrow{x}) \partial_\mu U^\dag(\overrightarrow{x})+O\left(\dfrac{1}{x}\right), \ \ \ \lim : \ x \to \infty \ ;
\end{equation}
those are semi-classical, (anti)self-dual solutions  of the equations of motion $G_{\mu \nu}= \pm \tilde{G_{\mu \nu}}$. They represent, in Minkowsky space, tunnelling events between two different vacua of QCD and they have a finite action contribution
\begin{equation}
S_{inst}=\dfrac{8 \pi^2 \vert Q\vert}{g^2} \ .
\end{equation}
\end{itemize} 
Since the infinite number of degenerate vacua are connected by some tunneling events, the correct way to express the ground state of the theory, in analogy with the Bloch's theorem, is a linear combination of all the separate topological vacua
\begin{equation}\label{theta_vacua}
\vert \theta \rangle=\sum_{Q=-\infty}^{\infty} e^{i\theta Q} \vert Q \rangle
\end{equation}
where $\vert Q \rangle$ indicates the vacuum with topological number $Q$ and $e^{i\theta Q}$ comes from the periodic structure of the vacua. From \eqref{theta_vacua} one understands that, in principle, there are an infinite number of possible ground states corresponding to all the values $\theta \in [0,2\pi)$. However, we shall see that any choice of $\theta$  isolates a sector of physical states completely disconnected from the others at different values of $\theta$.
If we consider the unitary gauge transformation $R$ that generates a shift of one in the topological vacuum  
\begin{equation}
R \vert Q \rangle = \vert Q+1 \rangle\ ,
\end{equation}
then we realize, from \eqref{theta_vacua}, that the ground state $\vert \theta \rangle$ is an eigenstate of $R$
\begin{equation}
R \vert \theta \rangle = e^{-i\theta} \vert \theta \rangle\ .
\end{equation}
Furthermore, any gauge invariant operator $\mathcal{O}$ must commutate with the gauge transformation $R$, i.e. $\left[ \mathcal{O},R\right]=0$. Hence,
\begin{equation}
0=\langle \theta \vert \left[ \mathcal{O},R\right] \vert \theta' \rangle = \left( e^{-i \theta'}-e^{i \theta}\right) \langle \theta \vert  \mathcal{O} \vert \theta' \rangle \ ,
\end{equation}
that, when $\theta \neq \theta'$, implies 
\begin{equation}
 \langle \theta \vert  \mathcal{O} \vert \theta' \rangle = 0\ .
 \end{equation} 
That means it is impossible for any physical (gauge invariant) operator to connect states with different $\theta$. Furthermore, since $\left[ \mathcal{H},R\right]=0$ also holds, the ground state $\vert \theta \rangle$ remains unchanged in time. In other words $\theta$ is a parameter of super-selection of the physical theory that, once selected, remains constant with no possibility of contact with states at different $\theta$.

Let us note that the ground state is not CP invariant as the topological vacua $\vert Q \rangle$ are not ($\text{CP} (Q) = -Q$). One can explicitly see this from \eqref{theta_vacua} 
\begin{equation}
\text{CP} \vert \theta \rangle=\sum_{Q=-\infty}^{\infty} e^{-i\theta Q} \vert Q \rangle \neq \vert \theta \rangle \ ,
\end{equation}
where the strength of CP violation is characterized by the angle $\theta$. However, experimental evidences on the neutron electric dipole moment (nEDM) limit the violation of CP in our world to a very small value. In particular, this can be translated in a constraint on the parameter $\theta$ \cite{Crewther:1979pi}
\begin{equation}
\theta<10^{-10} \ .
\end{equation}
This unnaturally small value of $\theta$, which is otherwise not restricted by theory, is known as the strong CP problem. One elegant solution was proposed by Peccei and Quinn \cite{Peccei:1977hh}, which makes the $\theta$-parameter vanish dynamically. However, the mechanism also requires the appearance of a Goldstone boson, the axion, which remains to be discovered.

\vspace{1cm}
\subsubsection{Axial anomaly and Atiyah-Singer index theorem}
For historical reasons, we are now going to briefly mention how the non-trivial vacuum structure of QCD is essential to explain the $U_A(1)$ axial anomaly in the context of chiral symmetry breaking.

Let us start by defining an anomaly in the context of quantum field theory. One talks about anomaly when a group of transformation $G$ leaves unchanged the action of the theory $G(S)=S$, but not the measure of integration $G(d\mu)=d\mu'$ of the generating functional :
\begin{equation}
Z=\int d\mu\; e^{-S/\hbar} \ ,
\end{equation}
so that, overall, $G(Z)= Z'\neq Z$. As a consequence, the Noether current, classically associated to the symmetry of the action in respect of the group $G$, will not be conserved any more. When that happens, one says that a classical symmetry is \textit{anomalously} broken at the quantum level. We see that in the limit $S \gg \hbar$ (classical limit) only the saddle point of the action $S_{min}$ contributes and, therefore, the integration over the measure in the field space does not matter any more. In this limit the conservation of the Noether current is restored, as it should be.  

QCD chiral symmetry is a physical example where an anomalous breaking of symmetry occurs. Let us recall the Lagrangian of $N_f$ families of massless fermions:
\begin{equation}\label{L fermion massless}
\mathcal{L}_{massless} \ = \ \sum_{f=i}^{N_f} \overline{\psi}^f(x) i \displaystyle{\not} \partial \psi^f(x) \ = \ \sum_{f=i}^{N_f} \overline{q}^f_L\,i\displaystyle{\not}\partial \,q^f_L + \overline{q}^f_R\,i\displaystyle{\not}\partial\, q^f_R \ , 
\end{equation}
where we have explicitly written the left and right components of the spinor $\psi^f = \left(\begin{array}{r} 
q^f_R \\
q^f_L
\end{array} \right) $. The Lagrangian is invariant under chiral rotation, in the $N_f$ fermions families, of both left and right components independently. The resulting symmetry is the group of transformations $U(N_f)_L \times U(N_f)_R$ called \textit{chiral symmetry}.

In QCD all the quarks are massive so, in principle, there is no exact chiral symmetry. However, the three quarks ($u,d,s$) have masses much smaller than the chiral symmetry spontaneous breaking energy scale  $\Lambda_\chi \sim 1$GeV. Therefore, they can be regarded as massless in first order approximation, in the sense that the spontaneous breaking of the chiral symmetry dominates over the explicit one. At first order in the quark masses, then, the chiral symmetry in QCD is
\begin{equation}
U(3)_L \otimes U(3)_R \ ,
\end{equation}
and can be decomposed into the irreducible representations : 
\begin{equation}
U(3)_L \otimes U(3)_R \rightarrow \ SU(3)_L \otimes SU(3)_R \otimes U(1)_V \otimes U(1)_A \ .
\end{equation}
The $U(1)_V$ symmetry corresponds to the baryon number conservation and the $U(1)_A$ is anomalously broken as we shall see in a bit.
As anticipated before, however, chiral symmetry spontaneously breaks at temperatures of the order of 1GeV. In particular the group 
\begin{equation}
SU(3)_L \otimes SU(3)_R\ ,
\end{equation}
is broken down to the vector subgroup
\begin{equation}
SU(3)_V\ ,
\end{equation}
which results in the classification of the hadrons into the SU($3$) irreducible representations. The remaining 8 broken generators produce as many Goldstone bosons, namely the octet of pseudo-scalar mesons ($ \pi^\pm , \pi^0 , K^\pm, K^0 , \eta $). Those mesons, despite being Goldstone modes, are massive as a result of the mass of the quarks that explicitly prevents chiral symmetry to be realized exactly.

The part which, historically, created most of the trouble was the $U(1)_A$ symmetry, i.e. the one associated with the transformation
\begin{equation}
q_L \rightarrow e^{i\theta\gamma_5} q_L \qquad q_R \rightarrow e^{-i\theta\gamma_5} q_R \ .
\end{equation}
That this symmetry was anomalous was first discovered by Adler \cite{Adler}, in the context of QED, and then extended by Jackiw and Bell for QCD \cite{Bell&Jakiw}. The realization of such a symmetry would protect the neutral pion from the decay into two photons $\pi^0 \rightarrow \gamma \gamma$, when we know that, on the contrary, this amplitude is finite in QCD 
\begin{equation}
\Gamma(\pi^0 \rightarrow \gamma \gamma) \approx 8.4\ \text{eV}.
\end{equation}
Furthermore, it is also impossible for it to be realized in the Nambu-Goldstone way since, in this case, one expects the presence of another light Goldstone boson with a mass $m\leqslant \sqrt{3} m_\pi \approx 230\ \text{MeV}$, which is not observed in nature. The only meson with the correct quantum numbers is the $\eta'$, however its mass is too big $m_{\eta'}=958\ \text{MeV}$. In the late 70's, Gerard 't Hooft \cite{'tHooft:1976up,'tHooft:1976fv} showed that the axial current of singlet 
\begin{equation}
j^5_\mu(x)=\sum_{i=1}^{N_f} \overline{\psi}_i(x) \gamma_\mu \gamma_5 \psi_i(x)
\end{equation}
was not conserved at the quantum level due to the presence of field configurations with non trivial topology. In particular, the anomaly explicitly depends on the topological charge
\begin{equation}\label{d_mu J_5}
\partial_\mu j^5_\mu(x) \ =\  \dfrac{N_f}{16 \pi^2} G^a_{\mu \nu} \tilde{G}^a_{\mu \nu}\ =\ 2 N_f Q(x) \ ,
\end{equation}
and cannot, therefore, be set to zero, resulting in the non conservation of the axial current. 

The $\eta'$ mass was also explained through the anomaly in connection with topology. More precisely the famous Witten-Veneziano formula relates it directly with the topological susceptibility in Yang-Mills theory
 \begin{equation}\label{formula Witten-Veneziano}
\chi^{(YM)} \ = \ \dfrac{f^2_\pi}{2N_f} (m^2_\eta + m^2_{\eta'} -2 m^2_K) \ .
\end{equation}

We shall briefly show how exactly the instantons are related to the divergence of the axial current. Let us consider the case of massless quarks in which the chiral unbroken symmetry is exact. In this case, the fermion propagator is the inverse of the Dirac operator $ S(x,y) = \langle x | (i\displaystyle{\not}D)^{-1} |y \rangle$. The Dirac operator can be expressed in term of its eigenfunctions $ i\displaystyle{\not}D \psi_\lambda(y) = \sum_\lambda \lambda \psi_\lambda(y)$, so that the fermion propagator reads :
\begin{equation}\label{ferm.prop.}
S(x,y) \ = \ \sum_\lambda \dfrac{\psi_\lambda(x) \psi_\lambda^\dag(y)}{\lambda} \ .
\end{equation}
One can then express the variation of the axial charge $Q_5=\int $ in terms of the fermion propagator \eqref{ferm.prop.}
\begin{equation}\label{DQ_5}
\begin{split}
\Delta Q_5 \ = \ &Q_5(t=+\infty)-Q_5(t=-\infty) \ = \ \int d^4x\ \partial_\mu j^5_\mu(x) \\
&= \ \int d^4x \ N_f \partial_\mu \Tr(S(x,x)\gamma_\mu\gamma_5) \\
&= \ N_f \int d^4x \ \Tr \left(\sum_\lambda \dfrac{\psi_\lambda(x) \psi_\lambda^\dag(x)}{\lambda} 2 \lambda \gamma_5 \right) \ .
\end{split}
\end{equation}
Now we note that $\gamma_5 \psi_\lambda$ is also eigenfunction of the Dirac operator :
\begin{equation}\label{autovalori doppi}
 \ \displaystyle{\not}D(\gamma_5 \psi_\lambda) \ = \ - \gamma_5  \displaystyle{\not}D \psi_\lambda \ = \ - \lambda \ \gamma_5 \psi_\lambda \ ,
\end{equation} 
with eigenvalue $-\lambda$. That means $\psi_\lambda$ and $\gamma_5 \psi_\lambda$ are orthogonal, which makes the integral \eqref{DQ_5} vanish for every $\lambda\neq0$. On the other hand, the zero modes $\psi_0$ give a finite contribution. Furthermore, one can choose the base in which chirality is well defined, i.e. $\gamma_5 \psi_0 = \pm \psi_0$ for right/left handed zero modes. In this way the integral \eqref{DQ_5} reduces to 
\begin{equation}\label{index theorem}
\Delta Q_5 \ = \ 2 N_f (n_L - n_R ) \ ,
\end{equation}
where $n_L$ and $n_R$ are the number of positive and negative chirality zero modes of the Dirac operator. The connection with topology arose when 't Hooft realized that the Dirac operator $i(\partial_\mu+A_\mu)\gamma_\mu$ has a zero mode solution when the field $A_\mu$ is the instanton \eqref{Ist BPRS}. This solution is
\begin{equation}\label{modo zero massless}
\psi_0(x) \ = \ \dfrac{\rho}{\left((x-x_0)^2+\rho^2\right)^{3/2}}\ \phi \ ,
\end{equation} 
where $x_0$ and $\rho$ are, respectively, center and size of the instanton, and $\phi$ is a space-independent spinor. However, since $\displaystyle{\not}D \psi_0=0$, it is clear that $\det\left( [\displaystyle{\not}D (A^I_\mu)]\right)=0$ along the instanton solution $A^I_\mu$. That means the functional integral of the vacuum
\begin{equation}
Z=\int DA_\mu\; \det\left( \displaystyle{\not}D [A_\mu]\right) e^{S(A_\mu)}\ ,
\end{equation}
will vanish exactly for such configurations. Another way to say it is that tunnelling between vacua of the theory is suppressed in the presence of massless fermions. In the chiral limit ($m_{ferm}=0$), in fact, we know that the topological susceptibility has to vanish $\chi^{chiral}_{top}=0$. However, the tunnelling amplitude is non-zero in the presence of \textit{external quark sources}, because zero modes in the denominator of the quark propagator may cancel against zero modes in the determinant.\\\\

Although the theory in the continuum is quite developed, exact calculations of many quantities (such as $\eta '$  mass, tunnelling amplitudes,  etc...) require a non-perturbative approach. In particular, the study of the theory dependence from the $\theta$-term is very challenging on the lattice since it is affected by the sign problem. Already the pure gauge theory SU(3) + \t -term cannot be simulated with standard Monte Carlo methods for real \t . Our goal, in this chapter, is to study pure gauge theory at real \t .

\vspace{2cm}
\section{Topology on the Lattice}
The nature of topological phenomena is highly non-perturbative. Some quantities could still be computed using approximated models like the diluted instanton gas approximation or in the large $N$ limit. For a completely non-perturbative study, however, the use of lattice field theory is required. Historically, the first main reason that brought interest in the study of topology on the lattice was the non-perturbative determination of the topological susceptibility for the Witten-Veneziano formula \eqref{formula Witten-Veneziano}.

At first, one might think that the lattice should not be able to retain the topological content of the configurations. In fact, lattice regularization requires the space to be discrete where every link can be continuously deformed into the trivial one $U_\mu(x)=\Id$. Therefore, the configuration space is simply connected; that means homotopy classes do not exist on the lattice and topology is, strictly speaking, always trivial. However, one must not forget that any operator defined on the lattice has a physical meaning only in view of its continuum limit. The topological quantities are no exception and so one just has to define a regularization of those operators such that they represent the proper physical quantities in the continuum limit, where the standard non-trivial topology is recovered.  In particular, it can be shown \cite{Luscher:1981zq,Phillips:1986qd} that for small values of the local action density
\begin{equation}\label{S<eps}
s_p=\sum_{\mu=1 }^3\sum_{\nu>\mu}^4 \left( \dfrac{1}{6}\text{Re }\Tr\left[ \Id - \Pi_{\mu \nu}(p)\right] \right) < \epsilon\ ,\ \ \ \ \ (\epsilon \approx 0.07)\ ,
\end{equation}
where $ \Pi_{\mu\nu} $ is the $ 1 \times 1 $ plaquette, the lattice field configurations fall again into well defined topological sectors. This regime is reached naturally when approaching the continuum limit where the value of the bare coupling becomes small ($g\rightarrow0$) and
\begin{equation}
\langle s_p \rangle = \dfrac{3}{8}\; g^2 + O(g^4)\ .
\end{equation}
However, it is not necessary to perform the continuum limit to meet the requirement \eqref{S<eps}. There are, in fact, various methods which main idea consists of smoothing the field configurations towards the local minimum of the Wilson action. This procedures "cool" down the configurations to values of $S< \epsilon$, getting rid of the short range fluctuations letting the long range topological modes emerge. We are going to discuss one of these methods in more detail later in this chapter.

We will start by introducing the local operator representing the density of topological charge on the lattice. As usual, more than one operator can be built that recover the correct naive continuum limit
\begin{equation}\label{lim q_L(x)}
q_L(x) \rightarrow a^4 q(x) + O(a^6) \ ,
\end{equation}
and they differ for orders $O(a^6)$.
One of the simpler and more common choice is the twisted double plaquette operator
\begin{equation}\label{q_L}
q_L(x) \ = \ - \dfrac{1}{2^4 \times 32 \pi^2} \sum_{\mu\nu\rho\sigma =\pm 1}^{\pm 4} \epsilon_{\mu\nu\rho\sigma} \Tr[\Pi_{\mu\nu}(x) \Pi_{\rho \sigma}(x)] \ ,
\end{equation}
where, again, $ \Pi_{\mu\nu} $ is the $ 1 \times 1 $ plaquette and $ \epsilon_{\mu\nu\rho\sigma} $ is the completely anti-symmetric Levi-Civita tensor extended to the negative directions with the prescription $ \epsilon_{\mu\nu\rho\sigma} =  - \epsilon_{(-\mu)\nu\rho\sigma} $. The presence of  $ \epsilon_{\mu\nu\rho\sigma} $ requires the plaquettes $ \Pi_{\mu\nu}$ and $\Pi_{\rho \sigma}$ to lie on two completely orthogonal hyperplanes. To get the bare, i.e. unrenormalised, topological charge on the lattice one has, then, simply to sum over all the lattice sites 
\begin{equation}\label{Q_L}
Q_L \ = \ \sum_x q_L(x) \ .
\end{equation} 
Taking the naive limit $a \rightarrow 0$, one can show \eqref{q_L} to classically recover the correct form of the continuum
\begin{equation}
q_L(x) \xrightarrow[a \rightarrow 0] \ a^4\; \dfrac{g^2}{32 \pi^2} \Tr \left[G_{\mu \nu}^a \tilde{G}^{a \mu \nu} \right] \ + \ O(a^6)\ .
\end{equation}
However, as anticipated before, the operator \eqref{q_L} needs to be correctly regulated on the lattice before it can represent the topological charge of the lattice fields. In particular a multiplicative renormalization is required
\begin{equation}\label{lim q_L(x)}
q_L(x) \rightarrow a^4 Z(g^2) q(x) + O(a^6) \ ,
\end{equation}
where $Z(g^2)$ is a finite function of the bare coupling $g^2$, obeying $Z(g^2) \rightarrow  1$ in the limit $g \rightarrow 0$. The value of $Z(g^2)$ can be computed in perturbation theory near the continuum fixed point and, for SU($3$), it is  \cite{Campostrini:1988cy,Campostrini:1989dh}
\begin{equation}\label{Z_q(beta)}
Z(g^2) \ = \ 1 -  0.908 g^2  \ + O(g^4) \ . 
\end{equation}
The presence of the additional renormalization \eqref{Z_q(beta)} is the reason why computing the bare topological charge \eqref{Q_L} will not result only in integer numbers, but any real value. The value of $Z(g^2)$ cannot be computed using perturbation theory far away from the continuum limit. In this case one has to employ some techniques that can extrapolate the topological content from a configuration in a non perturbative way.
Apart from using the fermionic definition of topology, via the index theorem, there are a number of techniques (cooling, smearing, gradient flow), that successfully deal with the gluonic operator \eqref{q_L}. Despite the differences between those methods, they all involve smoothing of the field in order to eliminate the short range quantum fluctuations and enhance the classical long range modes. In fact, they all result in the recovery of an almost integer value for the topological charge \eqref{Q_L}. 

Problems with lattice renormalization of topological observables first arose  when trying to compute the topological susceptibility on the lattice as a way to validate the Witten-Veneziano formula for the $\eta'$ mass. The expression for the topological susceptibility on the lattice is the topological charge two point function  
\begin{equation}
\chi_L \ = \ \sum_x \langle q_L(x)q_L(0) \rangle = \dfrac{\langle Q^2_L \rangle}{V} .
\end{equation}
This quantity will be affected, other than from the multiplicative renormalization, also from an additive one due to the contact terms arising when $x=0$ :
\begin{equation}
\chi_L \ = \ Z^2(g^2)\; a^4 \chi + M(g^2) \ . 
\end{equation}
The additive renormalization $M(g^2)$ contains the mixing terms of $\chi_L$ with all the other operators with the same quantum numbers and dimension $\dim_O \leq \dim_{\chi_L}=4$, i.e. the trace of the energy tensor and the identity
\begin{equation}
M(g^2) \sim A(g) \langle  G^a_{\mu \nu} G^a_{\mu \nu}\rangle +B(g) \ .
\end{equation}
In general, those renormalizations completely dominate the value of $Q_L$ and $\chi_L$ even in the scaling region of the continuum limit (80$\%$-90$\%$ of the real value). That is why particular care has to be put in eliminating those factors. 

From now on we are going to use the following notation for the topological charge on the lattice
\begin{equation}
\begin{split}
&Q \ \ \text{is the renormalized topological charge} \ , \\
&Q_L\ \ \text{is the bare one}\ ,
\end{split}
\end{equation}\label{QL=ZQ2}
where two definitions are related via \eqref{lim q_L(x)}
\begin{equation}
Q_L=Z(g^2)\ Q \ .
\end{equation}

The renormalization factor $Z(g^2)$ can be non-perturbatively estimated using one of the smoothing methods that allows to sample the the correct value of the topological charge
\begin{equation}\label{Z_q}
Z(g^2) \ = \ \dfrac{\langle QQ_L \rangle_{\theta=0} }{\langle Q^2 \rangle_{\theta=0}  } \ ,
\end{equation}
where the expectation value of $Q^2$ has to be chosen because  $\langle Q \rangle_{\theta=0} =0$.

\vspace{1cm}
\subsection{Gradient Flow for SU(3)}\label{sec:GradFlowSU3}
We shall now briefly describe the gradient flow method in relation to topology \cite{Luscher:2010iy,Luscher:2010we} . It has been shown in \cite{Bonati:2014tqa}, for SU(3) gauge theory, that the smoothing process obtained with gradient flow is equivalent to the one obtained with standard cooling. However, the first one seems to be more suitable to be extended to the SL($3,\mathbb{C}$) group in order to be employed together with complex Langevin dynamics. This is because most of the other methods (like stout smearing, action cooling) require the re-projection of some quantities onto the original unitary manifold, which is not easily extendible in the case where the original manifold is  SL($3,\mathbb{C}$). On the contrary, the gradient flow does not present this inconvenience as it does not involve any projection and, therefore, can be extended more naturally to the complexified manifold.
The gradient flow equation, in the continuum, is 
\begin{equation}\label{WilsonFlow}
\begin{split}
&\dot{V}_{\mu}(x,\tau)\ = - g^2 \ \left[  \partial_{x,\mu} S(V(\tau)) \right] V_\mu(x,\tau)  \\
&V_\mu(x,0)\ =\ U_\mu(x)\ ,
\end{split}
\end{equation}
where with $\dot{V}_{\mu}(x,\tau)$ we indicate the differentiation of the link $V_\mu(x,\tau)$ with respect to the flow time $\tau$, $S(V(\tau))$ is the Wilson action at the time $\tau$ and $\partial_{x,\mu}$ is the Lie derivative with respect of the link $V_\mu(x,\tau)$. We should note that \eqref{WilsonFlow} is very similar to the infinitesimal classical, i.e without statistical noise, Langevin equation. They differ, though, because the coupling $g$ cancels from the equation \eqref{WilsonFlow} in such a way that the evolution in gradient flow time is independent from the coupling. Since the variation of the field $\dot{V}_{\mu}(x,\tau)$ is proportional to minus the gradient of the action, it is clear that along the gradient flow, the action decreases monotonically, $\dot{S} \leq 0$, smoothing the gauge field in accord with \eqref{S<eps}.

The discretized version of \eqref{WilsonFlow} is :
\begin{equation}\label{WilsonFlow discr}
\dfrac{V_{\mu}(x,\tau+\epsilon)-V_{\mu}(x,\tau)}{\epsilon}\ =\ - \left( \sum_a \ Tr\left[ \lambda_a \ (\Gamma_\mu(x)-\Gamma^{-1}_\mu(x)) \right] \ \lambda_a \right) V_{\mu}(x,\tau)
\end{equation}
where $S(V(\tau))$ has been explicitly taken to be the Wilson action and
\begin{equation}
\Gamma_\mu(x)\ =\ \sum_{|\nu|\neq \mu, \nu = -4}^{4}\ \Pi_{\mu \nu}(x)
\end{equation}
is the sum of all the plaquettes containing the link $U_\mu$. There are several ways to numerically integrate the equation \eqref{WilsonFlow discr}; more refined algorithms will have better convergence in $\epsilon$ but will be computationally slower. We choose, for simplicity, to integrate via exponentiation, obtaining an expression analogous to Langevin without noise :
\begin{equation}\label{WF2}
V_{\mu}(x,\tau+\epsilon)\ =\ \exp \left[- \epsilon \left(\sum_a \ Tr\left[ \lambda_a \ (\Gamma_\mu(x)-\Gamma^{-1}_\mu(x)) \right] \ \lambda_a \right)\right] V_{\mu}(x,\tau)
\end{equation}
up to orders $O(\epsilon^2)$. 
\begin{figure}[!t]
\includegraphics[scale=0.6, angle =0]{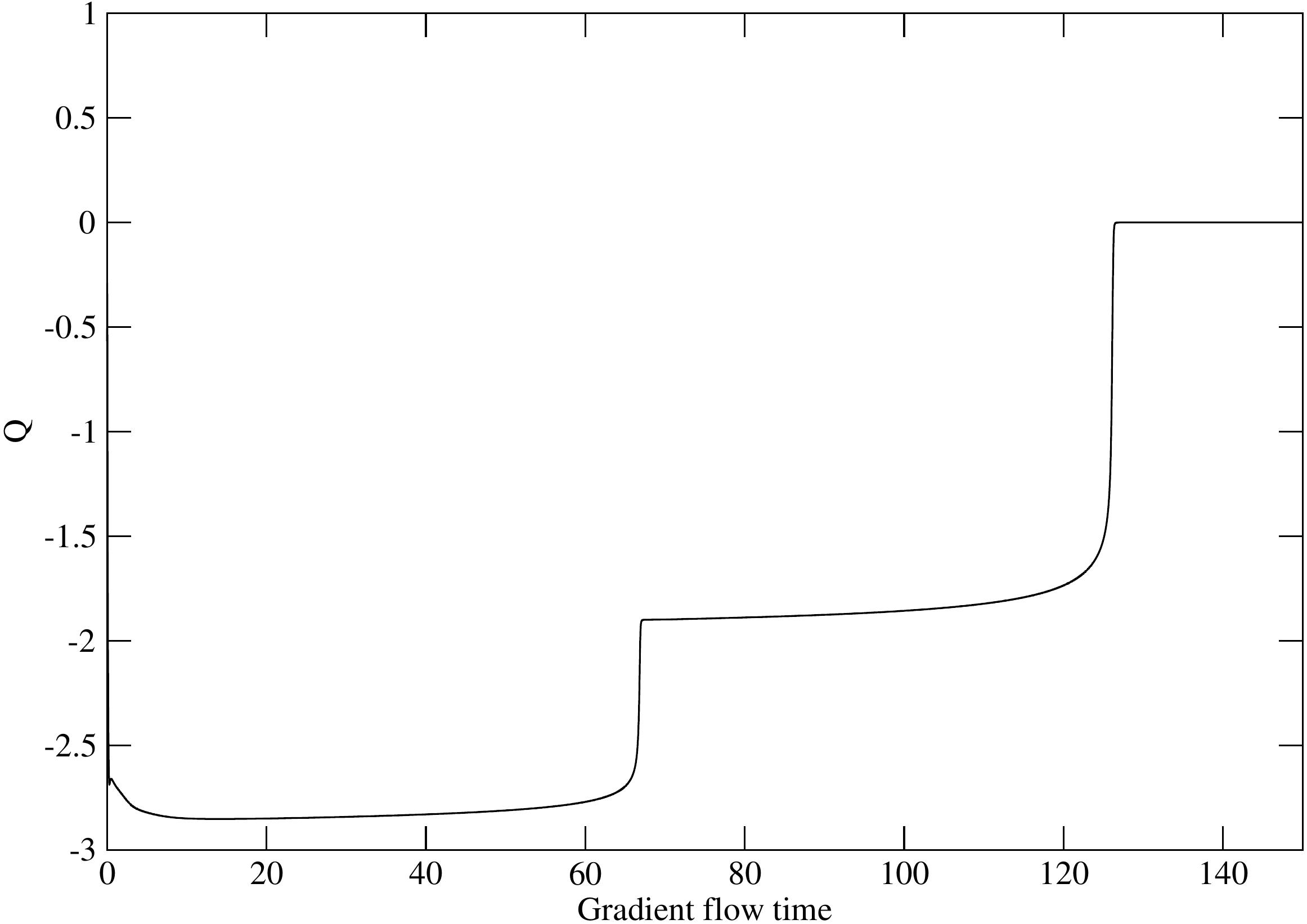}
\vspace{0cm}
\caption{History of the topological charge of a SU(3) configuration (with topological charge $Q=-3$) along the gradient flow, for a $12^4$ lattice at $\beta=5.9$.}\label{fig:WF_config_Q}
\end{figure}
In Fig.\ \ref{fig:WF_config_Q} we show the application of the gradient flow to a SU(3) configuration.
\clearpage
As one can see, after some thermalization time, the value of the topological charge stabilizes on an (almost) integer plateau. Eventually it will fall down onto an other (almost) integer plateau and so on until the configuration reaches the real classical minimum of the Lattice Wilson action, the trivial one with all links $U_\mu(x)=\Id$ and $Q=0$. In this specific example we can see how the topological content of the configuration consists in 2 anti-instantons, one with $Q=-1$ and the other with $Q=-2$. 
 Let us point out that the fact that the plateaus are not exactly integers is just a finite size effect. As soon as the the instanton's size becomes much smaller than the physical volume considered, the values of the plateaus reproduce the correct integers.
\begin{figure}[!t]
\includegraphics[scale=0.6, angle =0]{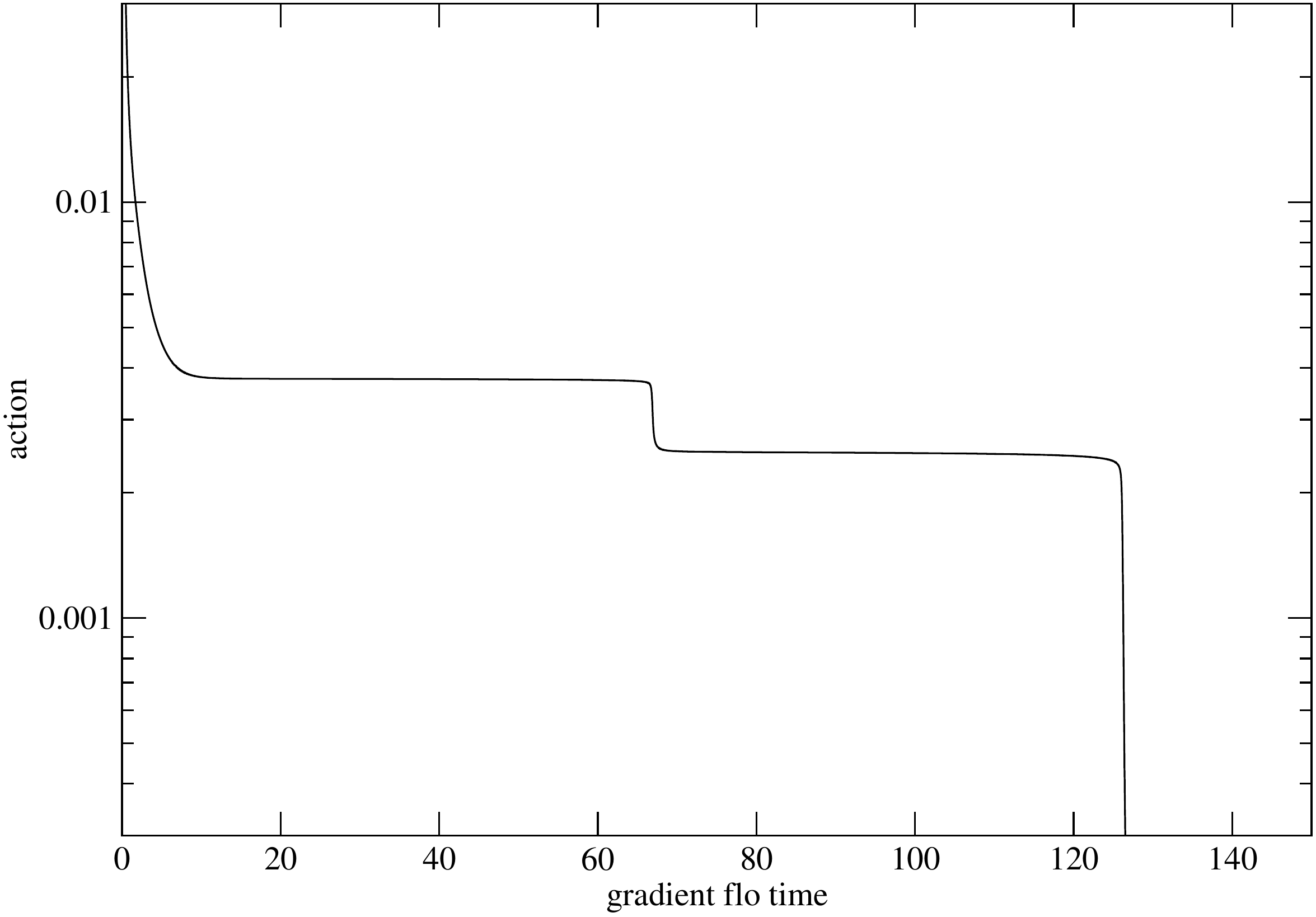}
\vspace{0cm}
\caption{History of the action $S$ for the same configuration of Fig.\ \ref{fig:WF_config_Q} . }\label{fig:WF_config_S}.
\end{figure}
We can also look at what happens to the action during the gradient flow time.  In Fig.\ \ref{fig:WF_config_S} we can see how the plateaux of the action represents the metastable minima of the Wilson action. As soon as an instanton is destroyed the system tunnels back to the minimum closer to the absolute minimum of the Wilson action with $Q=0$.

\clearpage
\section{Complex Langevin dynamics for SU(3) gauge theory with a $\theta$-term}
We shall introduce now the theory we want to study. In the introduction to this chapter, we mentioned why the strong CP problem is still a hot topic to be studied and how the introduction of an axion field might be an elegant solution for it. The Peccei-Quinn argument, however, requires the $\theta$ parameter to be a dynamical variable of the theory and, because of that, one needs to be able to perform simulations at finite $\theta$. 

Other than that, the dependence of QCD on $\theta$ is of theoretical and phenomenological interest by itself. In fact,  derivatives of the vacuum free energy $F(\theta)$ in respect to \t , computed at $\theta=0$, enter various aspects of hadron phenomenology. For example, we saw that the topological susceptibility $\chi_{top}$ enters the Witten-Veneziano equation for the $\eta'$ mass. The susceptibility $\chi_{top}$ can be seen both as the second moment of the topological charge distribution and as the linear response of the $\langle Q_{top}\rangle$ to the parameter $\theta$.

In the continuum, the Lagrangian of pure gauge theory with the $\theta$-term reads
\begin{equation}\label{L_theta}
\mathcal{L}_\theta \ = \ \dfrac{1}{4}G_{\mu \nu}^a(x) G_{\mu \nu}^a(x) - i \theta \dfrac{g^2}{64\pi^2} \epsilon_{\mu \nu \rho \sigma} G_{\mu \nu}^a(x) G_{\rho \sigma}^a(x) \ ,
\end{equation}  
where in the second part we recognise the topological charge density introduced in \eqref{Q_top}. Noticing that $\epsilon_{\mu \nu \rho \sigma} G_{\mu \nu}^a(x) G_{\rho \sigma}^a(x)= 4\ \epsilon_{0 \nu \rho \sigma} G_{0 \nu}^a(x) G_{\rho \sigma}^a(x)$, it is easy to see that the Wick rotation has no effect on the topological charge. This fact is quite relevant since it results in the  $\theta$-term being imaginary also in Euclidian space
\begin{equation}\label{Z_theta}
Z(\theta) \ = \ \int \left[DA \right] \ e^{-S_{YM}} \ e^{- i\theta Q_{top}} \ = \ \exp [-V  f (\theta)] \ ,
\end{equation}
where $f (\theta)=\frac{ F (\theta)}{V}$ is the free energy density. From the large-$N$ limit, it has been conjectured \cite{Witten:1980sp,Witten:1998uka} that the free energy F$(\theta)$ is a multibranched function of the type
\begin{equation}\label{Witten multibranched F}
$F$(\theta)=\ N^2 \min_k\ $H$\left( \dfrac{\theta+2 k \pi }{N}\right). 
\end{equation}
This function is periodic in $\theta$, but (if H is not constant) it cannot be smooth and, at some value of $\theta$, there has to be a jump between two different branches. In particular only the three points $\theta=0$ and $ \theta \pm \pi$ leave \eqref{Witten multibranched F} invariant under a CP transformation ($\theta \rightarrow - \theta$) . Furthermore, the absolute minimum of the free energy must be at $\theta=0$  where the integrand of the Euclidean space path integral is real and positive ($Z_{max}=e^{-V f_{min}}$). 

It would be very interesting to verify this conjecture directly. However, this is not possible because the complex integration measure \eqref{Z_theta} prevents, in general, the theory to be studied on the lattice using standard Monte Carlo methods. However, some information on the free energy has been extrapolated, around $F(0)$, using imaginary $\theta$ \cite{Panagopoulos:2011rb,D'Elia:2012vv,D'Elia:2013eua}
\begin{equation}
\theta_R \equiv -i \theta_I \ , 
\end{equation}
and then analytically continue the results at real $\theta_R$. One might expand the free energy density around the minimum at $\theta=0$ 
\begin{equation}
\mathcal{F}(\theta) \ = \ f(\theta) - f(0) \ = \ \dfrac{1}{2} \chi \theta^2 s(\theta) \ ,
\end{equation}
where $\chi$ is the topological susceptibility and $s(\theta) $ is a dimensionless function of $\theta$, which can be expanded as
\begin{equation}\label{s(theta)}
s(\theta) \ = \ 1 + b_2 \theta^2 + b_4 \theta^4 + ...\ .
\end{equation} 
As said before, the coefficients $\chi, b_2,b_4,$etc.\ are of great phenomenological interest and they are related, through derivatives of the free energy, to the connected momenta of the topological charge distribution.  From \eqref{Z_theta}, in fact, one finds :
\begin{equation}
\dfrac{d^k}{d\theta^k} f(\theta)\ = \ -(-i)^{k} \dfrac{\langle Q^k \rangle^\theta_c}{V} \ ,
\end{equation}
and, in particular, this equation can be used to relate the topological charge expectation value to the coefficients :
\begin{equation}\label{Q_med theta}
\langle Q \rangle_{\theta} = i\ V \dfrac{d}{d\theta} f(\theta )  =   i\ V \chi  \theta (1+2b_2\theta^2+ 3 b_4 \ \theta^4+...) \ .
\end{equation} 
It is clear, then, that the topological charge distribution is a very interesting observable to be studied at finite $\theta$.  

As we mentioned before,  the definition of the topological charge on the lattice is affected by the renormalization factor $Z(\beta)$ \eqref{Z_q}. When simulating at finite \t , this will not only affect the measurements of topological observables, a problem solvable using cooling techniques, but also the evolution of the system itself. Since the action depends on the bare topological charge $S_{\theta}=i \theta Q_L$, it is necessarily affected by its renormalization. The common strategy, in the case of finite \t , is to calculate $Z(\beta)$ at $\theta=0$ and then carry out the simulation at finite \t \ absorbing the renormalization into the parameter
\begin{equation}\label{S_thetaL}
  S_{\theta}=i \theta_L Q_L \ =\ i \theta_L Z(\beta) Q \ =\ i \theta Q\ .
\end{equation}   
Here $Q$ is the proper integer topological charge and we called \tl\ the bare parameter assigned in the simulation. Eq \eqref{S_thetaL} tells us that the actual value of the coupling with the topological charge is the renormalized 
\begin{equation}\label{theta_L}
\theta=Z(\beta) \theta_L.
\end{equation}
One needs, then, to rescale one's results in order to be dependent on the correct renormalized parameter \t .

In the following we are going to introduce the complex Langevin dynamics for pure gauge in the presence of a finite $\theta$-term, and we are going to show our results.

\vspace{1cm}
\subsection{complex Langevin dynamics}
Let us discuss the set up and recall the CL equation for gauge theories
\begin{equation}\label{CL eq}
\begin{split}
&U_{\mu x}(t+\epsilon) = R_{\mu x}(t)U_{\mu x}(t)\ , \\ 
&R_{\mu x}(t)= \exp\left[ -i \sum_a \lambda_a  \left( \epsilon D_a S[U_{\mu x}] + \sqrt{\epsilon} \eta_{a \mu x} \right)  \right] \ ,
\end{split}
\end{equation}
where both $U_{\mu x}$ and $R_{\mu x}$ belong to \SLtree, as it is evident from \eqref{CL eq} when the action $S[U]$ is complex.
Again, $\eta_a$ is \textit{real} Gaussian noise satisfying the relations
\begin{equation}
\begin{split}
&\langle \eta_{a \mu x} \rangle \ =\ 0 \ ,\\
&\langle \eta_{a \mu x}(t)\; \eta_{b \mu x}(t') \rangle  = 2  \delta(t-t') \delta_{ab} \ . 
\end{split}
\end{equation}
The lattice action is the discretized version of \eqref{L_theta}, i.e. Wilson action plus the topological charge coupled with the parameter $\theta$
\begin{equation}
\begin{split}
S\ =\ \sum_x &\left[ \beta \sum_{\mu\nu} \left( 1 - \dfrac{1}{2 N} \left[ \Tr(\Pi_{\mu\nu})+\Tr(\Pi_{\mu\nu})^{-1} \right] \right)\right. \\
 & \ \ \ \left. -\ \dfrac{i\ \theta}{2^4 \times 32 \pi^2} \sum_{\mu\nu\rho\sigma =\pm 1}^{\pm 4} \tilde{\epsilon}_{\mu\nu\rho\sigma} \Tr[\Pi_{\mu\nu}(x) \Pi_{\rho \sigma}(x)] \right] \ ,
 \end{split}
\end{equation} 
where with $\Pi_{\mu\nu}$ we indicate the plaquette. It is useful, at this point, to fix some notation,  with which we are going to indicate the following :
\begin{equation}
\begin{split}
&\theta \ \ =\ \text{renormalized \t, equivalent to the one in the continuum}\ , \\
&\theta_L \ =\ \text{bare (unrenormalized) \t\ on the lattice} \ , \\
&\theta_I \ =\  \text{imaginary bare \t}\ ,  \\
&\theta_R  \ =\ \text{real bare \t}\ .
\end{split}
\end{equation}

We shall now briefly discuss our strategy. The results shown in this section are mainly presented in the papers \cite{Bongiovanni:2013nxa,Bongiovanni:2014rna}.  Before starting to simulate the complex action at real \tr, a lot of care has to be put in the preparatory tests. As discussed in the last chapter \ref{cap:Langevin}, in order to control the exploration of the complexified configuration space of complex Langevin dynamics, gauge cooling is needed. 
\begin{itemize}
\item Particularly important is the behaviour of CL at \textit{imaginary} \ti\ where the action, starting from a \SUtree\ configuration, should remain real during all the evolution. As it is well known, also in the case of Monte Carlo methods, machine rounding errors tend to kick the dynamics off the \SUtree\ manifold. The standard procedure that prevent this to happen is called \textit{reunitarization} and consists in projecting back, every now and then, each link into the original \SUtree. Reunitarization can be applied also to Langevin dynamics, as long as the action is real, granting its convergence (as seen in Cap.\ref{cap:Langevin} for convergence of real Langevin dynamics). However that would not give us any information on the stability of CL in the case when the action is complex and reunitarization is not applicable. What we do, instead, is to let free CL to explore the complex space, even at \ti, and just use gauge cooling (see Cap.\ref{cap:Langevin}) to control it. In this way we can study the stability of CL in reproducing the correct distribution. Furthermore, it can be observed (and it has been shown \cite{Seiler:2012wz}) that the main concern for the correct convergence of CL, comes from the $\beta$ dependence of the Wilson action rather than from the parameters that trigger the sign problem ($\theta$ in this case). Making sure CL converges correctly with the real action is, then, a strong hint for its convergence even with complex action, at least in a similar range of parameters.

\item The second point we want to make sure of is for the observables to behave smoothly in the transition from \ti\ (real action, \ttwo $<0$) to \tr\ (complex action \ttwo $>0$).  After we ascertained CL can be controlled with gauge cooling at \ti , a smooth (without jumps) behaviour around $\theta^2=0$ indicates convergence also in the complex manifold, at least for small \tr .

\item Third and most important argument is the satisfaction of the criteria of correctness discussed in Cap.\ref{cap:Langevin}. The compactness of the observables in the complex directions is always the most important property to be verified in order to ensure convergence of CL . 
\end{itemize}

\begin{figure}[!h]
\includegraphics[scale=0.47, angle =0]{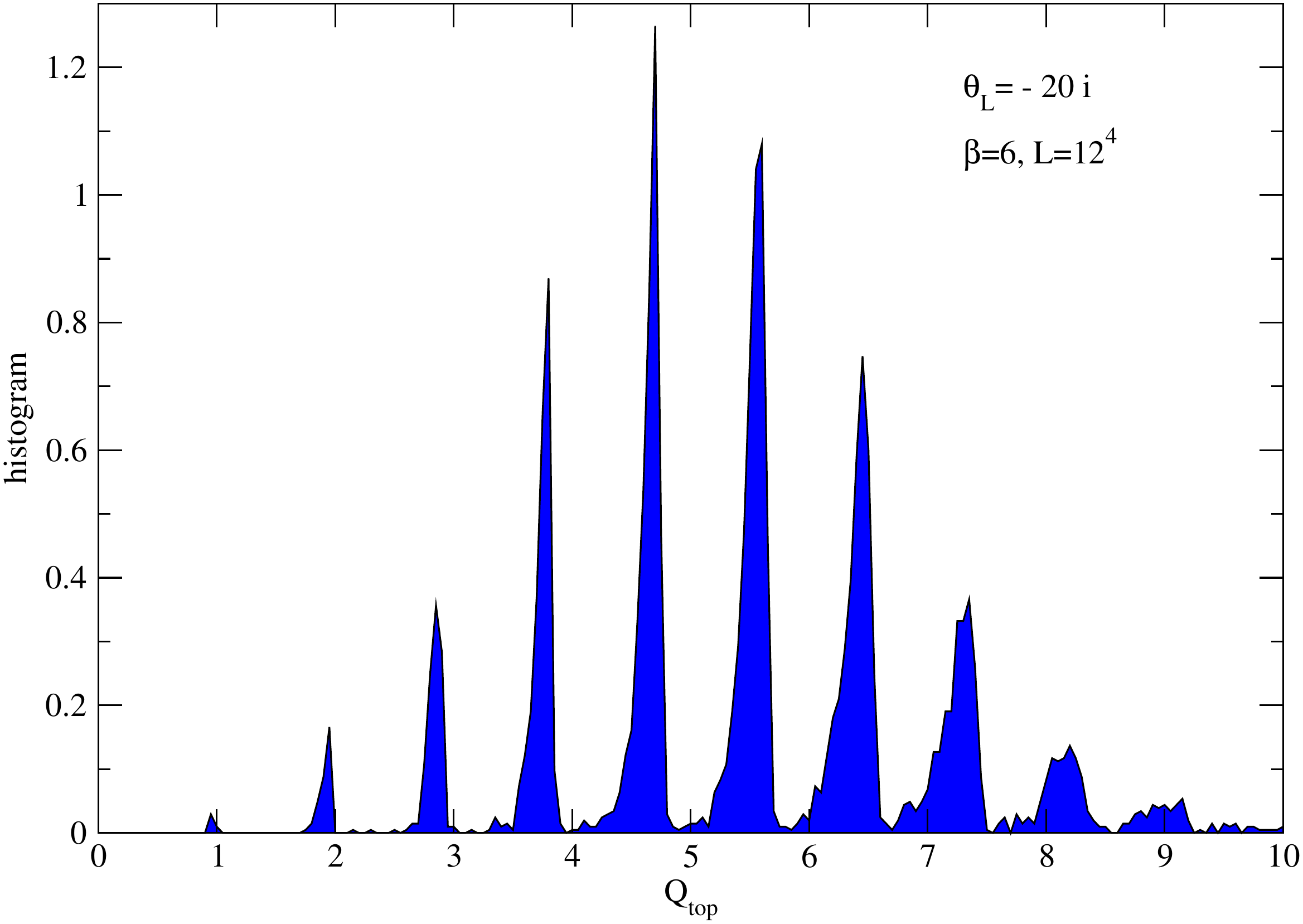}
\includegraphics[scale=0.47, angle =0]{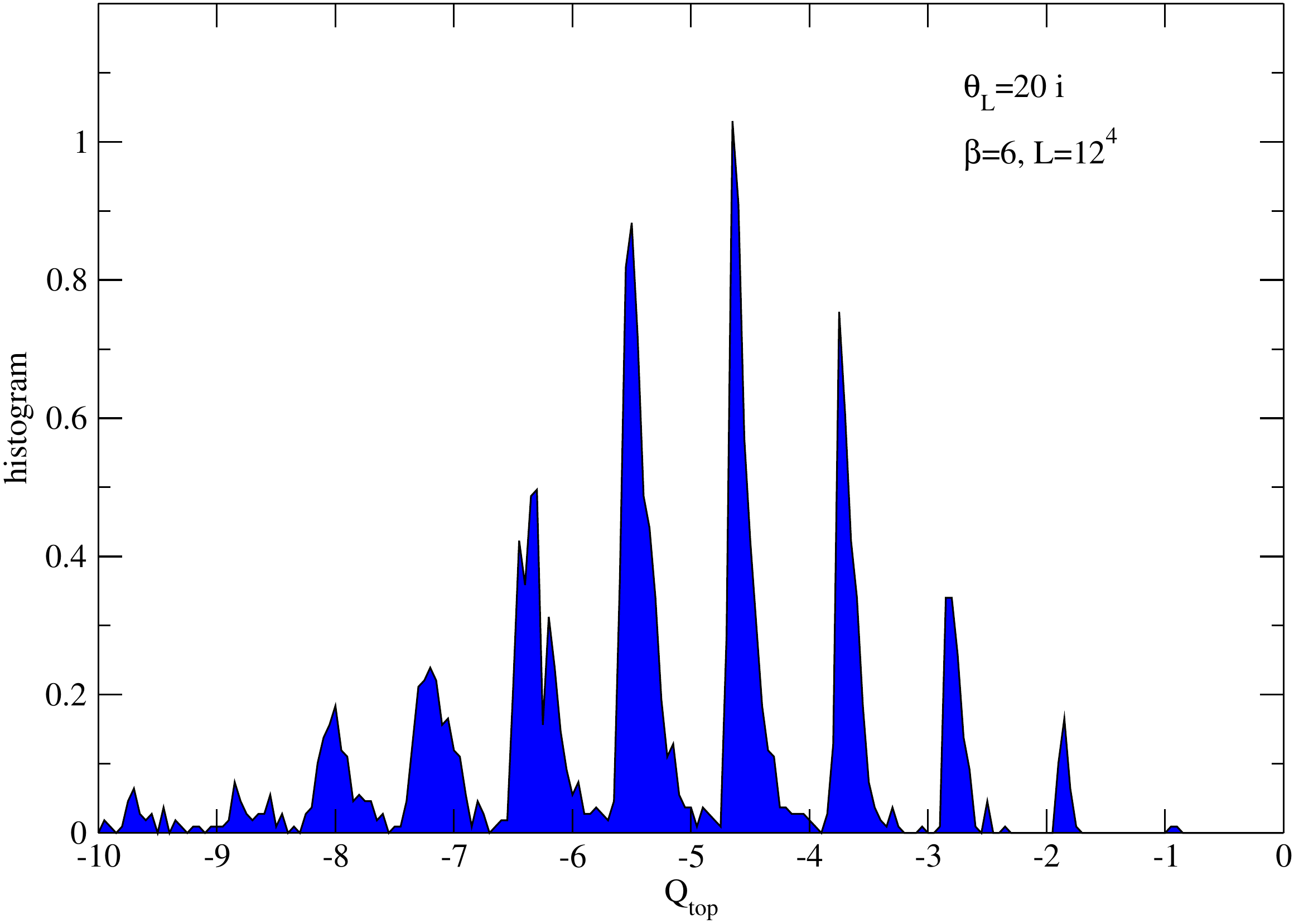}
\vspace{0cm}
\caption{ Distribution of $Q_{top}$ at \ti $=-20i$ (above) and \ti $=20i$ (below), on a $12^4$ lattice at $\beta=6$, obtained by generating configurations using complex Langevin dynamics with gauge cooling and subsequently applying gradient flow to recover the discrete values of the charges. The spacing between the peaks is a bit smaller than one as a finite volume effect, and this is responsable for the peaks being shifted in respect to the integers. }\label{fig:distrib_Q_th20}
\end{figure}

We shall proceed with the first point. As a proof of good convergence of CL at \tl\ we choose to reproduce the topological charge distribution and confront it with the one obtained via the Hybrid Monte Carlo (HMC) method.
In Fig.\eqref{fig:distrib_Q_th20} we show the result of such a test,  at bare \tl $=\pm 20 i$ and $\beta=6$ on a $12^4$ lattice.  Here configurations are generated using complex Langevin dynamics, including gauge cooling, to control the process.  We subsequently use gradient flow to smooth them and recover their topological content. We observe the expected response as the sign of \tl\ is flipped, i.e. $\langle Q \rangle_{\theta}$ is compatible with $-\langle Q \rangle_{-\theta}$ as predicted by \eqref{Q_med theta}. Also we checked the average value of the topological charge and we found agreement with the one computed with the HMC algorithm
\begin{equation}
\begin{split}
&  \theta_I=20\ : \ \ \ \langle Q\rangle_{CL} = -5.38 (\pm0.06) , \ \ \ \langle Q \rangle_{HMC} = -5.42 (\pm 0.09) \ , \\
& \theta_I=-20\ : \ \ \langle Q \rangle_{CL} = 5.46 (\pm 0.06) , \ \ \ \  \langle Q \rangle_{HMC} = 5.48 (\pm 0.09) \ .
\end{split}
\end{equation}
That is enough to make us confident CL converges to the right results at \ti , at least for values of $\beta$ not much smaller than $6$ .

\begin{figure}[!h]
\includegraphics[scale=0.7, angle =0]{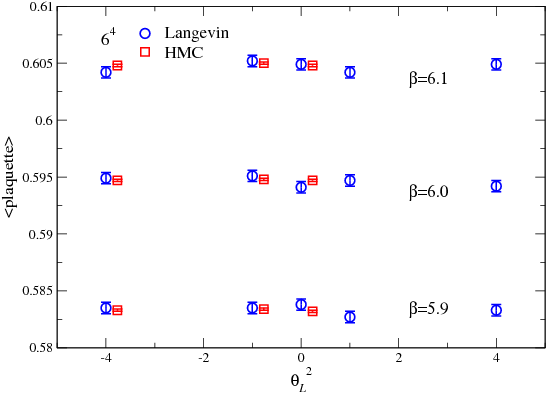}
\vspace{0cm}
\caption{expectation values of the plaquette in SU(3) Yang-Mills theory in the presence of a \t -term , for real and imaginary \tl, using CL and HMC (imaginary \tl only), on a $6^4$ lattice at three $\beta$ values. The HMC data has been shifted horizontally for clarity. }\label{fig:plaq_vs_th^2}
\end{figure}

To proceed with our testing plan, we compute the plaquette at 3 values of $\beta$, for the bare parameter going from imaginary $\theta^2_L < 0$ to real $\theta^2_L > 0$. In Fig.\ref{fig:plaq_vs_th^2}, we can see the smooth  transition of the plaquette from imaginary \t\ to real ones. Good agreement with Monte Carlo results is also shown at imaginary \tl . As expected, the plaquette does not depend much on \t , so that it keeps staying on the same value also at $\theta^2_L > 0$. Nevertheless, this is a very useful result because it shows the good behaviour of CL when the action is complex. 

\begin{figure}[!h]
\includegraphics[scale=0.7, angle =0]{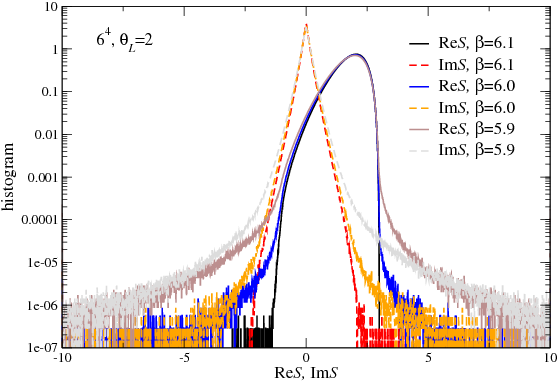}
\vspace{0cm}
\caption{Normalized histogram of the distribution of the complex action for real \tl =2 and $\beta = 5.9,6,6.1$ .  }\label{fig:plaq_distrib}
\end{figure}

As mention before, the key point to trust the simulation at \tr\ is the distribution of the observables in the complex plane. In Fig.\ref{fig:plaq_distrib} we show the normalized histograms of the values of the real and imaginary part of the action (Re$S$ and Im$S$). The plot shows, in a logarithmic scale, the decay of the observable in the complex plane (often referred to as 'skirt' of the observable). The more tight the skirts are, the better CL criteria of correctness are satisfied. We can see that at larger $\beta$, gauge cooling is very effective in controlling the skirts of the distribution while it becomes more and more problematic at smaller $\beta$. Eventually the skirt will become too wide to be controlled and CL will have problems of convergence. Nevertheless in the cases presented, even for $\beta=5.9$, CL reproduces the correct values for the action, as the results at small \tr\ lie on the analytical continuation of the ones at \ti\ (Fig.\ref{fig:plaq_vs_th^2}). 

Let us stress further that, at this stage, we are not yet interested in measuring physical quantities. In this sense, even if some observable is likely to be affected by finite-size effects in a $6^4$ volume at $\beta=6.1$, we are only interested in comparing CL and Monte Carlo at the same volume and $\beta$ and get the same results.

\clearpage
\subsection{Topological charge distribution}
Having completed the preliminary tests, we shall now go back to our goal to study the topological charge dependence on \t . The behaviour we would like to reproduce is the one shown in equation \eqref{Q_med theta}.  Unfortunately to do so we run into the problem of renormalizations of \t .  We mentioned before, \eqref{theta_L}, that it would be enough just to rescale \t\ in the action with the renormalization factor $Z(\beta)$ computed at $\theta=0$ (since it only depends on $\beta$). However, this procedure is not correct when both the action and the observable measured are affected by the same renormalization.  In particular, if the observable is $Q_L$ itself, this can be seen from the partition function  
\begin{equation}\label{Q_thetaL}
\langle Q_L \rangle \ = \ \dfrac{1}{Z(\theta_L)} \int \left[DU \right] \ Q_L\ e^{-S_{W}} \ e^{- i\theta_L Q_{L}} \ ,
\end{equation}
where the noise of $Q_L$ measured autocorrelates with the one of $Q_L$ in the action. In this case, one first has to measure the correct renormalized topological charge $Q$ and, only after, he is allowed to rescale \tl\ to get the correct result. However, computing the renormalized topological charge at real \t\ would require the implementation of some smoothing technique in \SLtree\ which is a hard task and is still a work in progress (we will show some results in the last section). Instead, what we are going to present here is a study of the bare lattice theory.

In the bare theory, the bare topological charge is still expected to follow the behaviour of \eqref{Q_med theta}, but with different values of the coefficients $\chi, b_2, b_4,..$ :
\begin{equation}\label{QL_thI&thR}
\begin{split}
&\langle Q_L \rangle_{\theta_I} = - \ V\dfrac{d}{d\theta_I} f(\theta_I) =   -\ V \chi_L \ \theta_I( 1\ -\ 2 b^L_2 \ \theta_I^2 + 3 b^L_4 \ \theta_I^4 + ...) \ , \\
&\langle Q_L \rangle_{\theta_R} = i\ V\dfrac{d}{d\theta_R} f(\theta_R) =   i\ V \chi_L \ \theta_R( 1\ +\ 2 b^L_2 \ \theta_R^2 + 3 b^L_4 \ \theta_R^4 + ...) \ .
\end{split}
\end{equation}

Here we explicitly wrote down the dependence of $Q_L$ on imaginary and real \tl\ and indicated with $\chi_L, b^L_2, b^L_4$ the bare parameters. Note that for real \t\  $\langle Q_L \rangle$ must be imaginary because the real part of \eqref{Q_thetaL} cancels with a CP transformation. 
 \begin{figure}[!h]
\includegraphics[scale=0.6, angle =0]{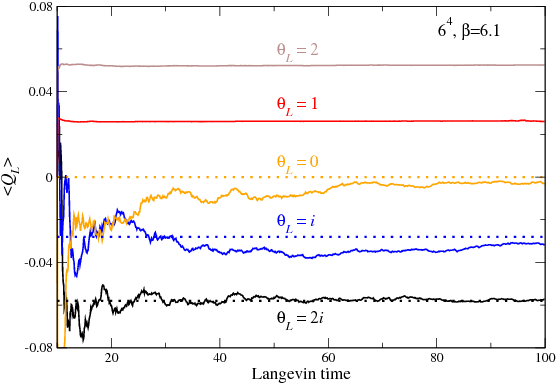}
\vspace{0cm}
\caption{CL running averages of the bare topological charge, for $\beta=6.1$. The imaginary part of $Q_L$ is shown for \tr\ and the real part for \ti . The dotted lines at \ti\ indicate results obtained with HMC algorithm.}\label{fig:Q_small_theta}
\end{figure}

In Fig.\ref{fig:Q_small_theta} we show the behaviour of the bare topological charge in the transition from \ti\ to \tr\ , for small values of the bare parameters \tl . The shaky lines are the running averages of $\langle \Re$e$(  Q_L) \rangle$ at \ti\ , where  $\langle \Im$m$(Q_L) \rangle$ is 0. We can see the averages converging to the values predicted by Hybrid Monte-Carlo. When \tl\ is real, we have the steady lines that represent $\langle \Im$m$(Q_L) \rangle$, while now $ \Re$e$(Q_L)$ fluctuates around 0, exactly as expected.

\begin{figure}[!h]
\includegraphics[scale=0.55, angle =0]{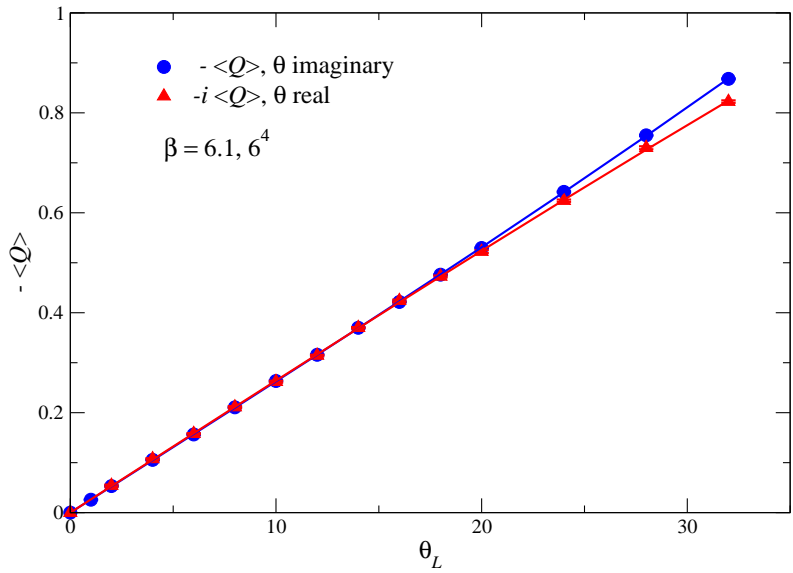}
\vspace{0cm}
\caption{Opposite curvatue in the dependence of $\langle Q_L\rangle$ on real and imaginary \tl, at $\beta=6.1$ .}\label{fig:QL_thI&thR}
\end{figure}

In general the renormalization is of the order of $Z(\beta)\simeq 0.1$ \cite{D'Elia:2012vv}, so that when we set $\theta_L=1$ we are really working at roughly $\theta \simeq 0.1$. That means in Fig.\ref{fig:Q_small_theta} \tl\ is actually quite small and we can mostly see the linear part of \eqref{QL_thI&thR}, proportional to the bare topological susceptibility $\chi_L$. To be able to observe the effect of higher powers of \tl\ in \eqref{QL_thI&thR} one needs larger \tl. In Fig.\ref{fig:QL_thI&thR} we show the dependence of $\langle Q_L\rangle$ on \ti\ (blue line) and \tr\ (red line) for bigger values of \tl. Again, imaginary and real parts of $Q_L$  are multiplied by a $"-"$ sign and an $i$, so that they can be plotted next to each other. Already from the plot it is evident that the two curves have opposite curvature at high \tl .  More precisely, by fitting the data with \eqref{QL_thI&thR} we obtain, for both curves 
\begin{equation}\label{<Q>th_fit}
V \chi_L = 0.026,\ \ \ \ \ \ \text{and}\ \ \ \  b_2^L \simeq 10^{-5} \ ,
\end{equation}
confirming the expected common linear response and an equal and opposite curvature, in perfect accord with the opposite sign of the cubic term in \eqref{QL_thI&thR}. 

We can now use our information about the \t\ dependence of $\langle Q_L\rangle$ to study the dependence of the bare topological susceptibility on $\beta$. Since an increasing value of $\beta$ means lower temperature (and also smaller physical volumes), we expect the susceptibility to decrease. To see this we just have to repeat the fit in \eqref{<Q>th_fit} at different values of $\beta$. 
\begin{figure}[!t]
\includegraphics[scale=0.55, angle =0]{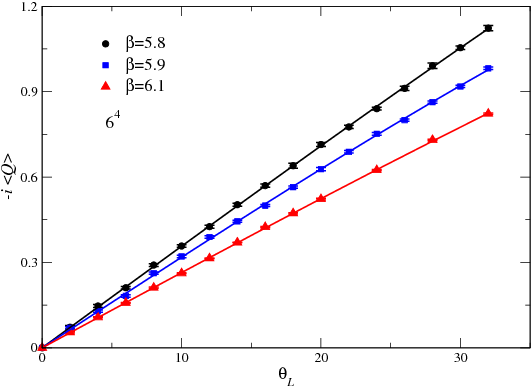}
\vspace{0cm}
\caption{Same as Fig.\ref{fig:QL_thI&thR}, at real \tl , for three values of $\beta = 5.8,5.9,6.1 $ .}\label{fig:QL_thR_variousBeta}
\end{figure}
In Fig.\ref{fig:QL_thR_variousBeta} we show just the data at \tr\ for fixed volume and increasing temperature (increasing $\beta$).
\clearpage
\begin{figure}[!t]
\includegraphics[scale=0.6, angle =0]{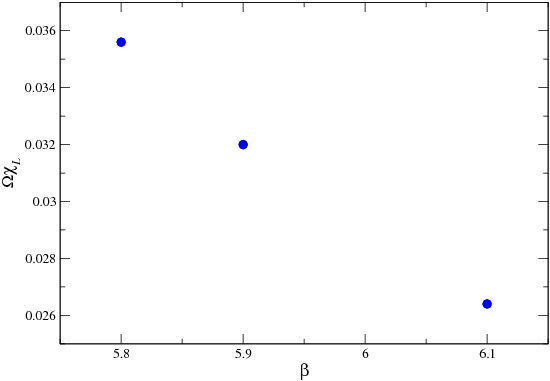}
\vspace{0cm}
\caption{Bare topological susceptibility $\chi_L$ as a function of $\beta$.}\label{fig:chiL_variousBeta}
\end{figure}
The values for the bare topological susceptibility times the volume, i.e. the linear term of the curves in Fig.\ref{fig:QL_thR_variousBeta},  are reported in Fig.\ref{fig:chiL_variousBeta}. Although they are not trivially connected with the values of the physical renormalized topological susceptibility, they must have the same qualitative dependence on the temperature. In fact, we can see that $\chi_L$ decreases as $\beta$ (temperature) increases, as is the case with the physical renormalized $\chi$.

\vspace{1cm}
\subsection{Topological charge on \SLtree\ manifold}
In the previous section we showed results for the Yang-Mills theory with bare topological charge. We already mentioned (Sec.\eqref{sec:GradFlowSU3}) the correct procedure to deal with the lattice renormalizations of topological quantities when the gauge group is SU(3). However, CL at real \t\ works in the \SLtree\ gauge group which greatly complicates things not only practically but also conceptually. The very definition of topology (reviewed in Sec.\eqref{sec:Top&inst_continuum}) relies on the compactness of the gauge group that generates, for the pure gauge fields, a precise mapping between the spatial infinity three-sphere $S_3$ and the group itself. As soon as the gauge group gets enlarged 
\begin{equation}
SU(3)\ \longrightarrow \ SL(3,C)
\end{equation}
it loses its property of compactness and the Yang-Mills part of the action does not have well defined classical minima any more. Consequently, the operator 
\begin{equation}
Q\ =\ \int d^4x\ F\tilde{F} 
\end{equation}
is not expected to assume integer values any more, nor to be invariant under small deformation of the fields . 

 Here we shall discuss some results on the subject that, however, are not complete yet. First of all we should keep in mind that SU(3) is still sub-group of \SLtree , so we should expect to have some sectors where some part of the topological charge is an integer. Whether the  actual topological charge is affected by this sector is another story and, in fact, we shall see that this is not the case. 

It is useful, in order to understand what is going on, to bring up an example which refers to a much simpler theory.  Let us consider a real scalar field theory with a topological term on a circle ($S_1$)  :
\begin{equation}
0 \leq x <  L\ ,\ \ \  0 \leq \phi\ < 2 \pi \  ,\ \ \ \phi\ =\ \phi +  2 k \pi\ \ ,\ \ \ k \in \mathbb{Z}\ ,
\end{equation}
where the partition function is
\begin{equation}
Z\ =\ \int D\phi\ e^{-S + i \theta Q} \ ,
\end{equation}
and the action and the topological charge are defined as :
\begin{equation}
S\ =\ \dfrac{1}{2} \sum_x (\partial_x \phi)^2 \ ,\ \ \ Q\ =\ \dfrac{1}{2 \pi} \sum_x \partial_x \phi \ .
\end{equation}
It is easy to see that the topological charge must be an integer classifying how many times the field $\phi$ wraps around itself on the circle. Furthermore, again for CP reasons, the expectation value of the topological charge must be purely imaginary 
\begin{equation}\label{Q phi}
\langle Q \rangle\ =\ \dfrac{1}{Z} \ \int\ D\phi\; e^{-S}\ i \sin(\theta Q)\ Q \ .
\end{equation}
In order to use complex Langevin dynamics, we complexify the field
\begin{equation}
\phi \ \ \rightarrow\ \ \phi_R\ +\ i\ \phi_I\ ,
\end{equation}
where $\phi_I$ is unbounded, and we can write separately the real and imaginary part of the topological charge 
\begin{equation}\label{Q_phi_complex}
 Q\ = \sum_x \partial_x \phi_R\ +\ i\ \sum_x \partial_x \phi_I \ =\  \ Q_R\ +\ i\ Q_I \ .
\end{equation} 

\begin{figure}[!t]
\includegraphics[scale=0.5, angle =-90]{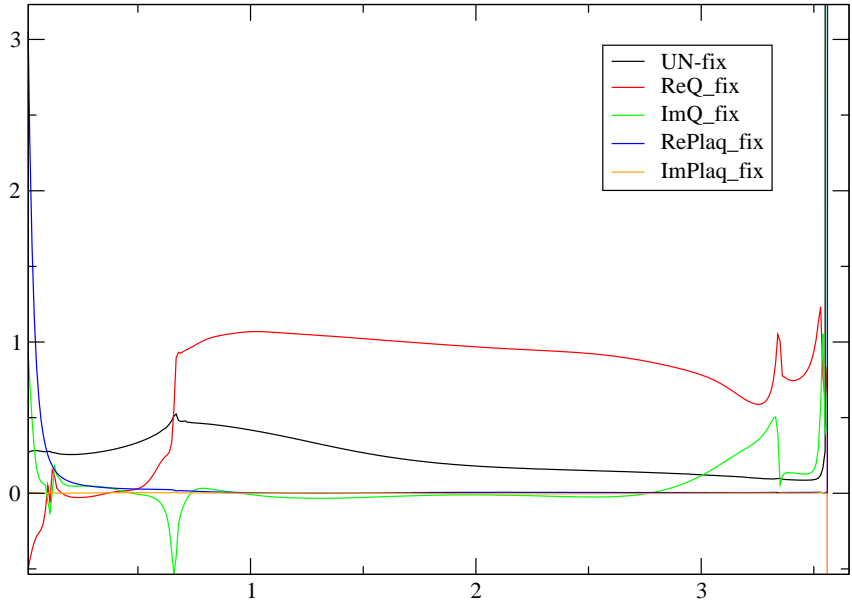}
\vspace{-1cm}
\caption{Wilson Flow on a  $SL(3,C)$  configuration (x axis is flow time) for  $\ \ \theta_L=20 , \ \ \ \beta = 6.1, \ \ L = 8^4 $.}
\label{fig:GradFlow SL(3,C) conf}
\end{figure}

The space still being a circle and the real part of the field $\phi_R$ periodic, we expect $Q_R$ to have the exact same topological meaning of $Q$ for the real theory.  However using CL (see Cap.\ref{cap:Langevin}), we know that  the observables are distributed according to the equilibrium Fokker-Planck probability distribution function, which is \textit{real} even when the fields are complex
\begin{equation}\label{<Q> CL}
\langle Q \rangle\ =\ \dfrac{1}{Z} \ \int\ D\phi\ P(\theta,\phi_I,\phi_R) (Q_R\ +\ i\ Q_I )\ .
\end{equation}
Since we know, from \eqref{Q phi}, that $\langle Q \rangle$ has to be purely imaginary for real \t , we have to conclude that $\langle Q_R \rangle = 0$. We conclude that with CL, the contribution to the average values of the topological charge at real \t\ comes entirely from the non-compact imaginary direction and, therefore, has no topological meaning in the enlarged manifold. Furthermore, the topological charge of the real subgroup, which still retains the usual topological properties, needs to cancel exactly for every real \t\ .

The case of \SLtree\ is completely analogous. Although the topological charge operator cannot be simply factorised as in \eqref{Q_phi_complex}, its properties must be the same. On the SU(3) subgroup it retains its properties and meaning, but needs to average to 0. On the other hand, on the complex manifold, $Q$ has no topological meaning but, provided good convergence of CL, it would average to the correct value.   

We would like to observe this on a \SLtree\ configuration. As anticipated, we are going to employ an extended version of the gradient flow discussed in Sec.(\eqref{sec:GradFlowSU3}). That is because, out of all the smoothing methods, the gradient flow seems the one which can get more naturally extended to \SLtree\ because it doesn't explicitly involves any projection on the real manifold. In Fig.\ref{fig:GradFlow SL(3,C) conf}, we show the effect of the gradient flow on a thermalized configuration at real $\theta_L=20$ . In particular, it is interesting to observe the real part of the topological charge operator (red line) reaching an integer plateaux, proof that topological sectors are still present in the SU(3) subgroup. The imaginary part of $Q$, on the other hand, does not stabilise on any integer value and just flows to (more or less) 0. 

Eventually, at flow time $t\simeq 3.6$, we can see that the gradient flow equation encountered an unbounded runaway direction, along which all the quantities diverged.

A consistent application of the gradient flow on the \SLtree\ manifold is still under work. Its development is essential to extract the correct imaginary part of the topological charge operator from the configurations and, therefore, to obtain the correct value of the renormalized topological charge at real \t .

\vspace{2cm}
In this Chapter we reviewed some aspects on the instantons and the structure of the vacuum of the continuum theory. We discussed the necessity to study topology, non-perturbatively, on the lattice and the procedures required to extrapolate meaningful values of the topological observables. We mentioned why it is relevant to understand the theory at finite \t\ and how this theory is affected by the sign problem on the lattice. We adopted complex Langevin dynamics as a solution for this and we showed our results at real \t\ for the bare theory. We explained the difficulties involved in the extraction of the renormalized topological quantities in the complexified manifold in which CL dynamics takes place. Lastly, we gave an account of our progress in the direction of solving those problems.

\begin{comment}

%% file: cap_LEFSCHETZ/Lefschetz_thimbles.tex
\tableofcontents
\end{comment}
\externaldocument{cap_Langevin/CLangevin.tex}

\chapter{Lefschetz thimbles}\label{cap:Lefschetz}
In this chapter we are going to introduce another method to deal with sign problem in lattice quantum field theory, i.e. integration along the \textit{Lefschetz thimble}. The starting idea is similar to complex Langevin dynamics. One is faced with an (highly) oscillatory integral over some real manifold, where the sign problem is severe, and tries to complexify the degrees of freedom so that in the new complex manifold the sign problem could be milder or absent. We saw in chapter \ref{cap:Langevin} how complex Langevin dynamics enlarges the system into the complex manifold. There the probability distribution function, generated by the Fokker-Plank equation, is real and, provided the criteria of correctness are satisfied and the action is holomorphic, leads the observables to converge to the right expectation value.

For the Lefschetz thimble, instead, the idea is similar to the one of the saddle-point approximation. However, here the integral is not approximated by the Taylor expansion of the complex function around the dominant stationary point. Instead, the original real domain of integration is deformed into the union of new paths, of the same dimension, i.e. the Lefschetz thimbles. Each thimble is a sub-manifold of the complexified space, with the property of passing through one critical point of the action. Along them the phase of the integrand is constant, although in general it has a different value on each of them. The union of all the thimbles represents a deformation of the original real domain. If the function is holomorphic, the integral over the new paths should have the same (Cauchy theorem) results as over the real manifold, with the advantage that the phase along the thimbles is constant. The sign problem, however, has not completely disappeared.  A residual sign problem is still present along each thimble, due to the curvature of the integration contour in the complexified space, i.e. the complex Jacobian coming from the parametrisation of the thimble.  A second sign problem appears in the case that more than one thimble contributes. In this case one has to take in account that the phases, although constant, are different along each thimble. This is typically referred to as a global sign problem.

Witten proposed to apply this method to quantum field theory \cite{Witten:2010cx,Witten:2010zr} and, since then, much progress has been made in the direction of application to lattice field theory \cite{Cristoforetti:2013wha,Mukherjee:2013aga,Fujii:2013sra}. Mostly studied have been the relativistic Bose gas at finite chemical potential \cite{Cristoforetti:2012su,Cristoforetti:2014gsa}, but also real time quantum mechanics and condense matter \cite{Tanizaki:2014xba,Mukherjee:2014hsa}.\\\\
Since both complex Langevin dynamics and Lefschetz thimble work in the complexified space, it is very interesting to compare them. In the following we are going first to review the basic theory of the Lefschetz thimbles and, then, compare results for U(1) 1-link model with a determinant and SU(2) 1-link model.

\vspace{2cm}
\section{Thimble equations}
First we are going to introduce the basics of the complexification on Lefschetz thimbles. Let us consider $n$ real degrees of freedom  $\textbf{x} = \lbrace x_1,x_2,...,x_n \rbrace \in \mathbb{R}^n$ that work as a domain for the complex action $S[\textbf{x}] \in \mathbb{C}$. The partition function for this model is defined, as usual, by the path integral 
\begin{equation}
Z=\int_{\mathbb{R}^n} d^n x\; e^{-S(\textbf{x})}  \ ,
\end{equation}
where $d^n x$ is the element of infinitesimal  volume.  As anticipated, the next step is to complexify the domain of integration, i.e. changing the variables $x_i \rightarrow x_i + i y_i$, while the action is extended to the holomorphic $S[z_i]=S[x_i+iy_i]$. Thanks to Picard-Lefschetz/Morse theory \cite{bott1982,Nicolaescu2011}, then, it is possible to select appropriate $n$ dimensional sub-manifolds, i.e. the \textit{Lefschetz thimbles}, immersed into the complex manifold, along which the integral has the same value as in the original real manifold. The new contour represent the \textit{downward} gradient flow associated to the real part of the action
\begin{equation}\label{eq_thimble_z}
 \dfrac{d z_i}{dt_i}= - \overline{ \left( \dfrac{\partial{S(\textbf{z})}}{\partial z_i} \right) }\ =\ -  \dfrac{ \partial{S(\overline{\textbf{z}})  }}{ \partial \overline{z}_i } ,
\end{equation}
where $t$ is the variable that parametrizes the thimble.  Equivalently the equations for the single components are
\begin{equation}\label{eq_thimble_xy}
\begin{split}
&\dfrac{d x_i}{dt_i}= -\text{Re}\, \dfrac{\partial{S(\textbf{z})}}{\partial z_i}\ ,\\
&\dfrac{d y_i}{dt_i}=\ \ \text{Im}\,  \dfrac{\partial{S(\textbf{z})}}{\partial z_i}\ ,
\end{split}
\end{equation}
The procedure is to start from the set $\Sigma$ of critical points $\lbrace \textbf{z}_\sigma \ \vert\  [\partial_{z_i} S[\textbf{z}]]_{\textbf{z}=\textbf{z}_\sigma} =0\rbrace $ and construct the sub-manifold solution of \eqref{eq_thimble_z}  $\mathcal{J}_\sigma$ that originates from those points. Along those trajectories the imaginary part of $S[\textbf{z}]$ is ensured to be constant and the integral with the real part is convergent :
\begin{equation}\label{S along thimble}
\begin{split}
&\dfrac{d}{dt} \text{Im}S = \dfrac{1}{2i} \left[ \dfrac{\partial S(\textbf{z})}{\partial z_i}\ \dot{z_i} - \dfrac{\partial S(\overline{\textbf{z}})}{\partial \overline{z_i}}\ \dot{\overline{z_i}}\right] =0\ , \\
&\dfrac{d}{dt} \text{Re}S = \dfrac{1}{2} \left[ \dfrac{\partial S(\textbf{z})}{\partial z_i}\ \dot{z_i} + \dfrac{\partial S(\overline{\textbf{z}})}{\partial \overline{z_i}}\ \dot{\overline{z_i}}\right] = -\left\vert  \dfrac{\partial S(\textbf{z})}{\partial z_i}  \right\vert^2 \leq 0 \ .
\end{split}
\end{equation}

It is useful to look at the one-dimension case, where the original manifold $\mathbb{R}$ gets enlarged into $\mathbb{C}$. Here we know (Cauchy theorem) that we can deform, with continuity, the original contour of integration as much as we want, as long as $S[z]$ is holomorphic and the new contour tends to $\mathbb{R}$ for $x \rightarrow \pm \infty$. We have no problem, then, to accept that the integral of $e^{-S[z]}$ along any complex path, respecting those two conditions, will have the same value as on $\mathbb{R}$. Furthermore, we can write the thimble equations more explicitly
 \begin{equation}\label{eq_thimble_1dim_xy}
\begin{split}
&\dfrac{\partial}{\partial t} x = - \dfrac{1}{2} \left( \dfrac{\partial u(x,y)}{\partial x} + \dfrac{\partial v(x,y)}{\partial y}\right)\ , \\
&\dfrac{\partial}{\partial t} y = - \dfrac{1}{2} \left( \dfrac{\partial u(x,y)}{\partial y} - \dfrac{\partial v(x,y)}{\partial x}\right)  \ ,
\end{split}
\end{equation}
where we used the holomorphicity of $S[x+iy]=u(x,y)+i v(x,y)$, with $u(x,y)$ and $v(x,y)$ real functions, and the property of the derivatives $\frac{\partial}{\partial z} = \frac{1}{2} \left( \frac{\partial }{\partial x} - i \frac{\partial}{\partial y} \right)$. Here one can show directly that the imaginary part of the action Im$S[z(t)] = v(x(t),y(t))$ is constant. Let us take its derivative along the thimble
\begin{equation}
\begin{split}
\dfrac{d}{dt} v(x(t),y(t)) \ & = \  \ \dfrac{\partial v }{\partial x} \dot{x} + \dfrac{\partial v }{\partial y} \dot{y} \  \\
& =\ -\dfrac{1}{2} \left( \dfrac{\partial v }{\partial x} \dfrac{\partial u}{\partial x} + \dfrac{\partial v }{\partial x} \dfrac{\partial v }{\partial y} + \dfrac{\partial v }{\partial y} \dfrac{\partial u }{\partial y} - \dfrac{\partial v }{\partial y} \dfrac{\partial v }{\partial x} \right) \ ,
\end{split}
\end{equation}

where we substitute the expression \eqref{eq_thimble_1dim_xy} for $\dot{x}$ and $ \dot{y}$. Using the Cauchy Riemann equations $\frac{\partial u}{\partial x}-\frac{\partial v}{\partial y} =\frac{\partial u}{\partial y}+\frac{\partial v}{\partial x}=0$, it follows that $\dot{v}(x,y) =0$ which means, of course, that $v(x,y)$ is constant. In 1 dimension, then, the thimbles method can be understood more intuitively using simple complex analysis. In higher dimension the argument is much more complicated but can be proven using Morse theory. 

Together with the Lefschetz thimble, one can create another set of sub-manifolds passing through each critical point $\textbf{z}_\sigma$ and associated, this time, to the \textit{upward} gradient flow. Those are path of steepest ascent $\mathcal{K}_\sigma$, generated by the equation   
\begin{equation}
 \dot{z}_i = +\ \overline{\partial_{z_i} S(\textbf{z})}\ ,
\end{equation}
 along which the Morse function  $-$Re$S[\textbf{z}]$ diverges. The unstable thimbles are essential to know which of the stable ones have to be taken into account to produce a complete equivalent the original real integration path. The prescription is
 \begin{equation}
 Z=\sum_{\sigma \in \Sigma} n_\sigma Z_\sigma
 \end{equation}
where $n_\sigma$ is the 'intersection piaring' 
\begin{equation}\label{intersection pairing}
n_\sigma=\langle \mathbb{R}^n,\mathcal{K}_\sigma \rangle \ .
\end{equation}
of the unstable thimble, passing through the critical point $\textbf{z}_\sigma$, with the original domain of integration. The product in \eqref{intersection pairing} counts the intersection points (or sub-manifolds) of two manifolds, and the intersection number $n_\sigma$ can be 0 or $\pm 1$, selecting which thimble contributes to the path integral and which not. Again, in the 1 dimensional case one can usually recognise the thimbles contributing just by by looking at the ones that represent a continuous deformation of the original real cycle.

The integral over thimbles can be explicitly written as
\begin{equation}
Z_\sigma = e^{-i\, \text{Im}S(\textbf{z}_\sigma)} \int_{\textbf{z}(t)\in \mathcal{J}_\sigma} dz(t)\ e^{-\text{Re}S[\textbf{z}(t)]} \ ,
\end{equation} 
where the constant imaginary part $e^{i\, \text{Im}S[\textbf{z}_\sigma]}$ has been factorized out of the integral and generates the global sign problem we anticipated before. The other, more subtle, residual sign problem comes from the integration measure $dz(t)$. Here, in fact,  one has to take in account the Jacobian coming from integration along the complex manifold
\begin{equation}
\mathcal{D}z(t) = \left\vert \dfrac{ \partial \textbf{z} }{ \partial \textbf{t} }  \right\vert d^n t \ ,
\end{equation}
which could generate a non trivial sign problem due to the curvature of the thimble. This sign problem, however, is supposed to be milder than the one at the beginning and, therefore, can be treated without compromising the measurements \cite{Fujii:2013sra}.

All the results for Lefschetz thimbles  (i.e. plots and numerical results) presented in this chapter are obtained by numerical integration with Mathematica of the Lefschetz equations \eqref{eq_thimble_xy} .

\section{Thimbles and Langevin dynamics}
Here we show a practical example for the application of the Lefschetz thimble method to a quartic model. We are also going to compare it with the complex Langevin dynamics approach to the same model. This result appears in the papers  \cite{Aarts:2014nxa}.

The action we want to examine is
\begin{equation}
S(z) = \frac{\sigma}{2} z^2+\frac{1}{4}z^4+hz,  \quad\quad\quad \sigma \in\mathbb{R},  h\in \mathbb{C}.
\end{equation}
where the linear term explicitly breaks parity symmetry $z\rightarrow -z$ and generates, in this case, the sign problem.  Here the original domain is $\mathbb{R}$ and its complexification is the complex plane.

The complex Langevin dynamics have been discussed in Cap.\ref{cap:Langevin}. Here let us just recall that the equation for $z$ reads
\begin{equation}\label{Lang eq z}
\dot z = -\partial_z S(z) +\eta,
\end{equation}
or explicitly
\begin{equation}\label{Lang eq xy}
\begin{split}
&\dot{x}  = - \text{Re}\, \partial_z S(z) +\eta, \\
&\dot{y} =  - \text{Im}\, \partial_z  S(z),
\end{split}
\end{equation}
where $\eta$ is real gaussian noise obeying
\begin{equation}
\begin{split}
&\langle \eta\rangle =0, \\
&\langle \eta^2\rangle =2.
 \end{split}
\end{equation} 
The Langevin equations \eqref{Lang eq xy} generate a \textit{real} probability distribution  function $P(x,y)$ over the complex plane,
\begin{equation}
Z = \int dxdy\, P(x,y) \ ,
\end{equation}
which is the solution of the Fokker-Plank equation associated to \eqref{Lang eq xy}. 

First of all we should note how the classical (i.e.\ without noise $\eta$) CL equation  \eqref{Lang eq z} is the complex conjugate of the thimble equation \eqref{eq_thimble_z}, i.e. the imaginary part has opposite sign. That leads to some important differences regarding stability of some trajectories. In particular, thimbles that flow to infinity in the imaginary direction coincide with run-away solutions of the classical CL equations. We saw (Cap.\ref{cap:Langevin}), however, that those kind of trajectories can always be avoided in the stochastic process (i.e.\ with noise $\eta$) by an appropriate choice of the CL integration stepsize. 

 Another obvious difference is that CL dynamics takes place in the whole complexified manifold, in this case the complex plane, doubling the degrees of freedom of the original problem. The Lefschetz thimbles, on the other hand,  are a complex sub-manifold of the complexified domain, but they have the same dimension of the original domain of integration. 
 
Those two points combined together also imply that the critical points, where $\partial_z S(\textbf{z}) \vert_{\textbf{z}=\textbf{z}_k}=0 $, have different properties in the two approaches. From the point of view of the Lefschetz thimble, a non-degenerate critical point \textit{has} to sit on a saddle point because, from \eqref{S along thimble}, it has to be a maximum for the stable thimble and a minimum for the unstable one.

i.e. the Hessian matrix  in these points $\partial^2_z S(z)\vert_{\textbf{z}=\textbf{z}_k^*}$ can be diagonalised and its eigenvalues are half positive and half negative. This property is essential to allow  both the stable and the unstable thimble to be associated with these points. 

For complex Langevin dynamics, instead, each critical point is either a relative maximum or a relative minimum. That means that the drift is either attractive or repulsive.
\begin{figure}[!t]
\includegraphics[scale=0.6, angle =0]{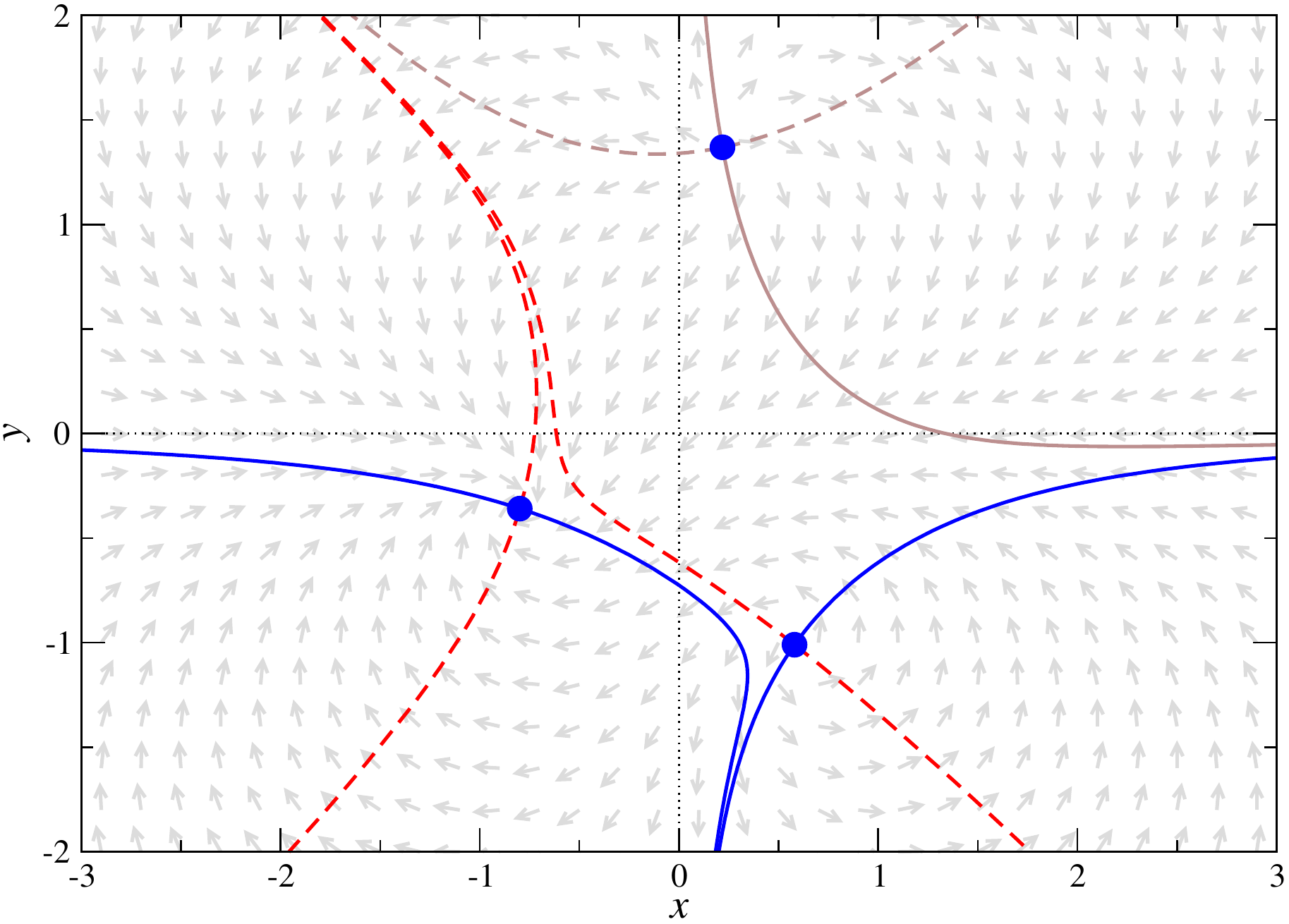}
\vspace{0cm}
\caption{Thimbles and Langevin flow in the quartic model with $\sigma=1$ and $h=1+i$: the blue circles denote the fixed points, the (normalised) arrows the classical Langevin drift, and full (dashed) lines the stable (unstable) thimbles. The two blue thimbles contribute. The third fixed point does not contribute.}\label{fig:quartic1}
\end{figure}
In Fig.\ref{fig:quartic1} we show both the Lefschetz thimbles (lines) and CL classical drift (grey arrows), while the blue dots are the three critical points
\begin{equation}
\begin{split}
&z_k = e^{2\pi i k/3}D - e^{-2\pi ik/3} \frac{\sigma}{3D},  \quad\quad\quad k=0,1,2, \\
&D = \left( -\frac{h}{2}+\frac{h}{2}\sqrt{1 + \frac{4\sigma^3}{27h^2}}\right)^{1/3}.
\end{split}
\end{equation}

From the Lefschetz thimble point of view each of the three points have one stable an one unstable direction. For CL, on the other hand, only the first one is attractive while the other two are repulsive (as one can see in Fig.\ref{fig:quartic1}).

Here we can see that the two thimble contributing, i.e.\ the ones whose connected unstable thimbles (red dashed lines) intersect the real axes, are clearly a deformation of the $\mathbb{R}$ contour with the limit at $x\rightarrow \pm \infty$ fixed. That fact that the two thimbles go to infinity in the $y$ direction should not worry since in that region of the complex plane the integrand vanishes. One can see  this by noticing that for large $z=r e^{i\theta}$ the dominant contribution to Re$S$ comes from the $z^4$ term. In particular,  Re$S$ is positive in the regions $\cos(4 \theta) > 0$, i.e. where $-\frac{\pi}{8}+ k \frac{\pi}{2}<\theta<\frac{\pi}{8}+ k \frac{\pi}{2}$, which, for $k=3$ that is exactly the region where the two thimbles go to infinity. This is general, i.e. the stable thimbles always in the region where Re$S \rightarrow + \infty$, otherwise the integral on it would not be finite, while the unstable ones always in the regions where Re$S\rightarrow - \infty$ . 

\begin{figure}[!t]
\begin{center}
\includegraphics[scale=0.5, angle =0]{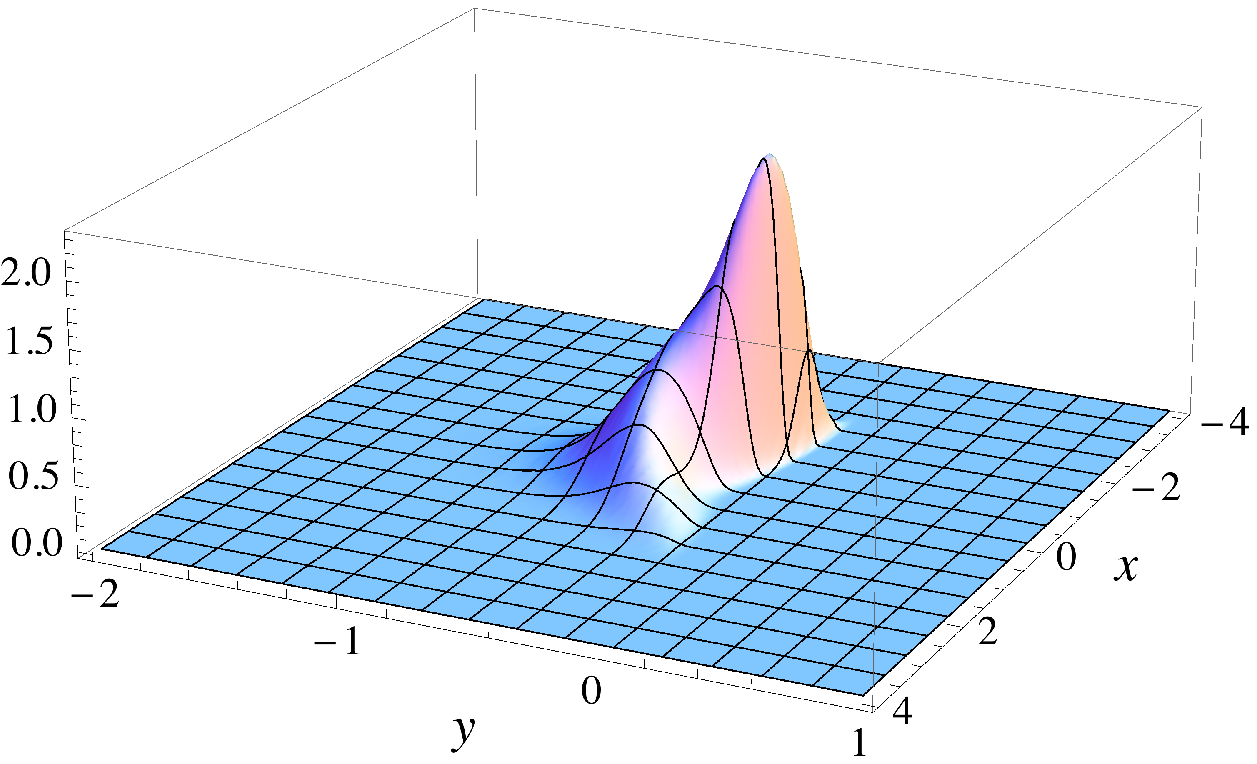}
\includegraphics[scale=0.38, angle =0]{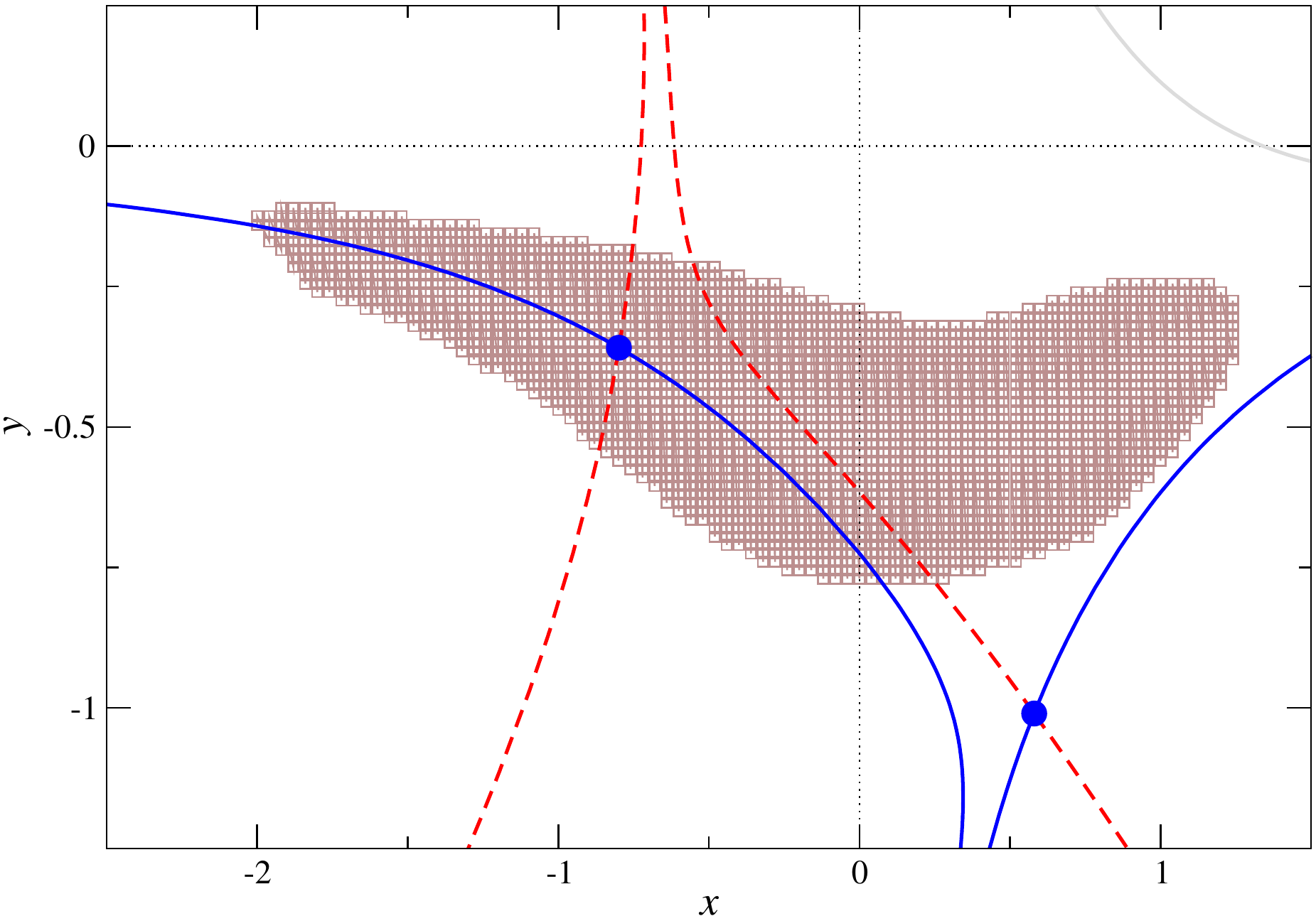}
\end{center}
\caption{Quartic model, with $\sigma=1$ and $h=1+i$. Histogram collected during a complex Langevin simulation (left) and a comparison with the thimbles (right). 
}
 \label{fig:quartic2}
\end{figure}
For what concerns CL, this model has been studied in detail in \cite{Aarts:2013uza} in the case of $\sigma \in \mathbb{C}$ and $h=0$. For the purpose of this section we just have to notice that two of the three critical points are repulsive, so that we expect CL dynamics to avoid the first two and concentrate somewhere near the third one. Furthermore, as mentioned earlier, thimbles that flows to $\pm \infty$ in the imaginary direction, correspond to runaway solution of the classical CL drift due to the complex conjugate equation of motion in the two models. Those trajectories, however, are avoided when the process is made stochastic. For the same reason, when the thimbles are parallel to the real axis, their attractive flow coincide with the CL one. 

We have now all the instruments to understand Fig.\eqref{fig:quartic2}. It shows Lefschetz thimbles in the complex plane together with the history of CL dynamics. We can clearly see that CL tends to follow the thimbles when they flow mainly in the real direction, while it strongly departs from them when they start to pick a relevant imaginary component. We can also observe the expected cluster of CL dynamics around the only attractive point (see histogram in the left part of Fig.\eqref{fig:quartic2}) while, of course, the repulsive ones are avoided.

We have verified that integration along the two contributing thimbles yealds to the right result. In particular $Z_1= 1.744 + i0.461$  and $Z_2 = 0.021 + i0.426$, which leads to the correct sum $Z = Z_1 + Z_2 = 1.765 + i0.887$.  For a numerical comparison with complex Langevin see the table (4.1).

We can conclude this section saying that knowing the Lefschetz thimbles structure of a complexified theory, at least in simple cases, gives very precise information about the CL dynamics distribution in the complex plane. The two methods, however, are far away from being in a one to one relation especially in the presence of a repulsive fixed point.

In the following we will extend this comparison to simple gauge models.

\vspace{2cm}
\section{U(1) model with determinant}
It is interesting to study a simple one-link abelian model, with link $U=e^{ix}$, with a QCD-like determinant in the action. This model has been studied in the paper \cite{Aarts:2008rr} specifically for the complex Langevin dynamics part. Here, instead, our goal is the comparison between CL dynamics and Lefschetz thimbles. The partition function of this theory is
\begin{equation}
 Z = \int_{\rm U(1)} dU \, e^{-S_B}  \det M
 = \int_{-\pi}^{\pi}\frac{dx}{2\pi}\, e^{\beta\cos x} \left[ 1+\kappa \cos(x-i\mu)\right],
\end{equation}
where $\beta$ is taken real and positive like the gauge coupling and the complex weight is introduced by the determinant $[\det M(\mu)]^* = \det M(-\mu^*)$, via the chemical potential $\mu$ like in QCD. We observe, already at $\mu=0$, that when $\kappa<1$ the weight is real and positive, while for $\kappa>1$ there is already a sign problem. Furthermore, after the exponentiation of $\det M$, the action we are left with is not holomorphic any more,
\begin{equation}\label{action U1}
S(z)  = -\beta\cos z - \ln \left[  1+\kappa\cos(z-i\mu)\right],
\end{equation}
which is reflected in the drift $\partial_z S(z)$ having poles where the determinant $1+\kappa\cos(z-i\mu)$ is zero :
\begin{equation}
\partial_z S(z)  = \beta\sin z + \frac{\kappa\sin(z-i\mu) }{ 1+\kappa\cos(z-i\mu)}.
\end{equation}
Violation of holomorphicity needs to be treated very carefully both in complex Langevin dynamics and in the Lefschetz thimble method. In the first case it invalidates the formal proof of convergence and it might lead to convergence to the wrong results \cite{Aarts:2010gr,Mollgaard:2013qra}. In the thimbles case it might compromise the deformation of the original contour of integration. However, as we shall see, the thimbles usually ends on the poles so that the integral over the contour is not affected by it. Moreover, in the case of multi-branched functions, one thimble is not expected to trespass in a branch different than the one where its critical point lies, as to do so it would need to cross a singularity.

\begin{figure}[!p]
\begin{center}
\includegraphics[scale=0.6, angle =0]{Figures/plot_full_b1_k0.25_m2-th-v3.eps}
\includegraphics[scale=0.6, angle =0]{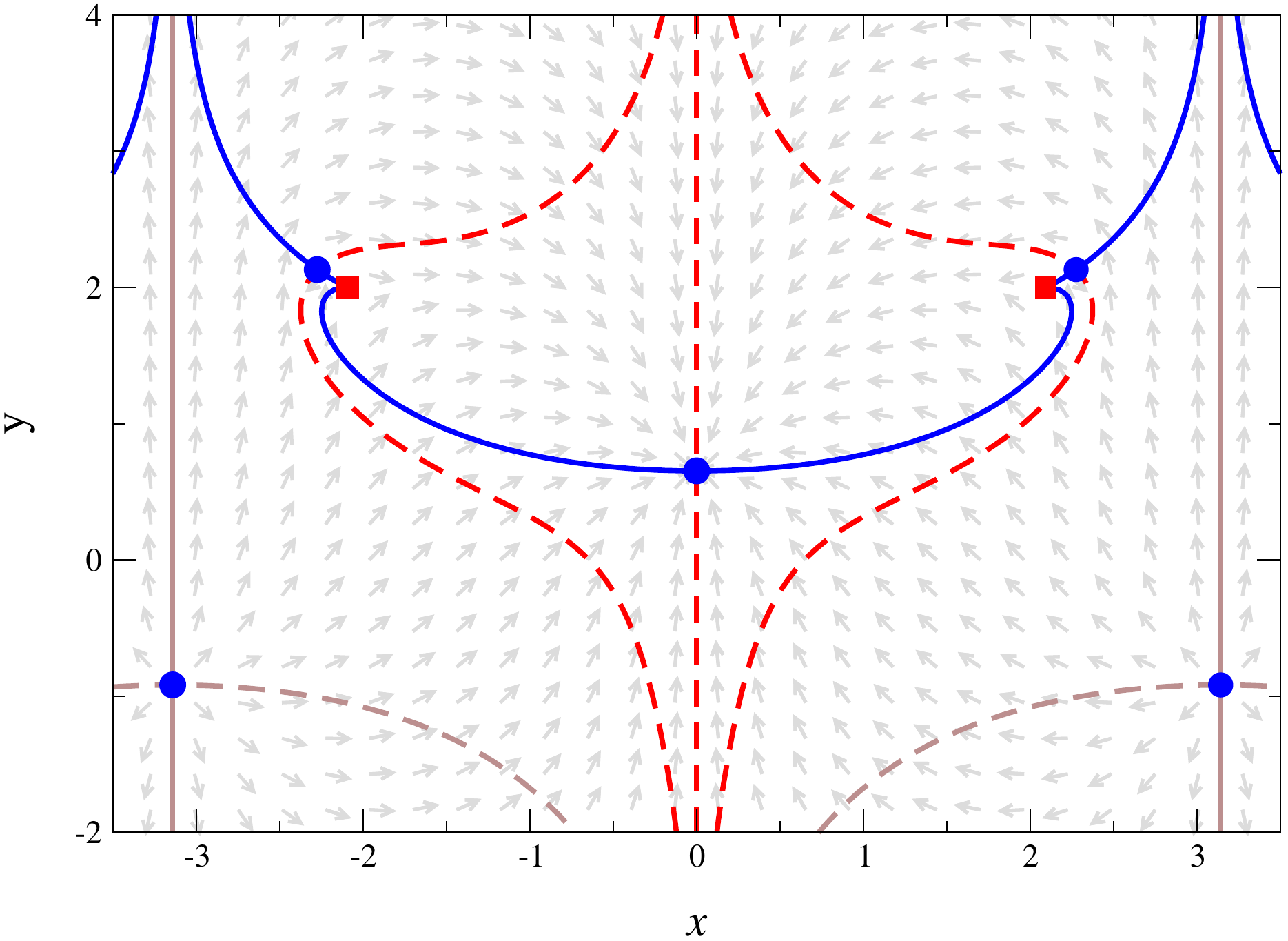}
\end{center}
 \caption{Thimbles and Langevin flow in the U(1) one-link model  with $\beta=1,\mu=2$ and $\kappa=1/2$ (up) and $\kappa=2$ (down): the blue circles indicate the fixed points,  the (normalised) arrows the classical Langevin drift, and full (dashed) lines the stable (unstable) thimbles.  Note that the blue lines indicate the contributing stable thimbles, with the red associated unstable ones, while the light brown lines indicate thimbles that do not contribute. The squares indicate where the flow diverges, Im$ S$ jumps, and the direction of the flow along the thimble changes sign. Only the blue thimble(s) contribute. 
}
 \label{fig:flow_full_4}
\end{figure}

Nevertheless, it is very important to test those two methods, and in general every method that aims to solve the sign problem, in the case where the action of the theory is non-holomorphic, since this is the case for full QCD. In this sense, this simple model is a very useful toy model since it can be solved analytically. In fact, the partition function and every power of the trace of the link $\langle \cos^n(x) \rangle$ are linear combination of modified Bessel functions
\begin{equation}\label{exact U1}
\begin{split}
&Z = I_0(\beta) + \kappa I_1(\beta)\cosh\mu, \\
&\langle \cos(x)\rangle=\dfrac{1}{Z} \left[ I_1(\beta)+ \dfrac{1}{2} \kappa \cosh\mu \left( I_2(\beta) + I_0(\beta)\right) \right].
\end{split}
\end{equation}
The goal will then be confronting our numerical results with the exact ones.

The critical points of the action are determined by
\begin{equation}\label{d/d_z S=0}
\partial_z S(z)=0
\end{equation}
and the singular points by
\begin{equation}
1+\kappa\cos(z-i\mu)=0,
\end{equation}
or explicitly
\begin{equation}\label{d/d_z S=inf}
\left\lbrace
\begin{split}
&\cos(x) \cosh(y-\mu) =-1/\kappa \\
&\sin(x) \sinh(y-\mu)=0 \ .
\end{split}
\right.
\end{equation}

Although the solutions to \eqref{d/d_z S=0} are quite complicated to solve exactly, one can still see that for $\kappa\leq 1$ the stationary points have to be at $x_c=0,=\pm \pi$, while for $\kappa>1$ solutions at different $x_c$ are possible. The same happens for the singular points \eqref{d/d_z S=inf}. Here we can see explicitly that for $\kappa\leq 1$ only solutions with $x_s=\pm \pi$ and $\cosh(y_s-\mu) =1/\kappa$ are allowed. Instead, when $\kappa>1$, the solutions $y_s=\mu$  and  $\cos(x_s)=1/\kappa$ become acceptable as well. The situation is captured in Fig.\ref{fig:flow_full_4}, where fixed points (singular point) are indicated with blue circles (red squares).

In this case, it is possible to gain some analytical insight about the thimbles. Again we need do separate the case of $\kappa \leq 1$ and $\kappa>1$.

\
\begin{itemize}
\item $\kappa\leq 1$   :\\
this is the case where critical points are only at $x_c=0,=\pm \pi$ and poles at $x_c=\pm \pi$. We will see that we one can tell which are the relevant thimbles just by looking at the action. Let us start with the critical point in $x_c=0$. Here we found the action \eqref{action U1} along the imaginary direction
\begin{equation}
S(0+iy)=-\left[ \beta \cosh(y) + \ln\left( 1+\kappa \cosh(y-\mu)\right) \right]
\end{equation}
to be completely real and unbounded from below as $y\rightarrow \pm \infty$. This means that Im$S(0+iy)=0$ and the integral is divergent, i.e. the $y$ axis is the unstable thimble. Since the $y$ axis definitely crosses the $x$ axis, we expect the stable thimble associated with $x_c=0$ to be relevant for the integral.

Similar is the case of the thimble associated to $x_c=\pm \pi$. The action here reads
\begin{equation}
S(\pm \pi+iy)= \beta \cosh(y) - \ln\left( 1-k \cosh(y-\mu)\right) .
\end{equation}
Since we are in the case of $k\leq 1$, Im$S(\pm \pi+iy)$ is again constant along the $y$ axis and its value only depends on $y$. When $\cosh(y-\mu)<1/\kappa$, in fact, Im$S=0$, otherwise  Im$S=\pm \pi$. The interesting fact here is that the value of Im$S$ jumps in correspondence of the singularities at $x_s=\pm \pi$ and $\cosh(y_s-\mu) =1/\kappa$.

 About the contribution of the thimble : we can see that in the regions $ x=\pm \infty, y \rightarrow \pm \infty$, the action $S\rightarrow +\infty$  so that the integral is convergent there and the thimbles are the stable ones. However, because they are exactly parallel to the $y$ axis, they cannot be a deformation of the original real axis, i.e. they cannot contribute to the integral. 
 
\vspace{0.5cm}
\item $\kappa > 1$   :\\
this is the case where some of the repulsive fixed points move away from $x=\pm\pi$, while also the singular drift is no longer at $x=\pm\pi$, but instead at $y=\mu$ and $x=x_s$ such that $\cos x_s = -1/\kappa$. Here the stable thimbles that flow to $y \rightarrow +\infty$ are not parallel to the $y$ axis any more. Furthermore, they manage to connect with the stable thimble connected to $x_c=0$ in the sense that they both end at the singularity, where the imaginary part of the action has a jump. In this way they belong to the deformation of the real axis, so that now they will need to be considered for the integral. Those thimbles, though, are too complicated to be expressed in an analytic form.
\end{itemize} 

This is illustrated in Fig.\ref{fig:flow_full_4}. In both cases the results, obtained by summing the integral over all the thimbles that contribute, agree with the exact result \eqref{exact U1}, provided both the residual phase factor and the global phase phase factor are correctly incorporated.

For what concerns complex Langevin dynamics the critical point in $x_c=0$ ia always attractive, while the others are repulsive. We can try to guess the CL scatter plot from the thimble structure, the same exercise we did for the quartic model.

\begin{figure}[!p]
\begin{center}
\includegraphics[scale=0.55, angle =0]{Figures/plot_full_b1_k0.25_m2-th-v4.eps}
\includegraphics[scale=0.55, angle =0]{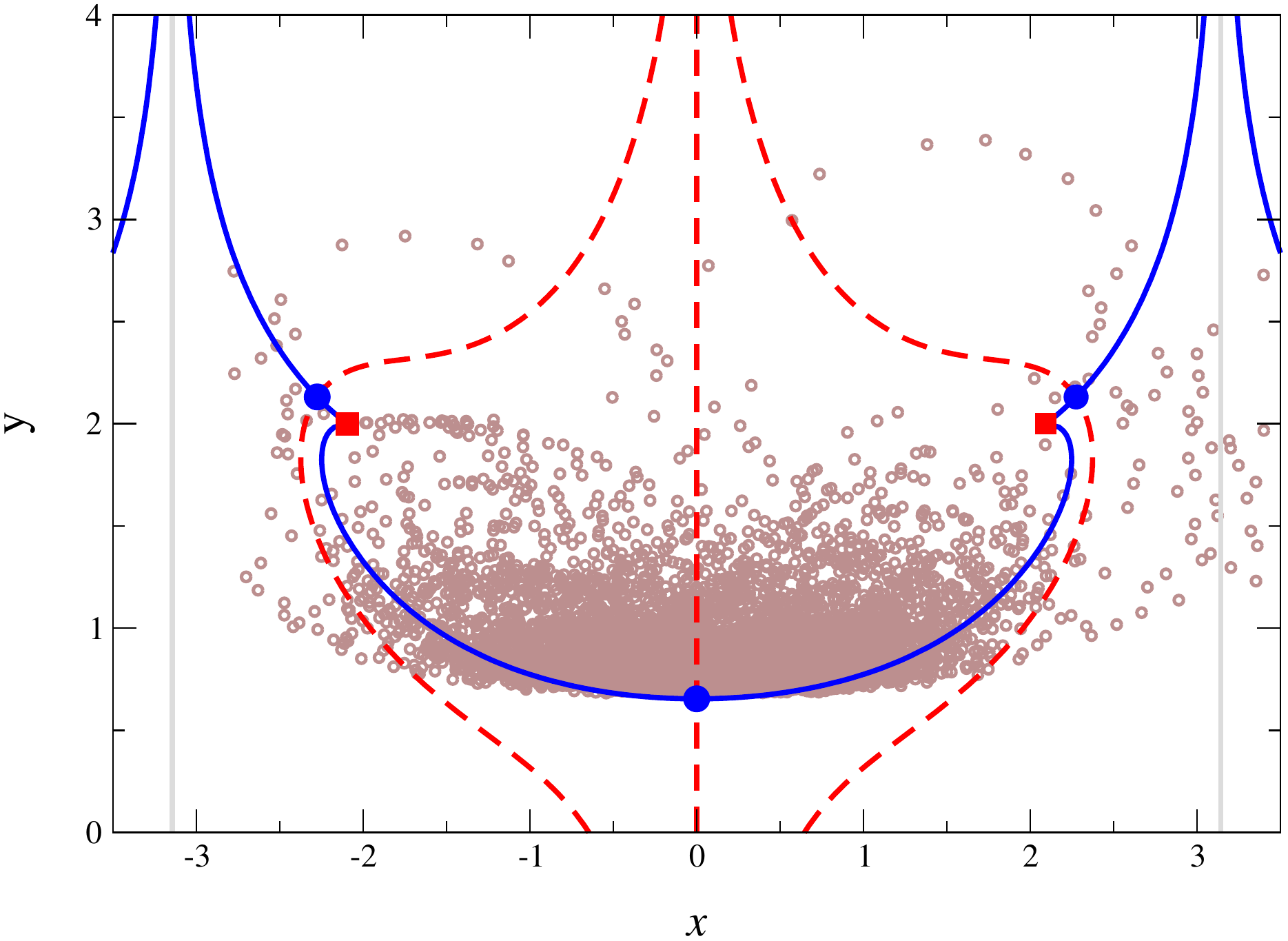}
\end{center}
 \caption{As in the previous plot, with scatter data from a complex Langevin simulation added.
}
 \label{fig:flow_full_5}
\end{figure}

\begin{itemize}
\item $\kappa\leq 1$   :\\
close the critical point at $x_c=0$ the thimble is mostly parallel to the $x$ axis, so CL classical drift has the same direction of the thimble flow. Furthermore, the latter approaches the critical point from above, that means the imaginary component of CL drift, i.e. the complex conjugation of the thimble flow equations, will point towards the positive $y$ direction. Then, we expect CL scatter plot to be clustered following the thimble around the attractive fixed point at $x_c=0$ and, more precisely, to be mostly above it. \\
The other stable thimbles at $x=\pm \pi$ are parallel to the $y$ axes, which corresponds to run-away solutions for CL. We expect then the latter to avoid those trajectories.\\
In Fig.\ref{fig:flow_full_5} (up) we see that, as expected, the CL dynamics stays over the main thimble around the fixed point at $x_c=0$. Moreover the compactness of its distribution and the fact that it stays away from the poles (red squares) leads to the correct results (see Table (4.1)) at least for low moments $\langle \cos^n(x)\rangle$, with small $n$  \cite{Aarts:2008rr} .

\vspace{0.5cm}
\item $\kappa > 1$   :\\
in this case the thimble connected with $x_c=0$ is much less flat in the $y$ direction. Also two of the stable thimbles, previously localised at $x=\pm \pi$,  move towards the centre and and join the main thimble. Altogether this necessarily leads to a much stronger classical complex Langevin drift in the $y$ direction. The resulting scattering plot (Fig.\ref{fig:flow_full_5} down) is, therefore, wider spread into the complex plane. Moreover, also the poles are closer to the attractive fixed point and, this time, CL dynamics manages to reach them. \\
In this case CL dynamics is found not to be able to reproduce the correct results (see Table (4.1)) even for the low momenta. Since the poles are immersed in the distribution the arguments in \cite{Aarts:2010gr} and \cite{Aarts:2009uq}, relying in holomorphicity, break down.
\end{itemize} 

To summarise, we find that in the presence of a determinant the flow has singular points. For Langevin dynamics, this leads to a breakdown of the formal justification and possible wrong results in practice. For the Lefschetz approach, we find that thimbles may end at singular points and the imaginary part of the action jumps by a constant.
Hence, we observed for the first time in\cite{Aarts:2014nxa}, if there is more than one contributing thimble, they connect either at $|z|\to\infty$ or at a singularity.

\clearpage
\section{SU(2) 1-link model}
We now extend the analysis to the case of a SU(2) one link model,
\begin{equation}\label{Z SU2}
Z = \int dU \exp\left[\frac{\beta}{2}\Tr U\right],\ \ \ \ \beta \in \mathbb{C}, \ \ \ \ \ U \in \text{SU(2)}.
\end{equation}
This time the sign problem is introduced by a complex $\beta$ instead of a determinant. This model is again analytically solvable due to the fact that the action only depends on the trace of the link $U$. It is, therefore, invariant under any change of basis of $ U \rightarrow R U R^{-1} $ and this allow to diagonalize  the link making its dependence from an angle $ \phi $ explicit
\begin{equation}
\begin{split}
&U(\phi,\overrightarrow{n})\ = \ e^{i \omega_a \sigma_a }\ =\ \cos(\phi)\Id +i \sin(\phi) n_a \sigma_a ,\\
&\Tr U=2 \cos \phi,
\end{split}
\end{equation}
where $\sigma_a$ are the Pauli matrices, $\phi=\sqrt{\omega_a \omega_a}$ and $n_a=\omega_a/\phi$. For convenience, we will refer to this formulation as the 'angle representation'.  
In this case one has also to consider the change in the Haar measure for the partition function 
\begin{equation}
\int dU \rightarrow  \int_{-\pi}^{\pi} d\phi \int \dfrac{d\Omega(\overrightarrow{n})}{4\pi} \ \sin^2(\phi) ,
\end{equation}
where $ \int \dfrac{d\Omega(\overrightarrow{n})}{4\pi} $ is the integration of the uniform angular measure on the unit sphere, in our case it is equal to $1$ because we only have dependence from $ \phi$.  This formulation allows the model to be solved analytically in terms of modified Bessel functions of the first kind
\begin{equation}
\begin{split}
&Z=\dfrac{1}{2} \left[ I_0(\beta)- I_2(\beta)\right], \\
&\langle \Tr U\rangle=2 \langle \cos(x)\rangle =\dfrac{1}{Z} \dfrac{4}{\beta}I_2(\beta) .
\end{split}
\end{equation}
Diagonalizing the link $U$ corresponds to a complete gauge fixing in the sense that the degrees of freedom of the model are reduced from three to one. A detailed study of the effect of coordinate changing on complex Langevin dynamics, including this model, has been done in \cite{Aarts:2012ft}. There it is shown how the introduction of a Haar measure, dictated by the coordinate change, can help the convergence of the dynamics and, therefore, it is a useful tool associated with CL.

The other main way to control CL dynamics is gauge cooling (see cap.\ref{cap:Langevin}). The latter does not require an explicit gauge fixing and it is employed when the dynamics takes place in the full gauge group.

Although, obviously, in both cases the model is the same, from the perspective of CL dynamics they are completely different. In the first case the space to be explored is the strip :  $-\pi< $ Re$\phi<\pi$,  $-\infty< $ Im$\phi<\infty$ of the complex plane , while in the second case it is the complex group SL($2,\mathbb{C}$) . In the following we will compare both approaches with the Lefschetz thimble method. The latter can be computed exactly in the angle representation and directly compared with CL in the same representation. Then, it can be mapped into the $\Tr U$ plane to be confronted with the case in which CL dynamics takes place into the whole SL($2,\mathbb{C}$) group.

\subsection{Angle representation}
In this approach the partition function \eqref{Z SU2} is written 
\begin{equation}
Z = \int_{-\pi}^\pi \frac{dx}{2\pi}\, \sin^2x\, e^{\beta\cos x} =  \int_{-\pi}^\pi \frac{dx}{2\pi}\, e^{-S(x)} , 
\end{equation}
with
\begin{equation}
S(z) = -\beta\cos z -2\ln\sin z.
\end{equation}
Similar to the determinant for the U(1) model, the Haar measure introduces a logarithm in the action which makes it non-holomorphic. As usual we need to identify the critical points and the poles of the drift
\begin{equation}
-\partial_z S(z)  = -\beta \sin z +2\frac{\cos z}{\sin z}.
\end{equation}
Hence, the fixed points are
\begin{equation}
\cos z_c^\pm = -\frac{1}{\beta}\left(1\pm\sqrt{1+\beta^2}\right),
\end{equation}
while the poles satisfy
\begin{equation}
\sin x \cosh y + i \sinh y \cos x=0,
\end{equation}
which are the points $x_s=0,\pm \pi$ and $y_s=0$.

\begin{figure}[!t]
\begin{center}
\includegraphics[scale=0.35, angle =0]{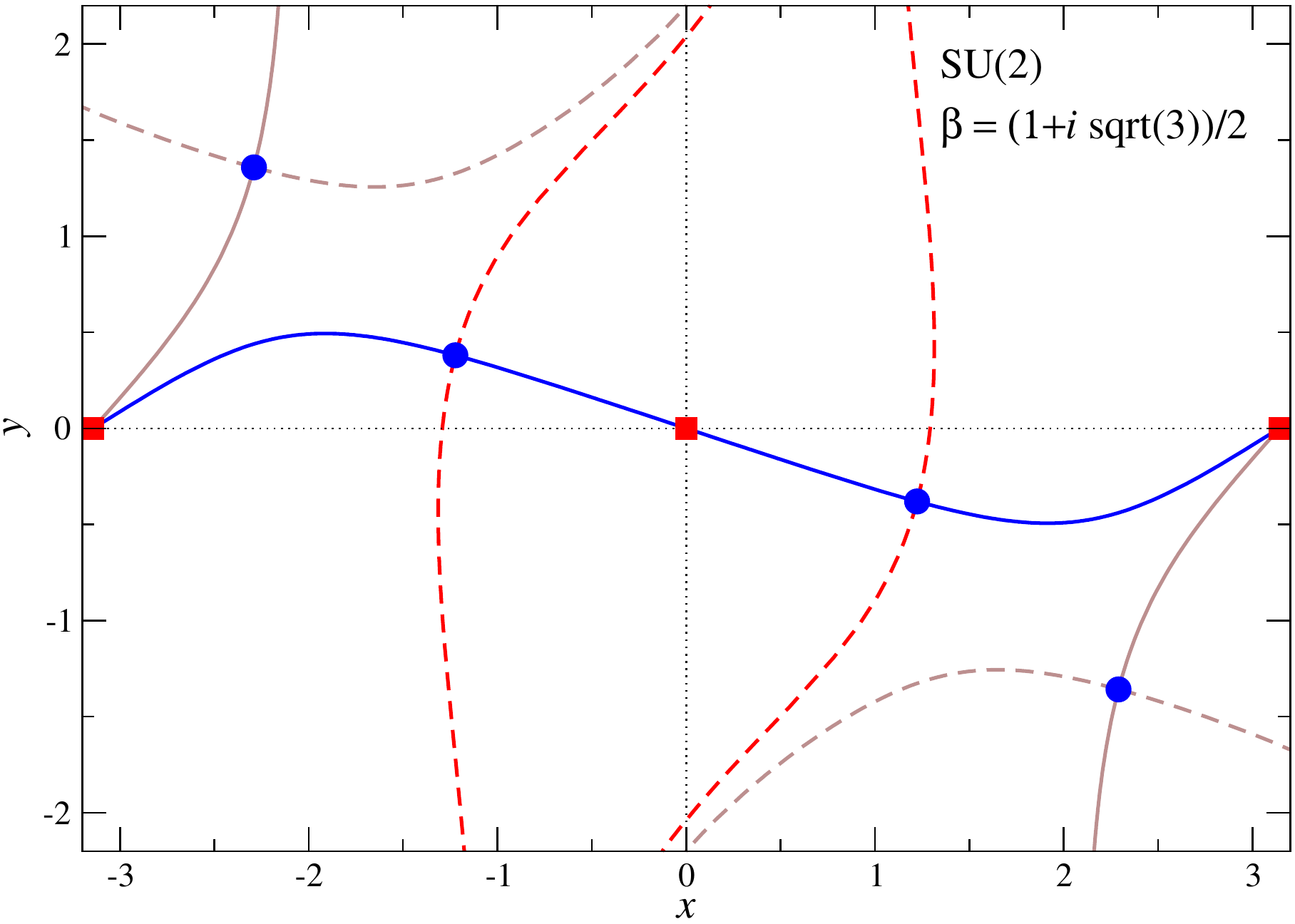}
\includegraphics[scale=0.35, angle =0]{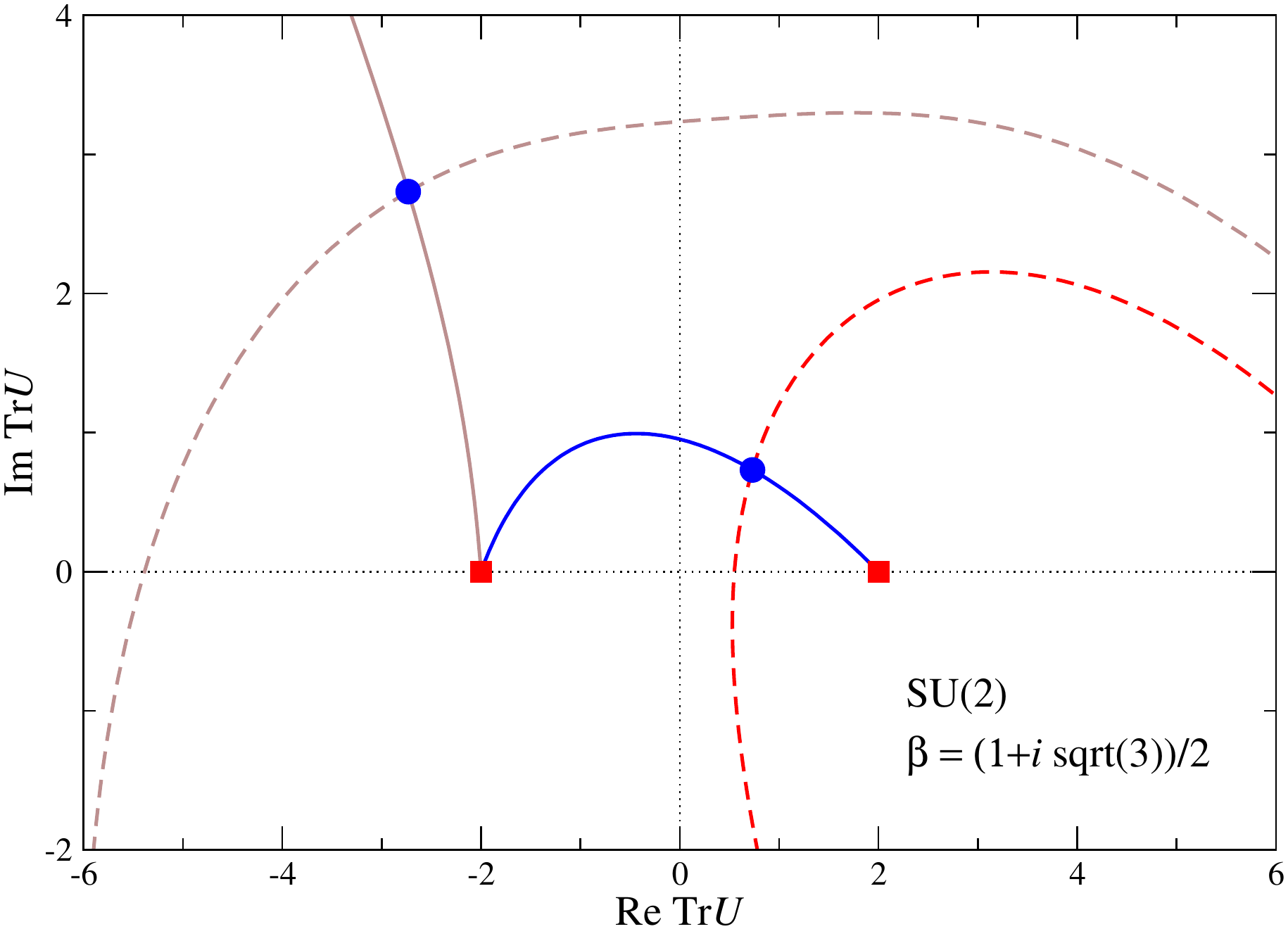}
\end{center}
\caption{
SU(2) one-link model, with $\beta=(1+i\sqrt{3})/2$: thimbles in the $xy$ plane (left) and the $\Tr U$ plane (right). 
Blue circles indicate fixed points, full (dashed) lines the stable (unstable) thimbles, and red squares the singular points. The blue thimble contributes.
}
 \label{fig:SU2-th}
\begin{center}
\includegraphics[scale=0.5, angle =0]{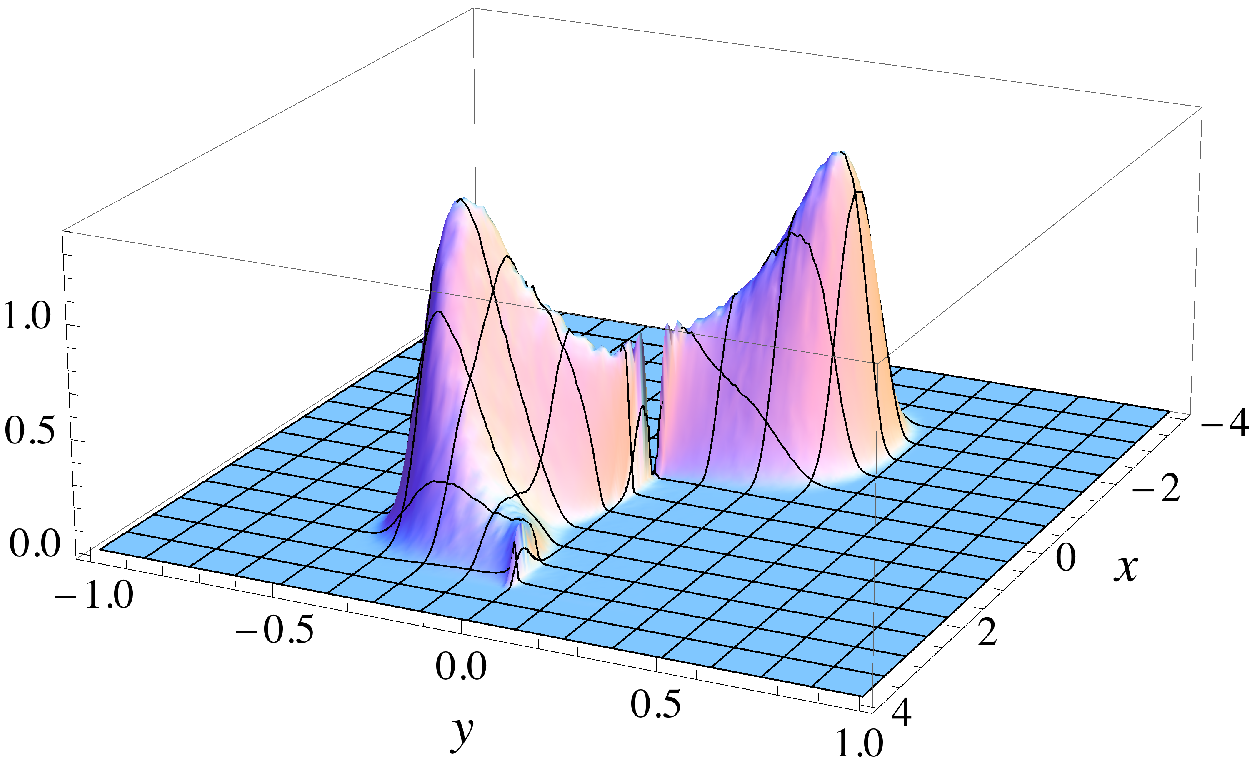}
\includegraphics[scale=0.5, angle =0]{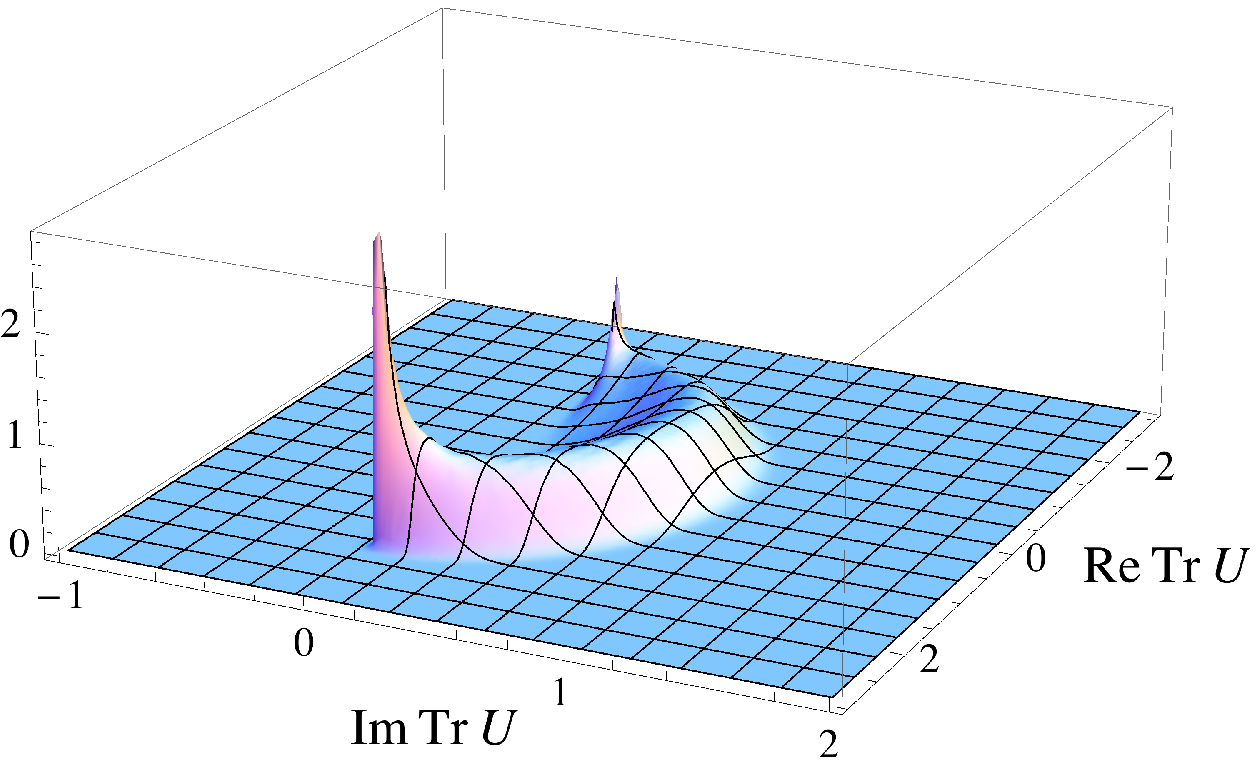}
\end{center}\label{fig:CL histogram}
\caption{
Histograms collected during a complex Langevin simulation in the gauge fixed formulation of the  SU(2) one-link model, with $\beta=(1+i\sqrt{3})/2$, in the $xy$ plane (left) and the $\Tr U$ plane (right).
}
 \label{fig:SU2-CL}
 \begin{center}
\includegraphics[scale=0.35, angle =0]{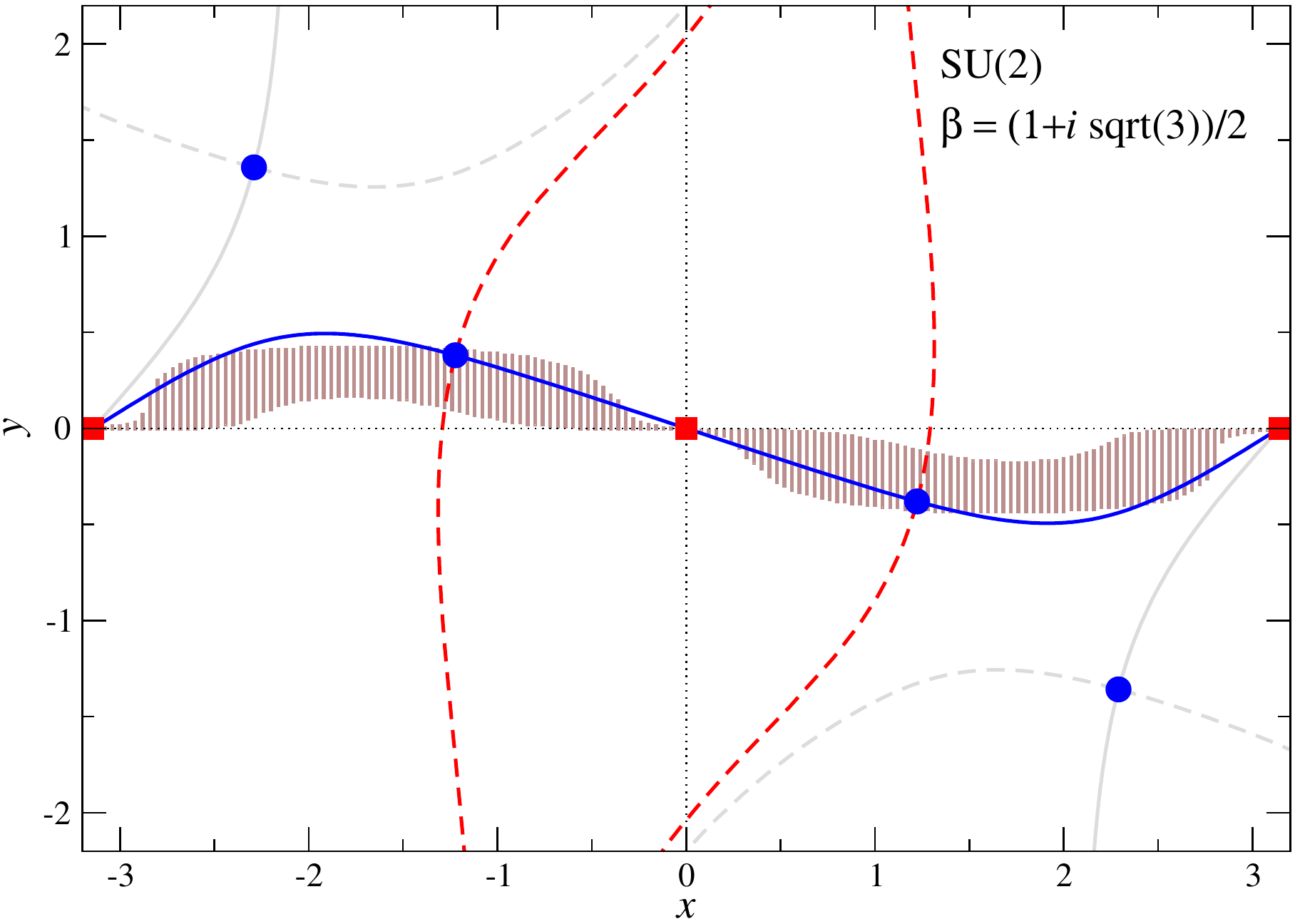}
\includegraphics[scale=0.35, angle =0]{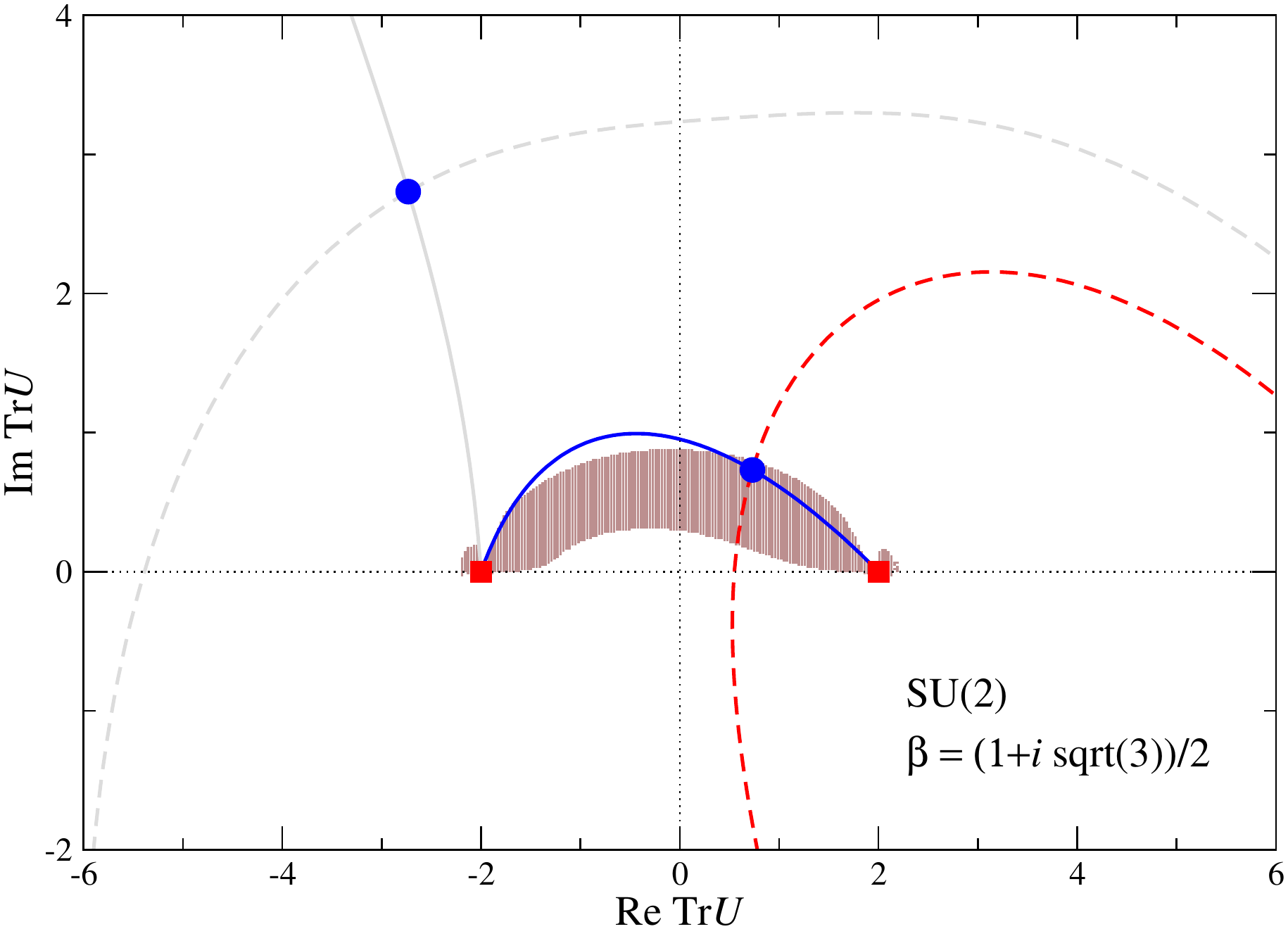}
\end{center}
\caption{
Comparison between the complex Langevin histograms and the thimbles, in the $xy$ plane (left) and the $\Tr U$ plane (right).
}
 \label{fig:SU2-th-CL}
\end{figure}
If we integrate numerically the thimble equations we obtain the picture in Fig.\ref{fig:SU2-th} (left). The plot respects the parity symmetry and, as expected, the stable thimbles contributing (solid blue lines) are the ones that form a deformation of the original contour $x \in \mathbb{R}, x \in (-\pi,\pi]$ . We can also map the results into the $\Tr U = 2\cos z$ plane, see Fig.\ref{fig:SU2-th} (right) to compare it with complex Langevin dynamics in the gauge formulation.
\clearpage
In Fig.\ref{fig:SU2-th-CL} we show a comparison between thimbles and the histogram of CL dynamics. We notice that CL stays very close to the thimble lines. Once more, we can appreciate how the complex conjugate drift leads CL dynamics to stay mostly below the thimble when this is concave, and mostly above when this is convex. Also they are connected with two repulsive fixed points, from the point of view of CL.

The two non-contributing thimble are ignored by CL. That is because they are far away from the two central attractive points and mostly oriented in the $y$ direction, i.e. they coincide with run-away  trajectories avoided by CL. 

We can clearly see how the distribution of CL dynamics into the complex plane is compact. Furthermore we can see from Fig.\ref{fig:CL histogram}(left) that the poles are avoided in the angle representation, while in the matrix representation for CL (right) the action has no poles to begin with. This, as expected, is reflected in the convergence to the right results (as we can see in Table (4.1)).

\subsection{Gauge dynamics with cooling}
As anticipated, we will study the same model but from a different approach for what concerns complex Langevin dynamics. This time we will not impose any gauge fixing and let CL free to explore the \SL2\ space. We will use, though, gauge cooling to control the dynamics. 

This approach is more expensive in terms of computing time, but has the advantage of being more easily generalised to the 4d case. Also if the gauge is not fixed, the Haar measure is not needed any more and the action is holomorphic.

Complex Langevin dynamics for gauge theories have been reviewed in Cap.\ref{cap:Langevin}. Here we will just briefly recall the evolution equation
\begin{equation}
U(t+\epsilon) = R(t) U(t), \ \ \ \ R = \exp \left[ i  \sigma_a \left(\epsilon K_a+\sqrt{\epsilon }\eta_a \right) \right] ,
\end{equation}
where $t$ is the (discretised) Langevin time, $K_a=-D_a S$ is the drift and $\sigma_a$ are the Pauli matrices. Application of \textit{gauge cooling} is necessary to control the width of the distribution in the complex direction. This process is orthogonal to the dynamics of the system and it can be implemented an arbitrary number of times between two consecutive CL updates. The single steps can be expressed as
\begin{equation}
U'(t) = \Omega_{gc}(\alpha) U(t) \Omega_{gc}^{-1}(\alpha),
\end{equation}
where parameter $\alpha$ regulates the intensity of the cooling. The number of steps is decided in such a way that the balance is optimal between effectiveness of gauge cooling and computer time usage.

\begin{figure}[!h]
\begin{center}
\includegraphics[scale=0.7, angle =0]{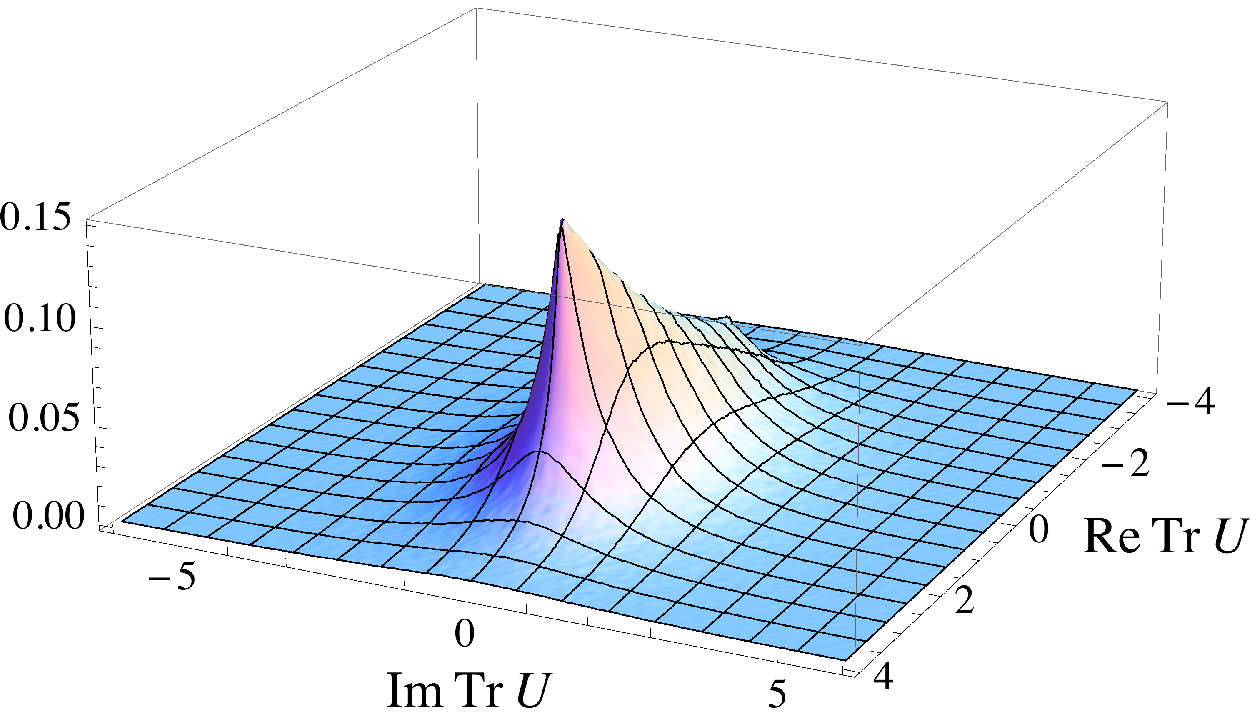}
\includegraphics[scale=0.5, angle =0]{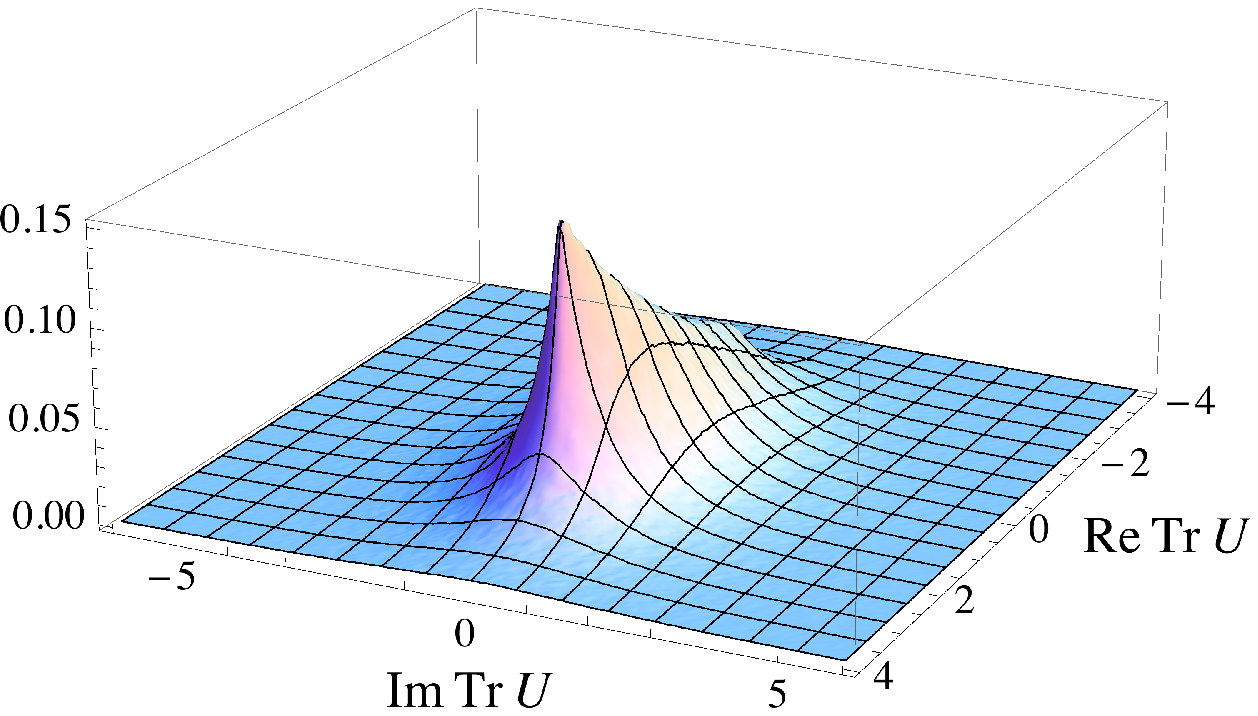}
\includegraphics[scale=0.5, angle =0]{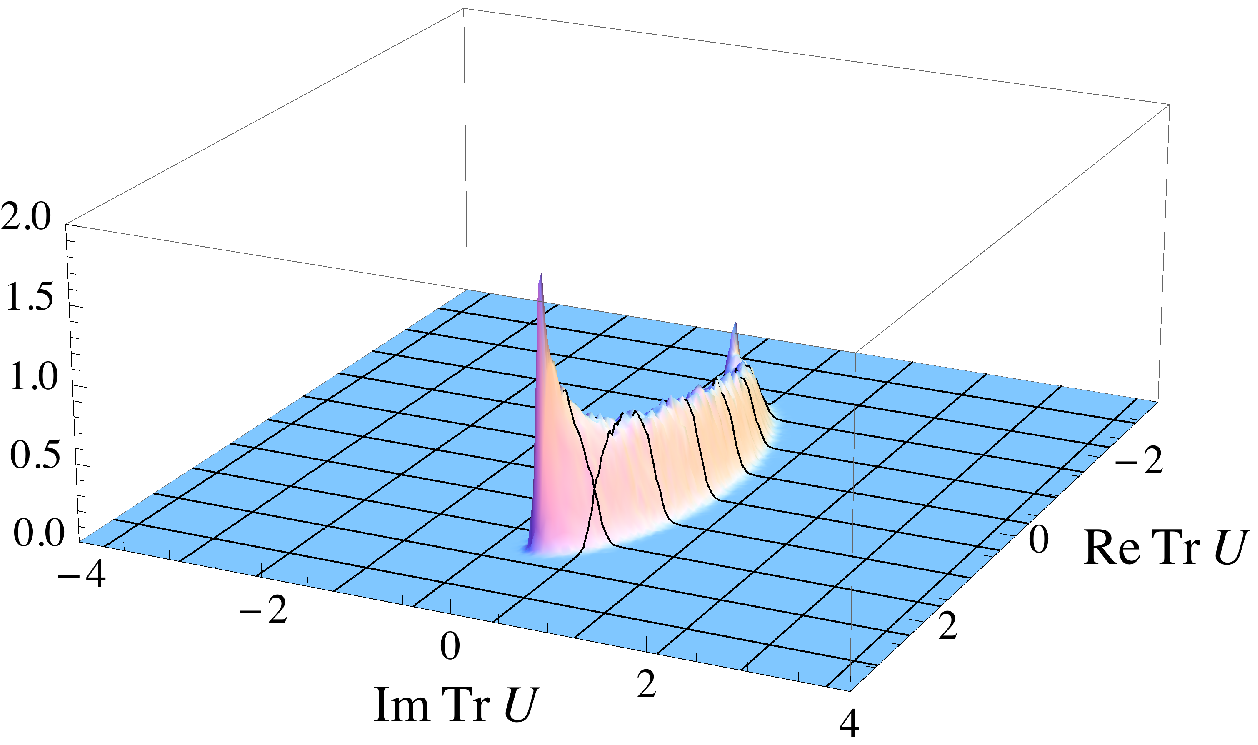}
\includegraphics[scale=0.7, angle =0]{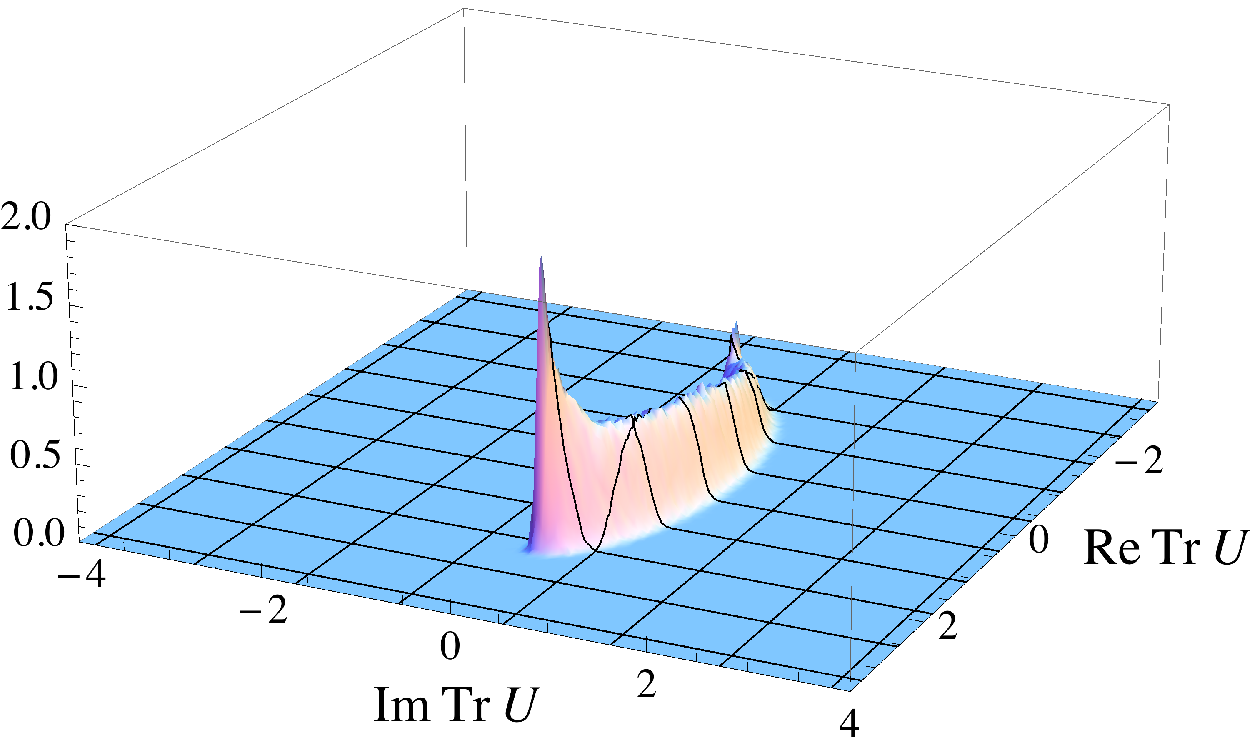}
\end{center}
\caption{
Histograms collected during a complex Langevin simulation in the matrix formulation of the SU(2) one-link model at $\beta=(1+i\sqrt{3})/2$, with gauge cooling, using 0, 1, 2, 4 (from top to bottom) gauge cooling steps.}
 \label{fig:SU2-CL-cooling}
\end{figure}

We can see in Fig.\ref{fig:SU2-CL-cooling} that increasing the number of cooling steps progressively improves the compactness of the distribution of $\Tr U$ in the complex plane. However one can clearly see a jump at 2 gauge cooling steps. That results in $\langle \Tr U \rangle_{CL}$ to converge to the right results (see Table (4.1)) for number of gauge cooling steps $\geq 2$.

\clearpage
Lastly we are going to analyse the case when $\beta^2=-1$ (and hence $\beta$ is purely imaginary). Here the fixed point  $z_d$, with $\cos z_d=-1/\beta=\pm i$, is degenerate in the sense that the Hessian $\vert \frac{\partial S}{\partial z^2} \vert_{z=z_d}=0$. The choice of imaginary $\beta$ is motivated by dynamics in real (Minkowskian) time, see e.g.\ Refs.\ \cite{Berges:2005yt,Berges:2006xc,Berges:2007nr} for complex Langevin studies. For degenerate fixed points the standard reasoning to justify the Lefschetz approach and construct the thimbles by numerical integration is not well defined. Here we give a brief analysis.

At the fixed point the action is real, with $S(z_d) = 1-\ln 2$.
We take $\beta=i$ and write
\be
\dfrac{1}{2}\Tr U = \cos z = u+i v.
\ee
Equating the imaginary part of the action,
\be
\text{Im} S = -u-\phi, \ \ \  \tan\phi = \frac{-2uv}{1-u^2+v^2},
\ee
to 0, then yields the thimbles, and we find
\be
v_\pm(u) = \frac{1}{\tan u}\left(u\pm\sqrt{u^2-(1-u^2)\tan^2u}\right),
\ee 
where  $0<u<1$ for $v_-(u)$ and $-1<u<0$ for $v_+(u)$. These two branches make up the stable thimble. 
The unstable thimble is given by $u=0$, for which the imaginary part of the action vanishes as well.

\begin{figure}[t]
\begin{center}
\includegraphics[scale=0.7, angle =0]{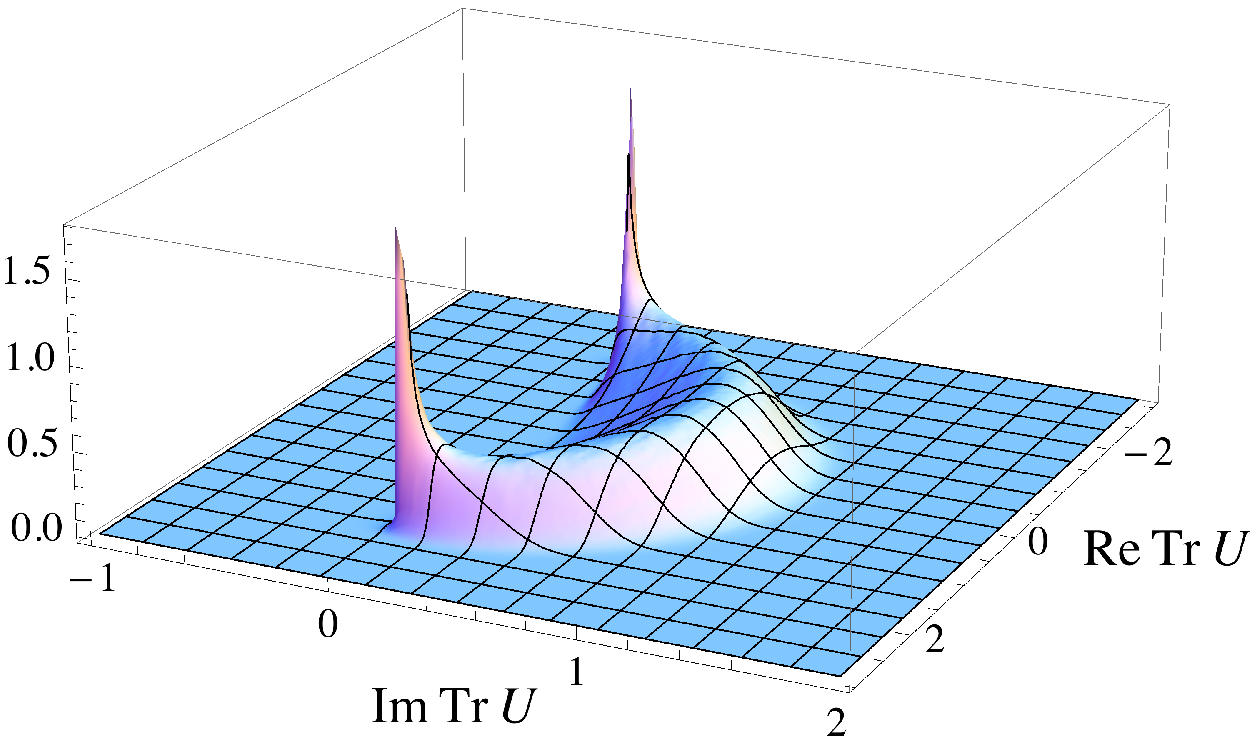}
\includegraphics[scale=0.7, angle =0]{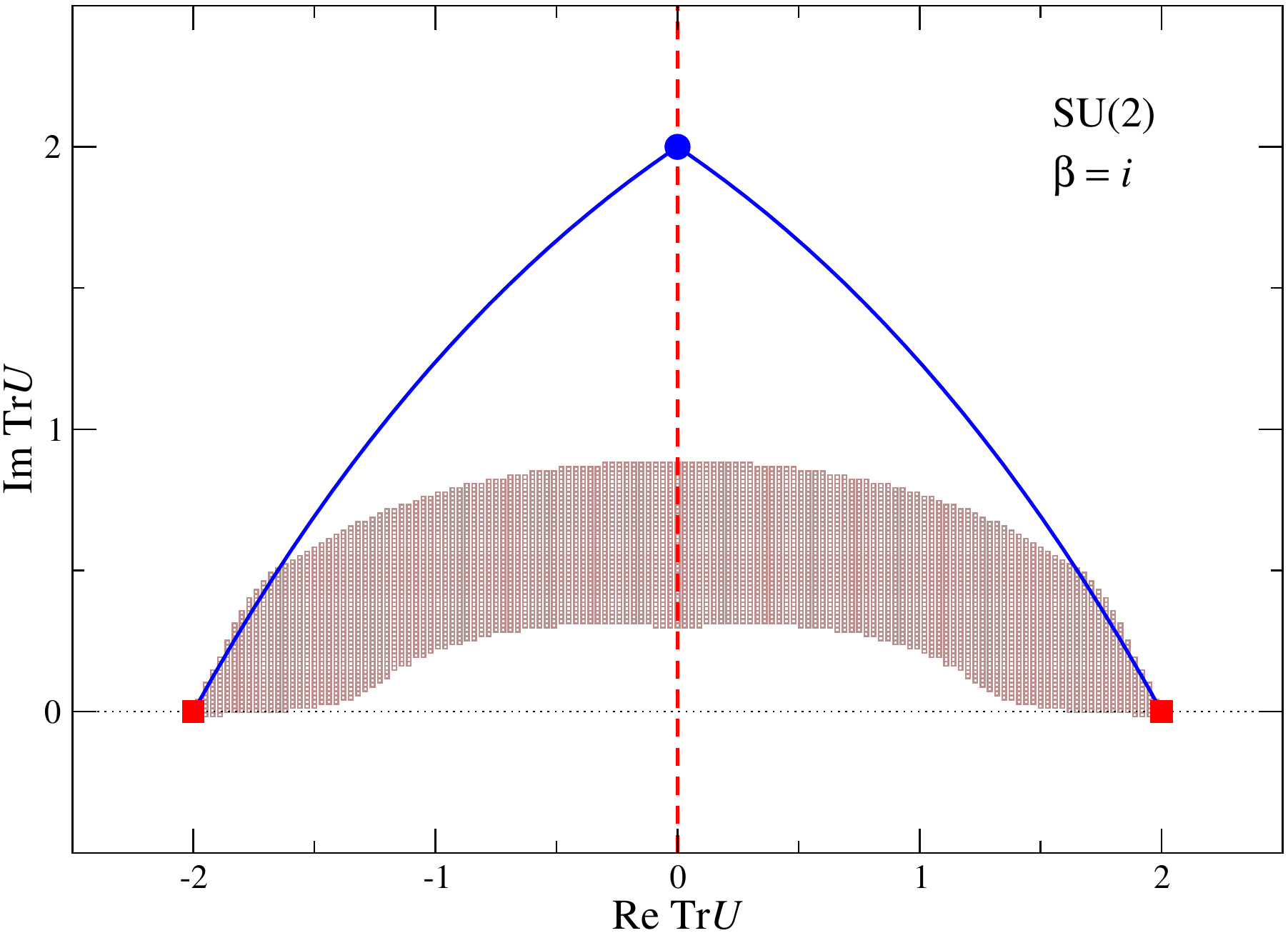}
\end{center}
\caption{
SU(2) one-link model at $\beta=i$. Histogram collected during a complex Langevin simulation in the  $\Tr U$ plane (left) and a comparison with the thimbles associated with the degenerate fixed point at $\Tr U=2i$ (right).
}
 \label{fig:SU2-th2-beta-i}
\end{figure}

The thimbles are shown in Fig.\ref{fig:SU2-th2-beta-i} (right), using the same colour coding as above. They are a deformation of the thimbles for general complex $\beta$, shown earlier, satisfying reflection symmetry in the Re$\Tr U=0$ axis. We have verified that integrating along the thimble, with the inclusion of the residual phase, gives the correct answer.
The Langevin histogram is shown in Fig.\ \ref{fig:SU2-th2-beta-i} (left). A similar histogram was obtained earlier in Ref.\ \cite{Berges:2007nr}. 
A comparison between the histogram and the thimbles is finally given in Fig.\ \ref{fig:SU2-th2-beta-i} (right). 
For this case we note that the distribution does not overlap substantially with the thimble and that again the distribution of the weight for both approaches is quite different.

\vspace{2cm}
In conclusion the aim of this chapter was to compare two of the most promising methods to deal with the sign problem in Lattice quantum field theory, i.e. complex Langevin dynamics and Lefschetz thimble. We examined several models that present a sign problem. Those were simple enough to be solved analytically but interesting enough to capture some important characteristics of more complicated gauge theories. We started with a quartic model to fix the ideas about Lefschetz thimbles and to set the parameters of comparison between the latter and complex Langevin dynamics. We proceeded by examining a one-link U(1) model with a determinant. Here the sign problem was introduced by chemical potential $\mu$, in analogy with full QCD. 

Lastly we looked at the case of one-link SU(2) model with complex $\beta$ parameter. This model could be approached in two different ways by CL. One way was to fix the gauge and study the dynamics in the lower dimensional parameter space. The other was to let CL free to explore all the complexified gauge degrees of freedom and control the dynamics with gauge cooling. Both approaches have been confronted with the thimbles method and found to converge to the right results.

About the comparison of the two methods, we found that the study of the Lefschetz thimbles is strictly connected with CL dynamics. The reason for this is that the two share the same critical points and their equation of motion are the complex conjugate of the other. Knowing the thimbles structures, therefore, allows one to have the exact information of the classical CL drift along the thimbles lines connecting the fixed points. 

The fixed point in the Lefschetz thimble formulation have always an attractive direction and a repulsive one. On the other hand, for CL they can be completely attractive or completely repulsive. However, knowing the thimbles flow near the fixed points allows to establish exactly what kind of fixed point they are in the CL formulation.

We also showed, when there are poles involved, that they are the endpoint of some thimbles while, for CL dynamics, they are quite dangerous and threaten the good convergence whenever they are immersed into the dynamics. 

Altogether we show that, starting from the Lefschetz thimble picture, one can make prediction on the CL dynamics with some accuracy.

  \begin{table}\label{Table: Lefsch}
  \hskip-1.5cm \begin{tabular}{ | l | l ||| l | l | l |}
  \hline
   Model & Observ. & Lefsc. thimbles & complex Lang. & Exact result \\ \hline\hline
  quartic mod.& $\langle z \rangle$ & -0.501475 - i 0.420789   & -0.4993(7) - i 0.4174(2) & -0.501475 - i 0.420789  \\ \hline 
  U(1) ($\kappa=0.5$)   & $\langle \cos z \rangle$  & 0.808709 & 0.814(5) - i 0.002(5) &  0.808709  \\ \hline
  U(1) ($\kappa=2$)      & $\langle \cos z \rangle$  & 1.05808 & 1.019(7) + i 0.021(9)&     1.05808     \\ \hline
  SU(2) (gauge fix.)        & $\langle \cos z \rangle$  & 0.271435 + i 0.43181  &  0.272(1) + i 0.4319(3) &  0.271435 + i 0.43181   \\ \hline
  SU(2) (gauge cool.)  & $\langle \cos z \rangle$  & \hspace{1.2cm} x  & 0.272(2)+ i 0.4315(5) &  0.271435 + i 0.43181     \\ \hline
  SU(2) ($\beta = i$)     & $\langle \cos z \rangle$  & i 0.5222229 & 0.001(1) + i 0.5222(4) & i 0.5222229 \\ \hline \hline
  \end{tabular}
  \caption{Comparison, between CL, Lefschetz thimble and exact result, of the numerical values for the models discussed in this chapter. The 'x' symbol, in the SU(2) (gauge cool.) line, means we didn't perform the Lefschetz thimble in the matrix representation, having the result already in the gauge fixed representation.}
  \end{table}

\begin{comment}

%% file: cap_DENSITY/Density_of_States.tex
\tableofcontents
\end{comment}

\externaldocument{cap_DENSITY/Density_of_States.tex}
\chapter{Density of States}
In the previous chapters we have discussed complex Langevin dynamics and Lefschetz thimbles as methods to deal with the sign problem in quantum field theory. In this last chapter we are going to review another promising approach to the same problem, i.e. the \textit{density of states} (DOS) approach. The idea is to construct the density of states of a given theory in the canonical, or grand-canonical, fashion. An efficient way has been developed in \cite{WangLandau:2001} for discrete systems, and then generalised to quantum field theories with continuous energy spectrum \cite{Azcoiti:2002vk,Fodor:2007vv,Langfeld:2012ah,Langfeld:2013xxa,
Langfeld:2014nta,Pellegrini:2014gha,Langfeld:2015qoa,Lucini:2014wga}. 
Standard Monte-Carlo methods are highly developed tools when it comes to measure expectation values of observables on theories with a real and positive Boltzmann weight $e^{-S}$. However, they do not allow direct computation of some extensive quantities such as the free energy or the partition function. Furthermore, as stressed many times before, they become inefficient when the theory is affected by a sign problem.

On the other hand, the density of state method is naturally suitable for calculating extensive quantities. Moreover, as we shall see in the following, it automatically reduces the sign problem of the full quantum field theory to one in a one-dimensional oscillating integral.

\vspace{2cm}
\section{Description of the method}\label{sec:Description of the method}
Let us consider the partition function
\begin{equation}
Z(\beta)=\int \mathcal{D} \phi \; \exp(- \beta S[\phi]),
\end{equation}
where $\phi$ is a generic field. As always, we can identify the Euclidean action $S[\phi]$ with the 'energy' of the field configuration. In analogy with the canonical ensemble in Statistical Mechanics one can define the number of states having energy $E$
\begin{equation}
\rho(E)=\int [\mathcal{D} \phi]\; \delta\left( S[\phi]-E \right),
\end{equation}
where $\delta()$ is the Dirac delta function.  The partition function can then be expressed as the integral of the density over the energy, weighted with the Boltzmann factor :
\begin{equation}\label{Z_beta}
 Z(\beta)=\int dE\; \rho(E) e^{- \beta E}.
 \end{equation} 
Any expectation value that can be expressed as an explicit function of the energy, can also be computed 
\begin{equation}
\langle O(E)\rangle = \dfrac{1}{Z} \int dE\;  O(E) \rho(E) e^{- \beta E}.
\end{equation}

In practice it is not possible to know $\rho(E)$ for every value of $E$ in a continuum system. Therefore, in the simulations, we will divide the energy spectrum in small intervals $[E_0-\Delta,E_0+\Delta]$ of with $2 \Delta$, with $E_0$ the central value of. Moreover, we would like to have an analytic expression for $\rho(E)$ inside each interval so that we can employ a  piecewise definition of the density of states
\begin{equation}
 \widehat{\rho}(E)= \sum_i \rho_i (E),
\end{equation} 
where $\rho_i$ is defined only inside the interval $[E_i-\Delta,E_i+\Delta]$.

In principle, if the intervals are small enough, it is perfectly acceptable to replace a regular function with its linear approximation, up to order $\Delta^2$.   However, we know that \rE\ can vary over many orders of magnitude depending on the range of energies considered. It is more convenient, then, to linearly approximate the natural logarithm of the DOS, in each interval, and then exponentiate the piecewise function to construct \rE.

The piecewise defined logarithm has the form
\begin{equation}
\ln \rho(E)=C(E_i) + a(E_i) (E-E_i)+ \mathcal{O}(\Delta^2), \ \ \ E \in [E_i-\Delta,E_i+\Delta]
\end{equation}
so that the DOS can be written as
\begin{equation}\label{rhoE}
\rho(E)=C_{exp}(E_i) e^{a(E_i) (E-E_i)}+ \mathcal{O}(\Delta^2), \ \ \ E \in [E_i-\Delta,E_i+\Delta].
\end{equation}
The parameter $a(E_i)$, the slope of the linear approximation of $\ln \rho(E)$, is the core focus of this approach and is the only quantity that has to be numerically computed. We shall discuss it in more detail below.  The factor $C(E_i)$ is a constant in each interval, whose function is to set $\ln \rho(E)$ at the right value in each interval.

\subsection{LLR algorithm}\label{LLR algorithm}
We shall now describe the LLR (Langfeld-Lucini-Rago or linear local relaxation) algorithm. It has specifically been created in order to compute the density of states of systems with a continuous energy spectrum.

As mention before, one is interested in computing the quantity $a(E_i)$ in the interval $[E_i-\Delta,E_i+\Delta]$. The way to achieve this is to insert the inverse of \eqref{rhoE} into the partition function, and tune the value of $a(E_i)$ to $a^*(E_i)$ in such a way that the density of states is annihilated in the interval, i.e. 
\begin{equation}\label{rho*e=1}
\rho_{exact}(E) e^{-a^*(E_i)E}=1,
\end{equation}
up to order $\Delta^2$ .
This automatically gives an estimate of the number of states itself 
\begin{equation}\label{rhoE 2}
e^{a^*(E_i) E}=\rho_{est}(E), \ \ \ \ E \in  [E_i-\Delta,E_i+\Delta].
\end{equation}
To carry out this process, we need to introduce a probe function $f(E)$, of which the expectation value can be measured in the interval
\begin{equation}\label{<f(E)>LLR}
F_i(a)=\langle \langle f(E) \rangle \rangle_{i,a}=\dfrac{1}{Z_a(E_i)} \int_{E_i-\Delta}^{E_i+\Delta} dE\; f(E) \rho(E) e^{-a E},
\end{equation}
where $\langle \langle \ \ \rangle \rangle_{i,a}$ means the average value in the interval $E_i$ at fixed parameter $a$, and $Z_a(E_i)$ is the partition function restricted to the same interval which also depends on the parameter $a$
\begin{equation}
Z_a(E_i)=\int_{E_i-\Delta}^{E_i+\Delta} dE\; \rho(E) e^{-a E}.
\end{equation}
Let us note that $Z_a(E_i)$ is just the normalization in the Monte Carlo restricted to the intervals, and has nothing to do with the partition function of the system which is defined in \eqref{Z_beta}.

At this point, the idea is to choose $f(E)$ in such a way that its expectation value is a good indicator to establish when $\rho(E) e^{-a^*(E_i)E}=1$. The simplest and more efficient choice is the energy itself or, more precisely, $f(E)=E-E_i$
\begin{equation}\label{DeltaE(a)}
\delta E(a(E_i))=\langle \langle E-E_i \rangle \rangle_a=\dfrac{1}{Z_a(E_i)} \int_{E_i-\Delta}^{E_i+\Delta} dE\; (E-E_i) \rho(E) e^{-a E}.
\end{equation}

From the point of view of the algorithm, we compute $\delta E(a(E_i))$ with a standard Monte Carlo in each interval $[E_i-\Delta,E_i+\Delta]$. Every time, before proposing a new configuration to the Monte Carlo,  we check if the energy $E_{new}$ of the configuration is inside the interval $[E_i-\Delta,E_i+\Delta]$ and if it is not, we reject the configuration. A slightly better implementation, that is not manifestly non ergodic, is to consider 'gaussian tails' attached to the extremities of each interval. In this way a configuration, which energy is out of the given interval, can still be proposed to the Monte Carlo with a probability of $\exp[E_{new}-(E_i\pm\Delta)]^2$, where the $\pm$ depends whether $E_{new}$ is bigger than $E_i+\Delta$ or lower than $E_i-\Delta$.\\\\
Going back to the description of the the LLR method, one can see that, when \eqref{rho*e=1} is satisfied, equation \eqref{<f(E)>LLR} implies that
\begin{equation}
\delta E(a^*(E_i))=0.
\end{equation}
We are, then, looking for an iterative process that changes \aE\ 
towards the zero of $\delta E(a)$. This is a classic one-dimensional root finding problem, with the complication that the value of our function, the energy, is known as a result of a Monte-Carlo simulation, i.e. it cannot be computed directly but only estimated via noisy observation.

Stochastic approximation methods are algorithms developed precisely for this purpose. The Robbins-Monro algorithm \cite{Robbins-Monro:1951} is the first of these class, and it is a simple adaptation of the famous Newton-Raphson root finding method for deterministic functions. According to the Robbins-Monro method, it is possible to achieve convergence in the root finding of a noisy function. 

The $n+1$ iteration of Robbins-Monro can be written as
\begin{equation}
a_{n+1}=a_n-d_n\ \delta E(a_n),
\end{equation}
where the coefficient $d_n$ must obey the constraints
\begin{equation}\label{c_n constraints}
\begin{split}
& \sum_{n=0}^\infty d_n =\infty, \\
& \sum_{n=0}^\infty d_n^2 < \infty.
\end{split}
\end{equation}
It is easy to see that any series 
\begin{equation}
\dfrac{d}{\sqrt{n}} < d_n\leq \dfrac{d}{n},
\end{equation}
satisfy the requirements \eqref{c_n constraints}, if $d$ is a finite constant. The faster is the decay of $d_n$ the faster the convergence \cite{Sacks:1958}
\begin{equation}
\lim_{n\rightarrow \infty} a_n = a^*.
\end{equation}
For this reason we might as well choose the fastest allowed
\begin{equation}
d_n=\dfrac{d}{n}.
\end{equation}

\begin{figure}[!t]
\begin{center}
\includegraphics[scale=0.45, angle =0]{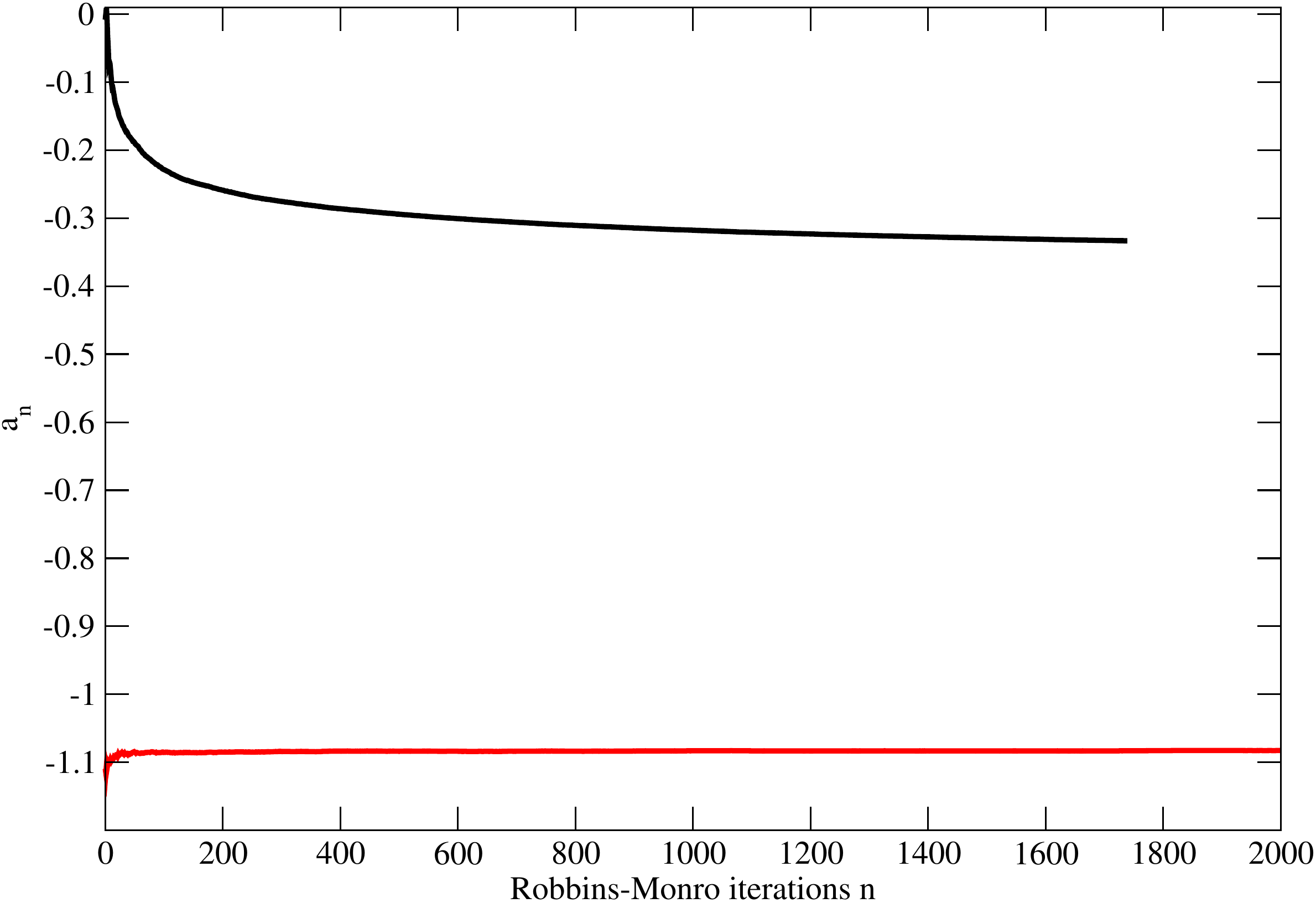}
\end{center}
 \caption{An example demonstrating the effect of the thermalization in the Robbins-Monro root finding method. In both cases the initial value is $a_0=0$. The black line represents the application of Robbins-Monro from the beginning, while the red line has been thermalized with 30 steps of Newton-Raphson.}
 \label{fig:RhoTherm}
\end{figure}

Qualitatively speaking, the coefficient $d_n$ has the function to kill the statistic fluctuations introduced by the noise. Furthermore, it is possible to prove that $\sqrt{n}(a_n-a^*)$ is asymptotically normal with variance
\begin{equation}
\sigma^2_a=\dfrac{d^2 \sigma^2_\chi}{2dF'(a^*)-1},
\end{equation}
where $\sigma^2_\chi$ is the variance of the noise. The optimal value for $d$ can be find simply by minimizing the variance $\frac{\partial^2 \sigma^2_a}{\partial d} =0$, that leads to
\begin{equation}
d=\dfrac{1}{F'(a^*)}.
\end{equation}
In our case, from \eqref{DeltaE(a)}, we found
\begin{equation}
F'(a^*)=\dfrac{\partial}{\partial a} \delta E(a) \vert_{a=a^*}=\dfrac{\Delta^2}{3} ,
\end{equation}
so that the explicit iteration of the Robbins-Monro reads
\begin{equation}
a_{n+1}=a_n-\dfrac{3}{(n+1)\Delta^2} \delta E(a_n).
\end{equation}

Let us conclude this part with an important remark.  Even though the proof of convergence of the Robbins-Monro method is valid for any initial value $a_0$, in practice starting from a random $a_0$ would introduce an offset impossible to eliminate due to the fact that at large $n$ the contribution of $d_n$ is suppressed. The solution of this problem is introducing some steps of \textit{thermalization} for the $a_n$. For example, a simple way we adopted is to start with the Newton-Raphson algorithm, i.e. $d_n=d$. In this case the $a_n$ are rapidly brought around the value $a^*$ since there is no suppression of the coefficient $d_n$. Once the $a_n$ fluctuate around a fixed value then one can switch to the Robbins-Monro method. An example of what we discussed can be seen in Fig.\ref{fig:RhoTherm}. Clearly starting from a value $a_0$ far away from $a^*$ prevents the Robbins-Monro to reach the right result in a sensible time (black line), even thought he correct asymptotic behaviour should still be reached. Let us note that the number of iterations in the Fig.\ref{fig:RhoTherm} is already two thousands. On the other hand, if we start with just 30 steps of Newton-Raphson as thermalization, the Robbins-Monro stabilizes very fast on the asymptotic value of $a^*$ (red line), which yields for the correct results.

Once the $a^*(E_i)$ are known for a range of energies large enough, the density of states can be build and the other observables, depending on $E$, can be easily computed. 

Clearly, this all procedures requires \rE\ to refer to a real weight $e^{-S}$ for \eqref{DeltaE(a)} to be able to be computed with a Monte Carlo. As we shall see in the following section, the way to extend the method to the complex case is to separate the imaginary part of the action from the real one, and compute the DOS relative only to the imaginary part.

\vspace{2cm}
\section{Relativistic Bose gas at finite chemical potential $\mu$}
In this section we are going to discuss the application of the LLR method to the case of a four dimensional complex scalar field theory at finite density, i.e. the Bose gas. A finite real chemical potential introduces a sign problem in this theory. This model has been extensively studied in literature with different approaches like complex Langevin dynamics \cite{Aarts:2008wh,Aarts:2009hn}, Lefschetz thimble \cite{Cristoforetti:2012su}, dual formulation on flux tubes \cite{Endres:2006zh,Endres:2006xu,Gattringer:2012ap} and extended mean field theory \cite{Akerlund:2014mea}. 

Because it is one of the simplest quantum field theories and because of the great amount of results present in the literature, the relativistic Bose gas at finite density is one of the best models to test new algorithms that deal with sign problem.

We will start by recalling the Euclidian action in the continuum
\begin{equation}\label{Scont 2}
S=\int d^4x\;  \left[ (\partial_4 -\mu)\phi^\dag (\partial_4+\mu) \phi + \partial_i \phi^\dag \partial^i \phi +m^2 \vert \phi \vert^2 + \lambda \vert \phi \vert^4 \right] ,
\end{equation}
 from which it is clear how the chemical potential enters as an imaginary temporal component of a gauge field $A_0$. This fact is important when it comes to the discretization on the lattice.

The \eqref{Scont 2} can be written as
\begin{equation}\label{Scont 1}
S=\int d^4x\; \left[ \partial_\nu \phi^\dag \partial^\nu \phi+(m^2-\mu^2)\vert \phi \vert^2 + \lambda \vert \phi \vert^4 +\mu (\phi^\dag (\partial_4 \phi)-(\partial_4 \phi^\dag) \phi) \right] ,
\end{equation} 
 where the chemical potential explicitly couples to the conserved charge $j_4=\phi^\dag \partial_4 \phi-\partial_4 \phi^\dag \phi$  associated to the global U(1) symmetry. Furthermore, it determines a shift in the mass parameter which, for finite $\lambda$ and $\mu$ large enough, is responsible for the symmetry broken phase (Mexican hat potential).

However, the naive discretization of \eqref{Scont 1} would not work because it loses this characteristic of the chemical potential.  Instead, the correct way to discretize the action is to write :
 \begin{equation}\label{Sdiscr1}
 S= \sum \left[ (2d+m^2) \phi_x^* \phi +\lambda (\phi_x^* \phi)^2 - \sum_{ \nu=1 }^4  (\phi_x^*e^{- \mu \delta_{\nu, 4}} \phi_{x+\widehat{\nu}}+\phi_{x+\widehat{\nu}}^* e^{\mu \delta_{\nu, 4}}\phi_x  )
  \right].
 \end{equation}
The sum is over the sites of an $N^3_sN_t$ lattice, with periodic boundary conditions in all the directions, $d=4$ is the number of dimensions, and $m$ is the bare mass. Let us note how the piece $\phi_x^*e^{- \mu} \phi_{x+\widehat{4}}+\phi_{x+\widehat{4}}^* e^{\mu} \phi_x $ is complex. We can rewrite \eqref{Sdiscr1} in terms of the real and imaginary part of the field $\phi=\frac{1}{\sqrt{2}}(\phi_1+i \phi_2)$
\begin{equation}\label{Sdiscr2}
\begin{split}
S=\sum_x &\left[ \dfrac{1}{2}(2d+m^2) \phi_{a,x}^2 + \dfrac{\lambda}{4} \phi_{a,x}^4- \sum_{ i=1 }^3\phi_{a,x}\phi_{a,x+\widehat{i}} \right. \\
&\left. \ \ \ \ \ -\cosh(\mu)\phi_{a,x}\phi_{a,x+\widehat{4}}+i \sinh(\mu) \epsilon_{ab}\phi_{a,x}\phi_{b,x+\widehat{4}} \vphantom{\frac{1}{2}}\right],
\end{split}
\end{equation}
where a summation over repeated indices is intended. The completely antisymmetric tensor $\epsilon_{ab}$ follows the prescriptions $\epsilon_{12}=-\epsilon_{21}=1$ and $\epsilon_{11}=\epsilon_{22}=0$. 

The operator number of particles on the lattice is
\begin{equation}
n_x=\dfrac{d}{d \mu} \ln Z \vert_{\mu=0}\ =\  \left( \delta_{ab} \sinh \mu - i \epsilon_{ab} \cosh \mu \right) \phi_{a,x} \phi_{b,x+\widehat{4}},
\end{equation}
which corresponds to the Noether current at $\mu =0$.

Since here we are mostly interested in the sign problem, we will fix, once and for all, the parameters $m=\lambda=1$, and every result we are going to show will follow the the same prescription.

\subsection{Generalised density of states}
Our goal is to build the density of states for this theory. First of all we will consider separately real and imaginary part of the action
\begin{equation}\label{S_r+ik S_i}
S[\phi]=S_R[\phi,\mu]+ i \kappa(\mu) S_I[\phi],
\end{equation}
where 
\begin{equation}
S_I[\phi]=\sum_x \left[ \phi_{1,x}\phi_{2,x+\widehat{4}}-\phi_{2,x}\phi_{1,x+\widehat{4}} \right].
\end{equation}
Then we will generalise the procedures reviewed in Sec.\ref{sec:Description of the method} in analogy with the Grand Canonical ensemble. Let us note that in \eqref{S_r+ik S_i} $\kappa(\mu)=\sinh(\mu)$ and $S_R[\phi,\mu]$ has a built-in dependence on $\cosh(\mu)$.

\begin{figure}[!t]
\begin{center}
\includegraphics[scale=0.5, angle =0]{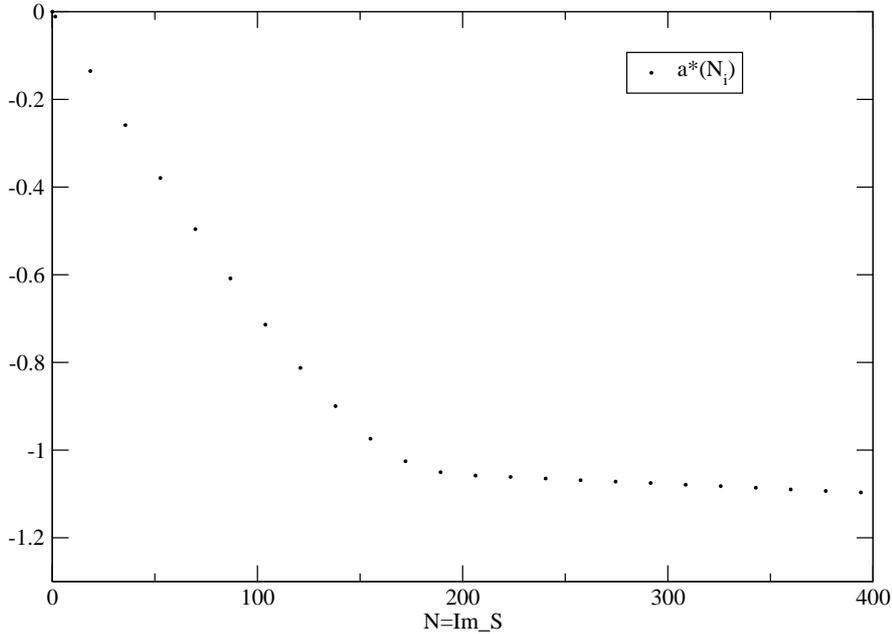}
\end{center}
 \caption{Distribution of the $a^*(N)$ for $L=8^4$ and $\mu = 0.9$.}
 \label{fig:aE_example}
\end{figure}

The idea is to elect $S_I[\phi]=N_I$ as 'imaginary' number of particle so that we can write the partition function
\begin{equation}\label{Z Ni 1}
Z=\sum_{N_I} \int d\phi\; \delta(S_I[\phi]-N_I) e^{-S_R[\phi,\mu]}  e^{ -i \kappa(\mu) S_I[\phi]} .
\end{equation}
It is important to note that, in this formulation, what we called $N_I$ is not the actual number of particles, but just the value of the imaginary part of the action. For this reason it will not take integer values and, in fact, can be any number $N_I \in \mathbb{R}$. Furthermore, $N_I$ is an extensive quantity and therefore can be written as $N_I=n_I \Omega$, where $\Omega$ is the 4 dimensional volume.

 We are, then, able to numerically compute the DOS at fixed $N_I$
\begin{equation}
\rho(N_I;\mu)=\int d\phi\; \delta(S_I[\phi]-N_I) e^{-S_R[\phi,\mu]}.
\end{equation}

Since the imaginary number of particle is a continuum, we can adopt for the DOS,  the same definition as for the Canonical case \eqref{rhoE} with the only difference that now the variable is $N_I$. To ease the notation, from now on we are going to rename $N\equiv N_I$. In complete analogy with the energy case, the logarithm of the DOS is a piecewise function defined in the intervals $N \in [N_i-\Delta,N_i+\Delta]$, where $N_i$ is the centre of the $i$-th interval. 

The partition function is then the Fourier transform of the DOS
\begin{equation}
Z=\int_{-\infty}^{\infty} dN\; \rho(N;\mu) e^{- i \kappa(\mu) N} .
\end{equation}

The LLR algorithm is also fundamentally the same as the canonical case. As before, we want to numerically compute the $a^*(N_i)$ in \eqref{rhoE 2}, and the observable we use in the process is the imaginary part of the action $N=S_I[\phi]$
\begin{equation}\label{DeltaN(a)}
\Delta N(a)=\langle \langle N-N_i \rangle \rangle_a=\dfrac{1}{Z_i} \int_{N_i-\Delta}^{N_i+\Delta} dN\; (N-N_i) \rho(N;\mu) e^{-a N}.
\end{equation}
As explained before, the root finding process for $a^*(N_i)$ is implemented separately in each interval  $[N_i-\Delta,N_i+\Delta]$.

Let us remark that under charge conjugation the imaginary part of the action changes sign $S_I[\phi]\rightarrow - S_I[\phi^\dag]$ while the real part remains the same, for this reason  the density of states has to be \textit{even} in $N$
\begin{equation}
\rho(N;\mu)=\rho(-N;\mu).
\end{equation}
For the same argument $a(N)$ has to be \textit{odd}
\begin{equation}
a(N)=\frac{d}{dN} \ln \rho(N;\mu)=-a(-N),
\end{equation}
which means that we really need only informations from $N\in [0,\infty]$, and the partition function can be written as
\begin{equation}\label{Z cos(k mu)}
Z=2 \int_{0}^{\infty} dN\; \rho(N;\mu) \cos(\kappa(\mu) N) .
\end{equation}

 In Fig.\ref{fig:aE_example} we show an example of the typical distribution of
\begin{equation}\label{a(N)=d/dN ln}
 a^*(N_i)=\frac{d}{dN} \ln \rho(N;\mu)\vert_{N=N_i}, \ \ \ \mu=0.9,
 \end{equation} 
for the Bose gas. We observe an initial approximately linear slope, which means the dominant contribution to $\rho(N;\mu)$ is gaussian, as we can see integrating \eqref{a(N)=d/dN ln}. Such gaussian behaviour for a generalised DOS has been already observed for the $\mathbb{Z}_3$ model with complex action in \cite{Langfeld:2014nta,Lucini:2014wga}. Here, though,  we can see the $a^*(N_i)$ suddenly change trend at a larger value of $N$, and stabilize to a less steep slope.

\begin{figure}[!t]
\begin{center}
\includegraphics[scale=0.5, angle =0]{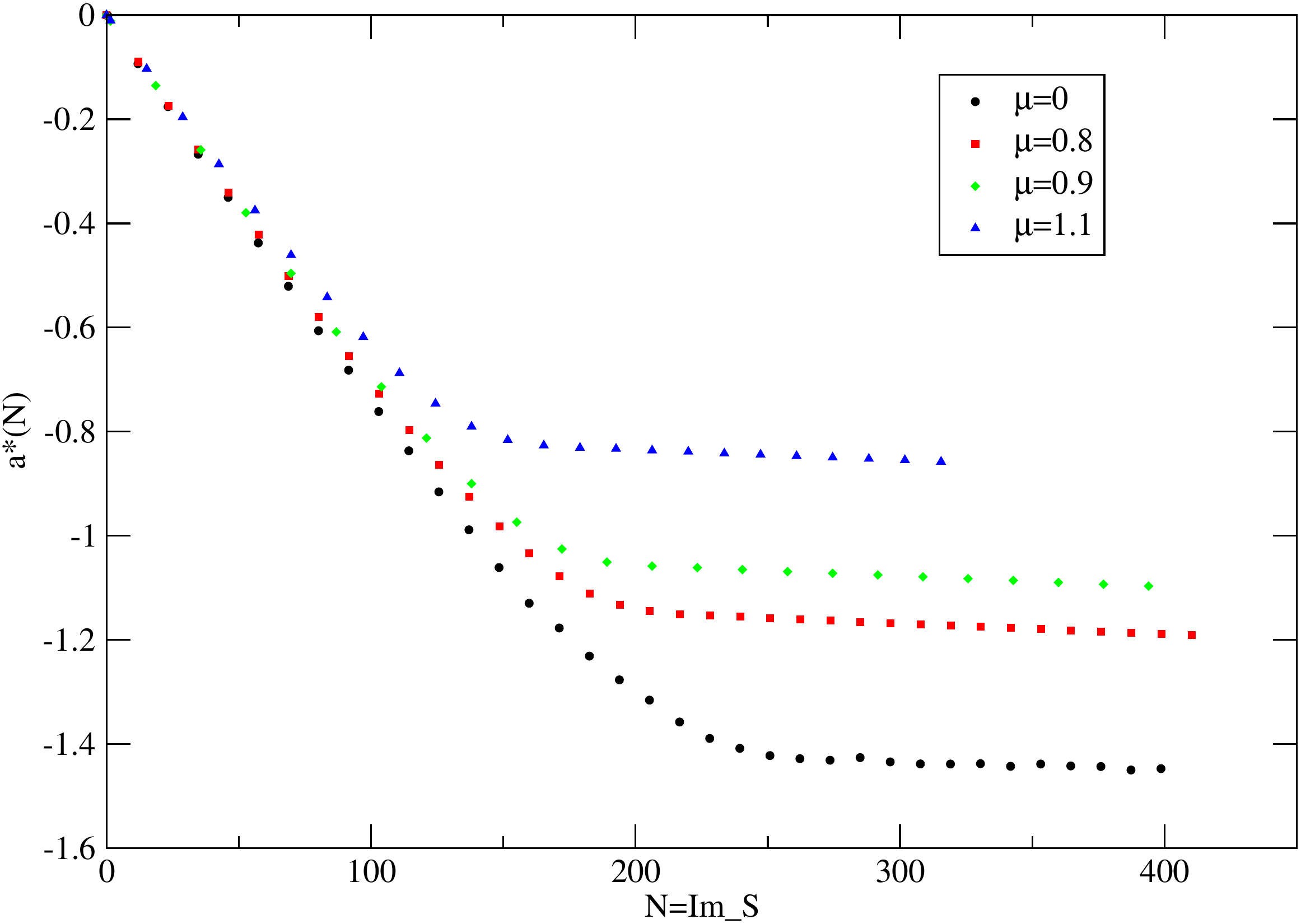}
\end{center}
 \caption{Dependence of $a^*(N)$ on the chemical potential $\mu$ for $L=8^4$.}
 \label{fig:aE_variousMu}
\end{figure}

This behaviour of $a^*(N)$ is quite general, and appears the same at different values of $\mu$ and volumes. In fact, it is useful to look, in Fig.\ref{fig:aE_variousMu}, at the dependence of $a^*(N)$ from the chemical potential to gain some insight on the severity of the sign problem. We can see how the initial linear behaviour is common at every chemical potential, but the higher is the value of $\mu$ the earlier they will depart from it to settle on the other slope. The latter has, again, the same inclination at every $\mu$ and it just differs only for the shift.

The behaviour in Fig.\ref{fig:aE_variousMu}, together with the obvious increase in frequency of the $\cos(\kappa(\mu) N)$ in \eqref{Z cos(k mu)}, contribute to the severity of the sign problem. In fact, let us suppose for a moment that the oscillation was constant at every $\mu$
\begin{equation}
   Z=\int dN \rho(N;\mu) \cos(\omega N),
 \end{equation}  
with
\begin{equation}
\dfrac{d}{d \mu} \omega =0.
\end{equation}
The DOS at zero chemical potential  $\rho(N;\mu=0)$, generated by the black dots in  Fig.\ref{fig:aE_variousMu}, would suppress the  oscillation $\cos(\omega N)$ more than $\rho(N;\mu=1.1)$ (generated by the blue triangles), because it keeps decaying with the fastest slope even after the one at $\mu=1.1$ is settled to the slower slope. Consequently $Z(\mu=0)$ would be bigger than  $Z(\mu=1.1)$ even if the oscillation was not increasing with the chemical potential, which means the dependence of the DOS from the chemical potential alone favours the sign problem.
 
In our case, furthermore, the $ \cos(\kappa(\mu) N)$ oscillates faster and faster increasing $\mu$, which obviously results in even more severe cancellations.

\vspace{2cm}
\section{The partition function from highly oscillating integral}
In this last section we are going to discuss how extract the value of the partition function knowing $\rho(N;\mu)$. At first sight, it would seem that \eqref{Z cos(k mu)} can be easily computed just by using some algorithm that operates a Fourier transform (FT). However this is, unfortunately, not the case because of the statistical noise imprinted in the DOS.  The argument for this can be formulated as follows.

We already know that our DOS is a fast decaying function, which is both in $L^1(\mathbb{R})$ and $L^2(\mathbb{R})$, since its main contribution is gaussian. Let us assume that we can consider $\rho(N;\mu)$ to be the sum of the 'true' distribution plus noise. We know (Plancherel theorem) that the Fourier transform $\widehat{f}$, of an $L^1(\mathbb{R}) \cap L^2(\mathbb{R})$ function $f$, is again  $L^2(\mathbb{R})$, i.e. $\widehat{f}$ is a fast decaying function in our frequency $\kappa=\sinh(\mu)$. On the other hand, the FT of the noise is again noise in $\kappa$, which will soon be dominant over the fast decaying signal, and will completely spoil the computation of the partition function. Unfortunately, this happens at relatively small values of $\mu$ so that prevents the observation of interesting phenomena like the Silver Blaze.

A way out seems to be fitting the $a^*(N)$. The only requirement is that the fitting function has to be $C^\infty$. In fact, using a function which presents an arbitrary discontinuity in the derivative would spoil the dependence of the partition function $Z$ from the chemical potential $\mu$. More precisely, a discontinuity in the $m$-th derivative of a function $f(x)$, dominates the FT:$f(x)=\widehat{f}(k)\sim k^{-m}$ with a polynomial decay.

\begin{figure}[!t]
\begin{center}
\includegraphics[scale=0.45, angle =0]{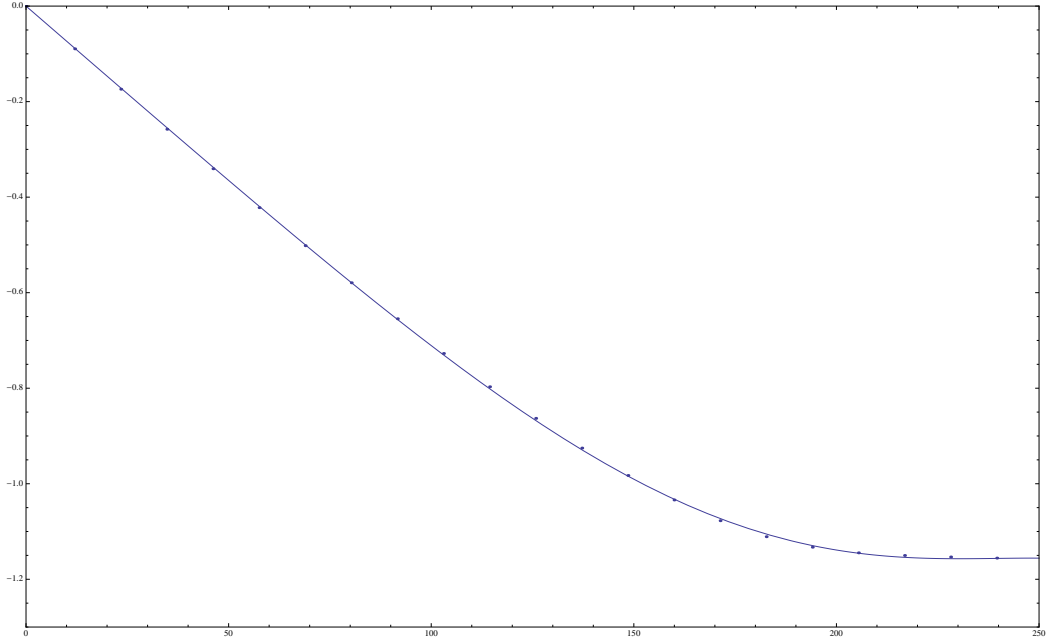}
\end{center}
 \caption{Fit of $a^*(N)$ with an odd polynomial of order 7 $a(N)=c_1 N+c_3 N^3+c_5 N^5+c_7 N^7$, for $\mu=0.8$ and $L=8^4$.}
 \label{fig:FitPol}
\end{figure}

A good choice for the fitting functions are the polynomials :
\begin{equation}
a^*(N)=\dfrac{d}{dN} \ln \rho(N;\mu) \simeq \sum_{i=0}^m c_{2i+1} N^{2i+1},
\end{equation}
where only odd powers are considered since we know that $a^*(N)$ is odd in $N$. We can see in Fig.\ref{fig:FitPol} that a polynomial of order seven $p(x)=a x+ b x^3 +c x^5 +d x^7$, is already good enough to fit the data very well. In practise, we use the Bayesian evidence for model comparison to determine the best polynomial fit that avoids over-fitting \cite{Spiegelhalter:2002,Johnson:2005,Bailer-Jones:2013}. 

Proceeding like this, we are able to shift the noise in the coefficients $c_{2i+1}$ in such a way that it does not affect the Fourier transform of
\begin{equation}\label{RhoPol}
\rho(N;\mu) = \exp\left( \sum_{i=0}^m \dfrac{c_{2i+1}(\mu)}{2i+2} N^{2i+2} \right),
\end{equation}
 which is, in fact, an $L^2(\mathbb{R})$ function. The noise appears in the result using some error propagating method (like Bootstrap or Jackknife).
 
 So, after the fit, we are left with the partition function 
 \begin{equation}\label{Zpol}
 Z=2 \int_{0}^{\infty} dN\; \exp\left( \sum_{i=0}^m \dfrac{c_{2i+1}(\mu)}{2i+2} N^{2i+2} \right) \cos(\kappa(\mu) N),
 \end{equation}
which is an highly oscillating integral. 

Problems of this kind are treated in mathematical literature (see \cite{IserlesBook:2006,Iserles:2005}), especially when the integral is restricted in an interval. In our specific case, we would like to adopt and confront two different methods :
\begin{itemize}
\item multi-precision numerical integration
\item Lefschetz thimbles method.
\end{itemize}
The first method is direct and consists in computing the FT in \eqref{Zpol} numerically. However, due to the cancellation in the integral, one has to be able to know the value of the function at any given point with a precision at least of the same order of magnitude of $Z$. The problem here is that not only the cancellations increase with $\mu$, but they become also exponentially more severe with the volume. The classic argument for this is to look at the average phase factor 
\begin{equation}
\langle e^{i \phi} \rangle = \dfrac{Z(\mu)}{Z_{pq}} \sim \exp\left( - V \dfrac{f}{T}\right),
\end{equation}
where $f$ is the free energy density and $Z_{pq}$ is the phase quenched partition function
\begin{equation}
\begin{split}
& weight = \vert weight \vert\; e^{i \phi} \\
& Z_{pq} = \int dN\; \vert weight \vert \ .
\end{split}
\end{equation}

If the precision we need to computing the integral \eqref{Zpol} numerically grows exponentially with the volume of our lattice, it means that  multi-precision will eventually not be enough. To give an idea, already at volume $8^4$ and $\mu=1$ the precision required is more or less 50 digits.

This method, then, do not represent an efficient solution to our problem per se, but it will be very useful to be compared with the other two at small volumes.

\subsection{Lefschetz thimble}

\begin{figure}[!t]
\begin{center}
\includegraphics[scale=1.4, angle =0]{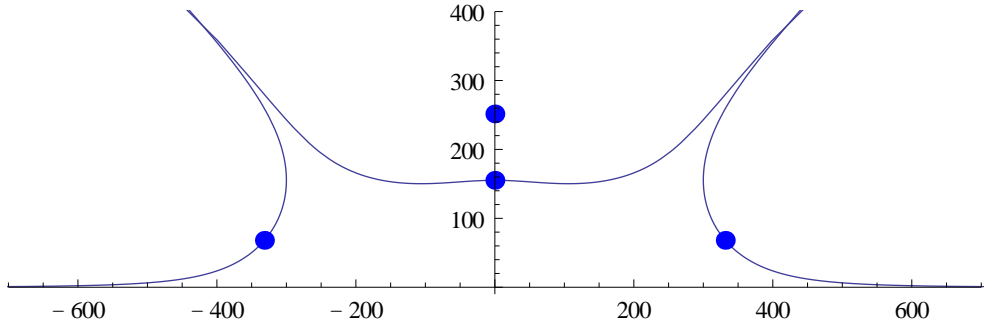}
\end{center}
 \caption{Contributing Lefschetz thimbles for order seven polynomial fit of the $a^*(N)$.}
 \label{fig:Lefschetz_thimble}
\end{figure}
Another very successful way to deal with oscillating integral is the Lefschetz thimbles method. We already described it in detail in Cap.\ref{cap:Lefschetz}, and here we will recall just the very main steps. 

After having performed the polynomial fit and, therefore, having an explicit form for \eqref{RhoPol}, we can define the action-like function
\begin{equation}
S[N]=-\sum_{i=0}^m \dfrac{c_{2i+1}(\mu)}{2i+2} N^{2i+2}+i \kappa(\mu)N,
\end{equation}
including in it the oscillating term. We proceed by complexify the variable $N\rightarrow z=x+iy$, and we look for the critical points of $S[z]$, i.e. $\lbrace z_\sigma \in \mathbb{C} : \partial_zS[z]\vert_{z=z_\sigma}=0   \rbrace$. Starting from each one of those critical points $z_\sigma$ we can build a Lefschetz thimble $J_\sigma$ with equations
\begin{equation}
\begin{split}
&\dot{x}=-\text{Re}\left(\partial_zS[z]\right),\\
&\dot{y}=+\text{Im}\left(\partial_zS[z]\right),
\end{split}
\end{equation}
where $t$ is the parametrization of the thimbles.  In this case, they are one-dimensional trajectories in the complex plane, along which the imaginary part of the action Im$S[z]$ is constant. In the end, we are able to write the partition function as the sum of integrals over the thimbles
\begin{equation}
Z=\sum_\sigma m_\sigma e^{-i \text{Im}S[z_\sigma]}\int_{J_\sigma} \frac{\partial z}{\partial t} dt\; e^{- \text{Re}S[z(t)]}\ ,
\end{equation}
where $m_\sigma$ the intersection number that tell us if the thimble contributes or not. The union of all contributing thimbles forms a deformation of the original real axis. In Fig.\ref{fig:Lefschetz_thimble} we show a typical example of new domain of integration composed by the contributing Lefschetz thimbles in the complex plane.

\begin{figure}[!t]
\begin{center}
\includegraphics[scale=0.55, angle =0]{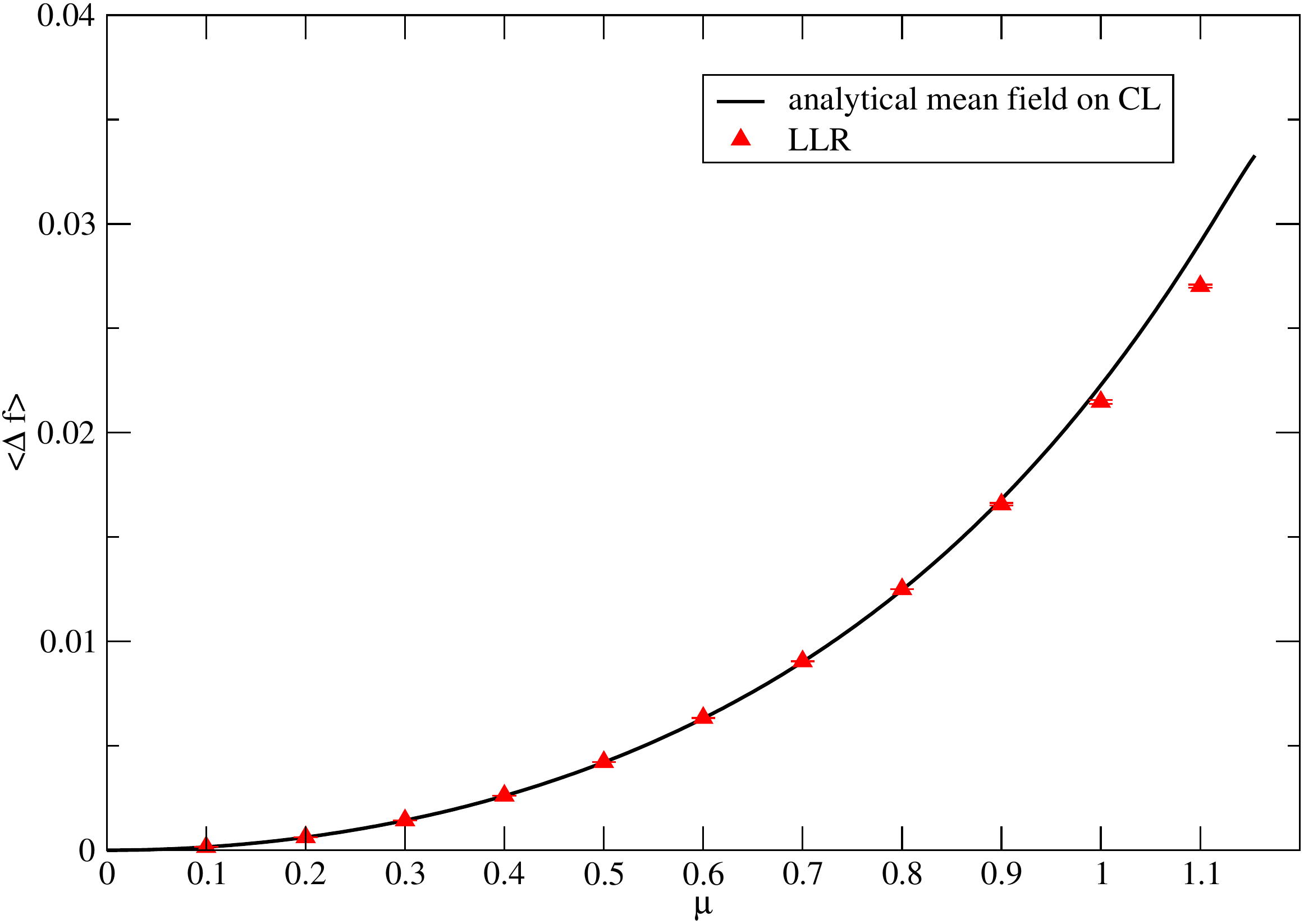}
\end{center}
\vspace{-0.7cm}
 \caption{Plots of the average free energy difference $ \langle \Delta f \rangle = -\frac{T}{V} \log \frac{Z(\mu)}{Z_{pq}}$. Comparison between LLR method for density of states (red dots) and analytical mean field theory result applied to complex Langevin dynamics (black line). Simulation on a $8^4$ lattice with the choice of parameters : $m=\lambda=1$.}
 \label{fig:LogPhase8}
\end{figure}

Along this new contour we are left with a global phase $e^{-i \text{Im}S[z_\sigma]}$, which is usually harmless even though it is different on each thimble and, therefore, has to be carefully taken into account. Together with the global phase, there is a residual sign problem coming from the Jacobian $\frac{\partial z}{\partial t}$. The latter, in general, is much milder than the original sign problem and is seldom a problem and also in our case it does not create any problem to the numerical integration along the thimbles.\\

\begin{figure}[!t]
\begin{center}
\includegraphics[scale=0.45, angle =0]{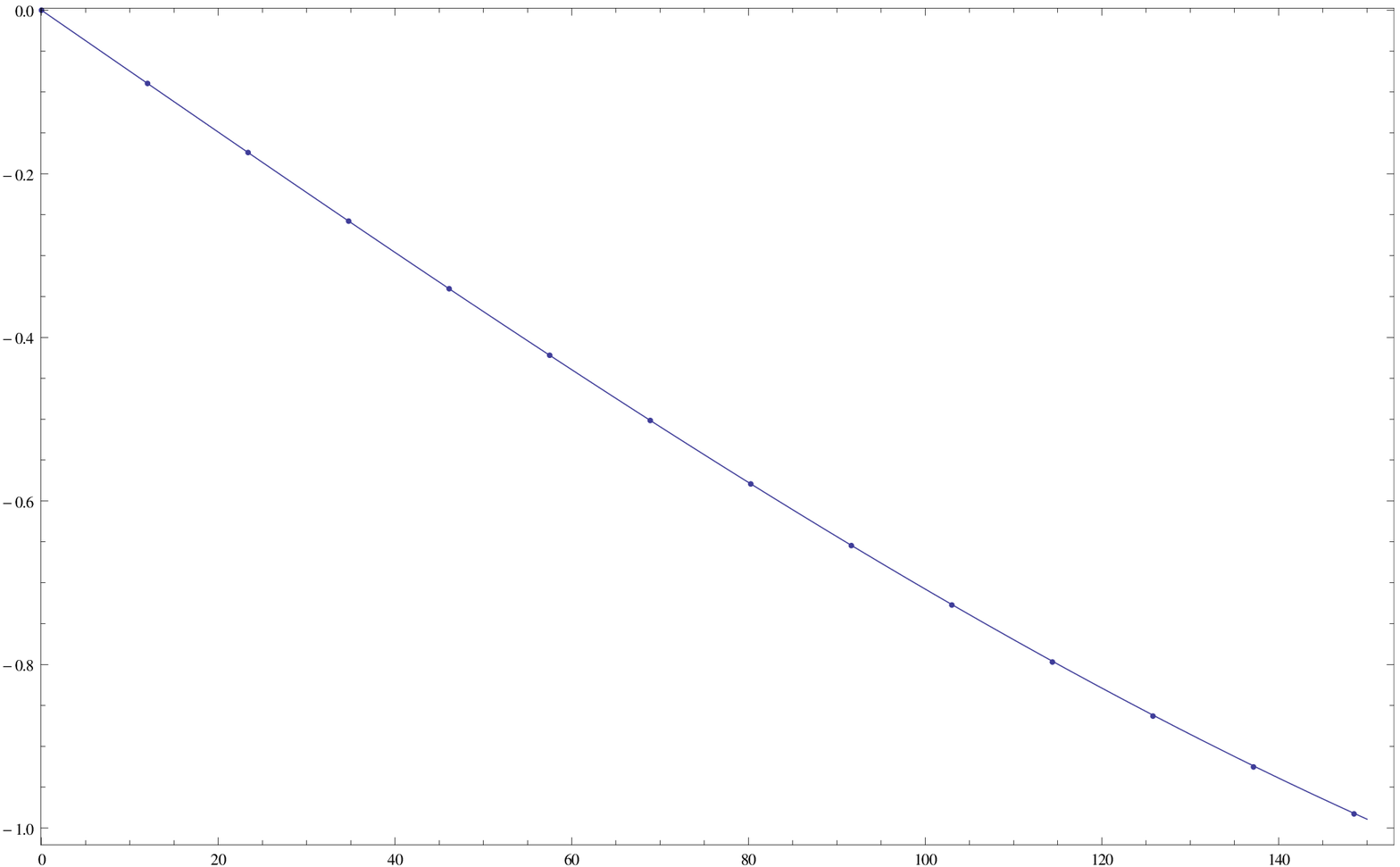}
\end{center}
\begin{center}
\includegraphics[scale=0.4, angle =0]{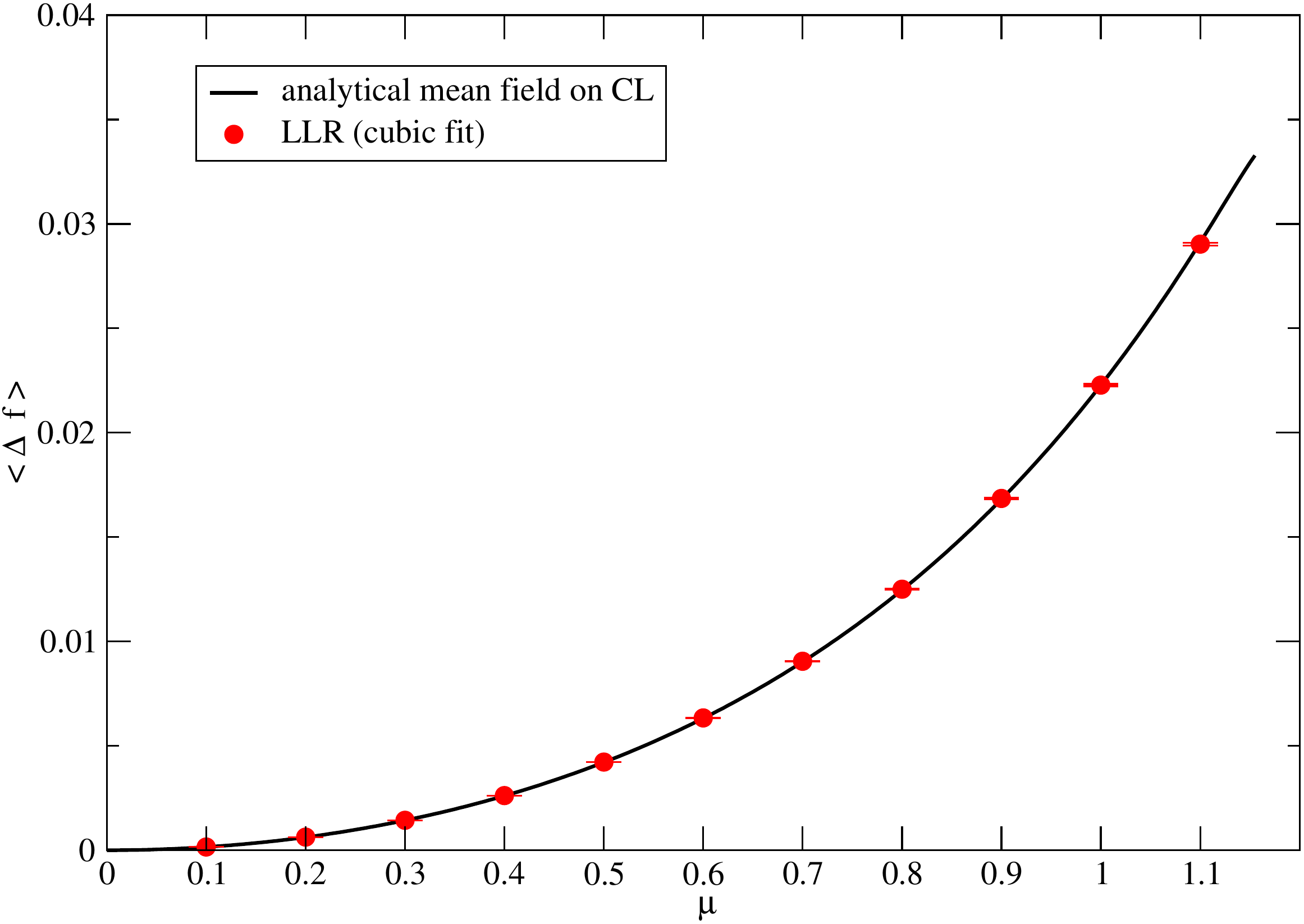}
\end{center}
\vspace{-0.7cm}
 \caption{\textit{up} : equivalent of Fig.\ref{fig:FitPol} but with a cubic fit of the data, $a(N)=c_1 N+c_3 N^3$, up until the change in slope, which in this case ($\mu=0.8$) happens around $N\simeq 150$. \textit{down} : same as Fig.\ref{fig:LogPhase8} where, this time, the DOS has been generated by the cubic fit of the $a(N)$ shown above.}
 \label{fig:LogPhase8_cubucFit}
\end{figure}

The results obtained with multi-precision and with Lefschetz thimble agree with each other. In Fig.\ref{fig:LogPhase8} we show the average free energy density difference between full and phase quenched theory $ \langle \Delta f \rangle = -\frac{T}{V} \log \frac{Z(\mu)}{Z_{pq}}$, for the  $8^4$ lattice. We confront our results (red triangles) with the analytical computation (black line) of mean field complex Langevin dynamics \cite{Aarts:2008wh}, which the author shared with us. We notice a small discrepancy starting from $\mu=1$. An explanation for that could be a systematic error introduced by the fit in our method or an inaccuracy of the mean field theory at high $\mu$ or both together. Along this line of thoughts, an interesting observation is that if we fit the data only up until the change in slope (as in Fig.\ref{fig:LogPhase8_cubucFit} up)  with a cubic polynomial $a(N)=c_1 N+c_3 N^3$, then we get values of $\langle \Delta F \rangle$ compatible with the mean field expected value (Fig.\ref{fig:LogPhase8_cubucFit} down).

We would, also, like to stress that the typical value of the partition function for the last values of $\mu$ is of the order of the $10^{-30}-10^{-40}$. That means the LLR method for computing the DOS is able to achieve a remarkable precision.

\vspace{2cm}
In conclusion, in this Chapter we described the density of states method, based on the LLR algorithm, as a candidate for the study of theories affected by the sign problem. We showed how this allows to reduce a complicated four-dimensional theory in a one dimensional integral. The oscillating integral, though, is still very problematic even in one dimension, and forces us to implement a fit of the $a^*(N)$ to get rid of the noise. This inevitably introduces a systematic error which we try to control with, for example, Bayesian methods. 

Lastly we showed some results for the average of the logarithm of the phase. We found good agreement with a mean field calculation up to $\mu=1$ where the two methods give slightly different results.

We also found that the DOS method is capable of extremely precise direct measurements as, for example, partition functions of the order of $10^{-40}$ or smaller.

\begin{comment}

%% file: cap_Conclusions/Conclusions.tex
\end{comment}

\externaldocument{cap_THETA/ThetaTerm.tex}
\addcontentsline{toc}{chapter}{\protect\numberline{}Conclusions}
\chapter*{Conclusions}
The topic of this thesis is centred on the sign problem in lattice field theory. In particular, our efforts have been concentrated on three different approaches, namely complex Langevin dynamics, Lefschetz thimble and density of states, and their application to various theories and toy models affected by the sign problem.

Particular focus has been put on complex Langevin dynamics. We dedicate Cap.\ref{cap:Langevin} to review stochastic quantization as an alternative to the usual path integral formulation of the QFT, and to discuss the formal proofs of correctness for complex Langevin dynamics. We also introduced the concept of gauge cooling as a method to control CL dynamics into the complexified space. We employed gauge cooling in the one dimensional Polyakov chain model, where the sign problem is triggered by a complex $\beta$ in the action. We showed that, indeed, gauge cooling is able to control the dynamics in the complex directions leading CL to converge to the right results even where, without gauge cooling, it would have failed. 

After that, we moved on to applying complex Langevin dynamics to the Yang-Mills pure gauge theory with a topological $\theta$-term. This probably represents the most challenging chapter (Cap.\ref{cap:thetaTerm}) of the thesis. We showed that the use of CL dynamics, helped by the gauge cooling technique, allows us to obtain results at real $\theta$, at least for the bare (unrenormalised) theory. We put particular care in testing the criteria of correctness to ensure the results at real $\theta$ are reliable. We produced, at real $\theta$, the behaviour of the bare topological charge $Q_{bare}(\theta)$, expected by analytical continuation from imaginary $\theta$ where the action is real and Monte Carlo methods can be employed. Lastly, we used this information to study the dependence of the bare topological susceptibility $\chi_{top}$ from $\theta$.

Another very interesting method developed in order to deal with the sign problem is the one based on the Lefschetz thimbles. The starting point of this method, as well as complex Langevin dynamics, is the complexification of the fields in the theory. Moreover, the equations of the thimbles are the complex conjugate of the classical, i.e. without noise, CL equations. A comparison which enhances the similarity and the differences between those two approaches is, therefore, very interesting. In Cap.\ref{cap:Lefschetz}, we compare CL dynamics and Lefschetz thimble in the study of a quartic model, U(1) one-link model with a complex determinant, and SU(2) non abelian one-link model with an imaginary $\beta$ parameter in the action. Here we showed the quartic model mostly as an introduction on the Lefschetz thimbles method. The second model is interesting because the determinant can be included as a non-holomorphic term in the action, in analogy with what happens in QCD at finite density. The third model presents similarities with gauge theories in real time dynamics, i.e. imaginary $\beta$. From the CL point of view, it can be tested both with complete gauge fixing and gauge cooling. Furthermore, the gauge fixed representation produces the Haar measure, which also can be included as a non holomorphic term in the action. This study led us to observe directly the connection between CL and Lefschetz thimbles and, furthermore, we could establish the role of the singularities in the thimbles approach, i.e. they are always the end point of a thimble.

The last approach to the sign problem we have dealt with is the density of states. We reviewed how, in this method, one numerically computes the canonical density of states relative to the action, simplifying the quantum field theory to a one dimensional problem.  We showed that, when the action is complex, one can regard the imaginary part of the action Im$(S)=N$ as the grand-canonical variable.  It is possible, then, to build a positive generalized density of states $\rho(N)$ by using only the real part of the action Re$(S)$. In the end, we are left with a one dimensional oscillating integral in $N$. We applied the generalized DOS method to the relativistic Bose gas theory at finite density. We considered two different ansatz for the fit of the DOS  and, in both cases, we showed our computation of  free energy density as a function of the chemical potential $\mu$.

\vspace{1cm}
In this thesis we described in detail three approaches to the sign problem. Of those three, Lefschetz thimbles and density of states method are somehow more recent and still in a testing phase. Which means promising results have been achieved in several toy models and simple field theories, like the Bose gas at finite density, but the step to full gauge theories and QCD has yet to be done. Complex Langevin dynamics, on the other hand, has already completed the testing part, developing formal criteria of correctness which guides the trustworthy of its results. Furthermore, its application to gauge theories with a sign problem has already started in the last couple of years, obtaining promising results even for full QCD. The most recent efforts are mainly aimed to deal with non holomorphic drifts and to achieve full control of CL dynamics in the whole space of the parameter, using gauge cooling or some alternative method. The successful completion of both this challenges would elect CL dynamics as the first consistent method to quantitatively study the phase diagram of any gauge theory affected by the sign problem, in particular QCD at finite density.


%% file: Bibliography/Bibliography.tex
\end{comment}
